\documentclass[prb,twocolumn,superscriptaddress,floatfix,showpacs]{revtex4}
\usepackage{graphicx,amsfonts,amssymb,amsmath, hyperref, enumerate,multirow,rotating}

\newif\ifhyper
\hypertrue
\ifhyper
\hypersetup{
   citecolor = {green},
   colorlinks = {true}, 
   urlcolor = {blue} 
}
\fi

\newcommand{\beq}{\begin{equation}}
\newcommand{\eeq}{\end{equation}}
\newcommand{\beqa}{\begin{eqnarray}}
\newcommand{\eeqa}{\end{eqnarray}}
\newcommand{\ket} [1] {\vert #1 \rangle}
\newcommand{\bra} [1] {\langle #1 \vert}
\newcommand{\braket}[2]{\langle #1 | #2 \rangle}

\def\bra#1{\langle#1\vert}
\def\ket#1{\vert#1\rangle}

\def\Longarrow{\protect\@lra}
\def\@lra{\relbar\joinrel\relbar\joinrel\relbar\joinrel%
          \relbar\joinrel\rightarrow}

\begin{document}

\title{Exploring corner transfer matrices and corner tensors \\ for the classical simulation of quantum lattice systems}

\author{Rom\'an Or\'us}
\affiliation{Max-Planck-Institut f\"ur Quantenoptik, Hans-Kopfermann-Str. 1, 85748
Garching, Germany}

\begin{abstract}

In this paper we explore the practical use of the corner transfer matrix and its higher-dimensional generalization, the corner tensor, to develop tensor network algorithms for the classical simulation of quantum lattice systems of infinite size. This exploration is done mainly in one and two spatial dimensions ($1d$ and $2d$). We describe a number of  numerical algorithms based on corner matrices and tensors to approximate different ground state properties of these systems. The proposed methods make also use of matrix product operators and projected entangled pair operators, and naturally preserve spatial symmetries of the system such as translation invariance. In order to assess the validity of our algorithms, we provide preliminary benchmarking calculations for the spin-1/2 quantum Ising model in a transverse field in both $1d$ and $2d$. Our methods are a plausible alternative to other well-established tensor network approaches such as iDMRG and iTEBD in $1d$, and iPEPS and TERG in $2d$. The computational complexity of the proposed algorithms is also considered and, in $2d$, important differences are found depending on the chosen simulation scheme. We also discuss further possibilities, such as $3d$ quantum lattice systems, periodic boundary conditions, and real time evolution. This discussion leads us to reinterpret the standard iTEBD and iPEPS algorithms in terms of corner transfer matrices and corner tensors. Our paper also offers a perspective on many properties of the corner transfer matrix and its higher-dimensional generalizations in the light of novel tensor network methods. 

\end{abstract}

\maketitle

\section{Introduction}

It is commonly accepted that understanding the properties of quantum systems of many particles is one of the most important and challenging problems in condensed matter physics. In this sense, much remains yet to be understood. For instance, and in spite of many efforts, it is still unclear what is the mechanism responsible for high-$T_c$ superconductivity in cuprates \cite{highTcCup}, not to mention iron-based \cite{iron} and organic \cite{org} superconductors, for which this mechanism is also a mistery to a great extent. Many other condensed matter phenomena beyond Landau's paradigm of phase transitions have also proven non-trivial to understand. In this respect, a great interest in \emph{exotic} (i.e. beyond Landau's paradigm) phases of matter has arised recently. Examples of this are, to name a few, topologically ordered phases (where a pattern of long-range entanglement prevades over the whole system) \cite{topo}, quantum spin liquids (phases of matter that do not break any symmetry) \cite{spinliq}, and deconfined quantum criticality (quantum critical points between phases of fundamentally-different symmetries) \cite{deconf1, deconf2}.

The standard approach to understand these systems is based on proposing some simplified, physical model that is believed to mimic the relevant interactions responsible for the observed physics. This is the case of e.g. the Hubbard and $t-J$  models for high-$T_c$ cuprate superconductors, as well as quantum Heisenberg antiferromagnets with frustrating interactions for some magnetic materials with a spin liquid ground state \cite{hubbardtJ}. Sometimes we are lucky, and these models are exactly solvable. In practice, this means that one can compute some (if not all) relevant properties analitically. But this is not usually the case and, in spite of their apparent simplicity, these models turn out to have such an outstandingly complex behavior that one needs to rely on alternative approaches. Quantum simulations, as proposed by Feyman \cite{quantumsim}, are certainly a possibility. Recent experimental results in this direction using e.g. ultracold atoms in optical lattices are in fact really promising \cite{optlat}. However, the current technological status does not allow yet to fully understand many interesting systems. Thus, one needs to rely on faithful methods to implement numerical (classical) simulations. 

From the point of view of numerical simulation algorithms, there has been increasing interest in recent years in the so-called \emph{tensor network methods} to simulate strongly correlated systems \cite{tn}. In these methods the wave function of the system is described in terms of a network of interconnected tensors (a tensor network). As such, tensor network techniques offer efficient descriptions of quantum many-body states of the system that are based on the amount of entanglement in the wave function. The amount and structure of entanglement is a consequence of the chosen network pattern and the number of parameters in the tensors. The most famous example of a tensor network method is probably the Density Matrix Renormalization Group (DMRG) \cite{dmrg1,dmrg2, dmrg3, dmrg4}, introduced by White in 1992, and which has been the technique of reference for the last 20 years to compute ground state properties of $1d$ quantum lattice systems. Recently, though, many important insights coming from quantum information science have motivated the appearance of a host of new tensor network methods. Nowdays it is easy to get lost in the zoo of names for all these methods, e.g. Time-Evolving Block Decimation (TEBD) \cite{tebd}, Folding Algorithms \cite{folding}, Projected Entagled Pair States (PEPS) \cite{PEPS}, Tensor-Entanglement Renormalization Group (TERG) \cite{TERG}, Tensor Product Variational Approach \cite{TPVA}, Weighted Graph States \cite{weight}, Entanglement Renormalization (ER) \cite{ER1, ER2}, String-Bond States \cite{StringBond1, StringBond2}, Entangled-Plaquette States \cite{EntangPlaq}, Monte Carlo Matrix Product States \cite{MCMPS}, Tree Tensor Networks \cite{TTN1, TTN2, TTN3, TTN4} and Continuous Matrix Product States \cite{CMPS}, just to name some of the most recent ones. Each particular method has its own advantages and disadvantages, as well as optimal range of applicability.  

Tensor network methods are also an interesting approach to many-body systems since they offer a lot of flexibility. For instance, with tensor networks one can study a variety of systems in different dimensions, of finite or infinite size \cite{iDMRG1, iDMRG2, iTEBD1, iTEBD2, iPEPS,TERG,VDMA1, VDMA2}, with different boundary conditions \cite{pbc1, pbc2}, symmetries \cite{sym1, sym2, sym3, sym4, sym5, sym6, sym7, sym8, sym9}, as well as systems of bosons \cite{boson1, boson2}, fermions \cite{ferm1, ferm2, ferm3, ferm4, ferm5, ferm6, ferm7, ferm8} and frustrated spins \cite{frusTN1, frusTN2, frusTN3, frusTN4}. Different types of phase transitions \cite{1stOrd} as well as the robustness of topological order to local perturbations \cite{RobustTopo1, RobustTopo2, RobustTopo3, RobustTopo4} have also been studied in this context. The fact that it is possible to develop algorithms for infinite-size systems is quite relevant, since this means that it is possible to approximate the properties of a given system directly in the thermodynamic limit and without having to rely on finite-size scaling extrapolations. This is achieved by cleverly exploiting the translational invariance of the system. Examples of methods using this approach are iDMRG \cite{iDMRG1, iDMRG2} and iTEBD \cite{iTEBD1, iTEBD2} in $1d$ (the 'i' means \emph{infinite}), as well as iPEPS \cite{iPEPS,iPEPS2} and TERG \cite{TERG} in $2d$. 

The possible variations of all these methods are, in practice, unending. In this respect, we feel that a useful tool still relatively unexplored in this context (albeit with exceptions) is the so-called Corner Transfer Matrix (CTM) \cite{CTM1, CTM2}. This was originally introduced by Baxter in 1968 in the context of classical statistical mechanics and exactly solvable models \cite{CTM681, CTM682}. The CTM has very nice properties, specially regarding its spectrum of eigenvalues, and has been a standard tool to find the exact solution of many classical $2d$ models such as the hard-hexagon and related models \cite{CTM1, CTM2}. However, its practical use goes well beyond analytical solutions, and numerical algorithms can be developed to approximate the properties of $2d$ classical lattice models based on clever truncations in the eigenvalue spectrum of CTMs. Baxter himself was the first to explore this possibility by means of a variational method \cite{CTM681, CTM682} that was an extension of the so-called Krammers-Wannier approximation \cite{KW}. In fact, Baxter's method can be understood as a precursor of DMRG but in the context of classical lattice models (this statement will be made more precise later on). The formal combination of CTMs and DMRG was later on put forward by Nishino and Okunishi in the so-called Corner Transfer Matrix Renormalzation Group (CTMRG) \cite{CTMRG1, CTMRG2}. Since then, numerical algorithms using CTMs have been widely used, mostly focusing on classical lattice models. From the point of view of quantum lattice systems, Ref.\cite{CTMRG1, CTMRG2} already discussed the possibility of simulating $1d$ quantum systems by using CTMs and a Suzuki-Trotter decomposition of the evolution operator \cite{suztrot1, suztrot2}. Nevertheless, we believe that CTMs have not yet been fully exploited as a tool in the context of the novel tensor network methods that are being developed for quantum lattice systems. 

Our aim in this paper is to cover in part this gap by exploring the applicability of the CTM to develop algorithms for the simulation of quantum lattice systems. We do this mainly in $1d$ and $2d$, which naturally leads us to consider the generalization of the CTM to higher dimensional systems. Following the convention from previous works \cite{CTM681, CTM682, CTM3d}, we call this generalization \emph{corner tensor}. More specifically,  in this paper we describe a number of numerical algorithms based on CTMs and corner tensors to approximate ground state properties of quantum lattice systems of infinite size. The methods that we present rely also on the use of Matrix Product Operators in $1d$ and Projected Entangled Pair Operators in $2d$, and naturally preserve the spatial symmetries of the system, including invariance under translations. This, in turn, is a significant difference with respect to some previous approaches for infinite systems that slightly break translational invariance \cite{iTEBD1, iTEBD2, iPEPS,iPEPS2}. Also, we will see that these methods produce, in a natural way, individual tensors for the 'bra' and 'ket' parts of local expectation values, in a way similar to the so-called single-layer picture \cite{single}. This does not seem to have remarkable consequences in $1d$ but, as we shall discuss, it has some interesting implications for the $2d$ algorithms when compared to other methods such as e.g. iPEPS \cite{iPEPS,iPEPS2}.

In order to prove the validity of our algorithms we provide preliminary benchmarking calculations for the spin-1/2 quantum Ising model in $1d$ and $2d$. Our methods are roughly comparable in accuracy to other well-established approaches such as the iDMRG \cite{iDMRG1, iDMRG2} and iTEBD \cite{iTEBD1, iTEBD2} methods in $1d$ and the iPEPS method and TERG in $2d$ \cite{iPEPS,iPEPS2,TERG}, thus offering a possible alternative to them. Moreover, we will also discuss that it is possible in principle to use some of the $2d$ results for the development of a $3d$ algorithm in combination with some tensor updates, in the same spirit as the algorithm in Ref.\cite{iPEPS2} for $2d$ systems. The computational complexity of all the proposed algorithms is also analyzed and, as we shall see, important differences are found depending on the dimensionality and chosen simulation scheme. 

For completeness, we also discuss briefly in Appendix B the case of periodic boundary conditions, as well as real time evolution. This discussion will lead us to a beautiful interpretation of the standard iTEBD and iPEPS algorithms in terms of CTMs and corner tensors. We also believe that the algorithms described in this paper will be useful for the pracical implementation and development of further tensor network methods in the future. 

This work is organised as follows: In Sec.II we present several preliminary concepts and generalities. These include notions on tensor networks (Sec.II.A), corner transfer matrices (Sec.II.B,C), Matrix Product States and Projected Entangled Pair States (Sec.II.D),  and time evolution with Matrix Product Operators and Projected Entangled Pair Operators (Sec.II.E). In Sec.III we present two simple algorithms based on CTMs and corner tensors to approximate ground state properties of infinite-size quantum lattice systems. More elaborated versions of these algorithms are presented in Appendix A. After discussing the general approach in Sec.III.A, we consider $1d$ systems in Sec.III.B and $2d$ systems in Sec.III.C. A summary of the proposed methods and their complexities is done in Sec.III.E. In Sec.IV we present preliminary benchmarking numerical calculations for $1d$ and $2d$ systems. Sec. V contains our conclusions and final remarks. Moreover, in Appendix B we discuss further possibilities, such as $3d$ quantum lattice systems, periodic boundary conditions, and real time evolution.

\begin{figure}
\includegraphics[width=0.45\textwidth]{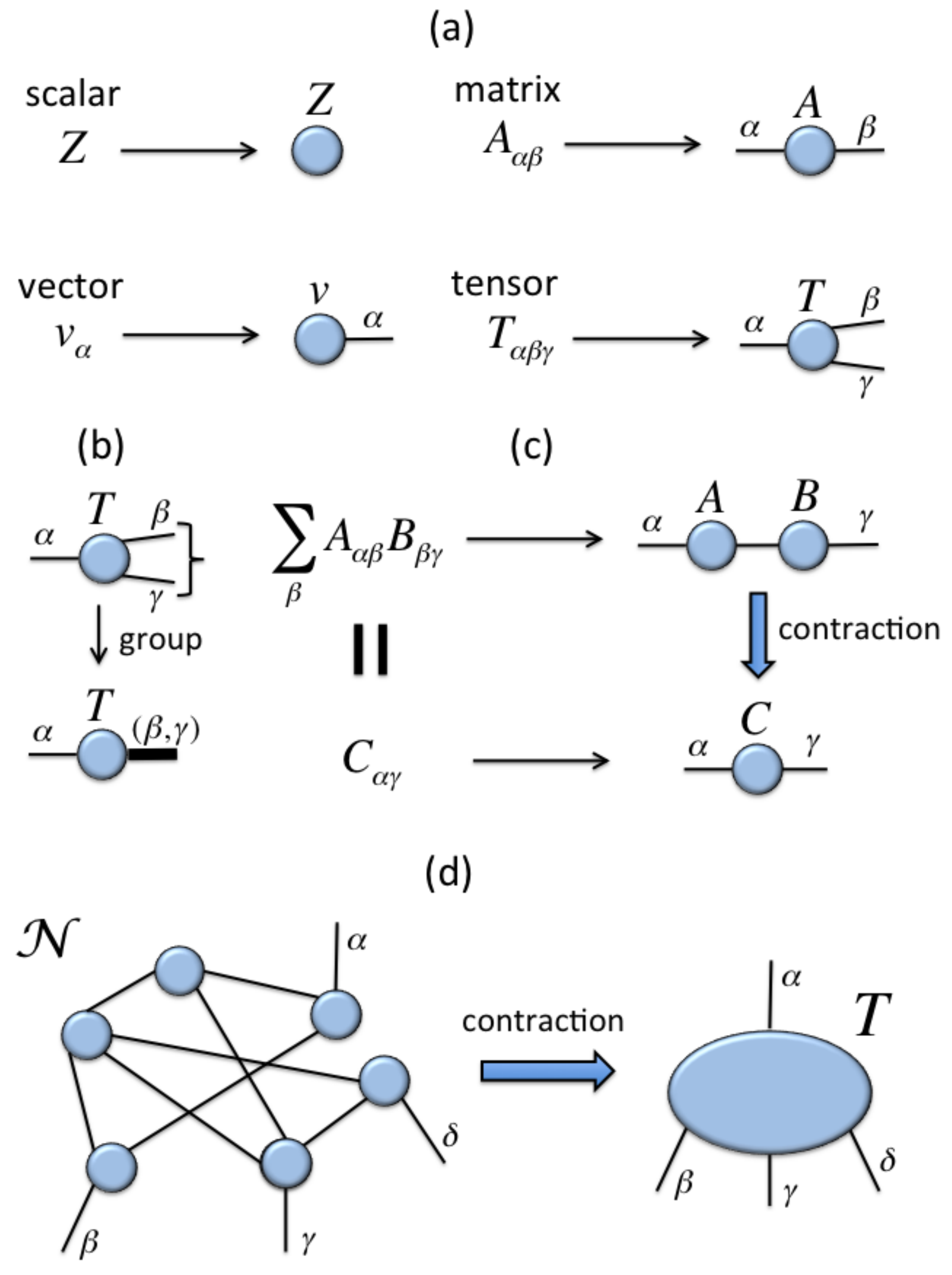}
\caption{(color online) (a) Tensor network diagrams representing a scalar, a vector, a matrix and a three-index tensor. 
(b) The grouping of two indices $\alpha$ and $\beta$ of a tensor produces another index $(\alpha,\beta)$, represented as a thick line. (c) Diagram representing a matrix multiplication, or \emph{contraction} of an index. (c) Diagram for tensor network $\mathcal{N}$, with four open (non-contracted) indices $\alpha, \beta, \gamma$ and $\delta$. The result of the contraction is a four-index tensor $T$.}
\label{fig1}
\end{figure}
\begin{figure}
\includegraphics[width=0.5\textwidth]{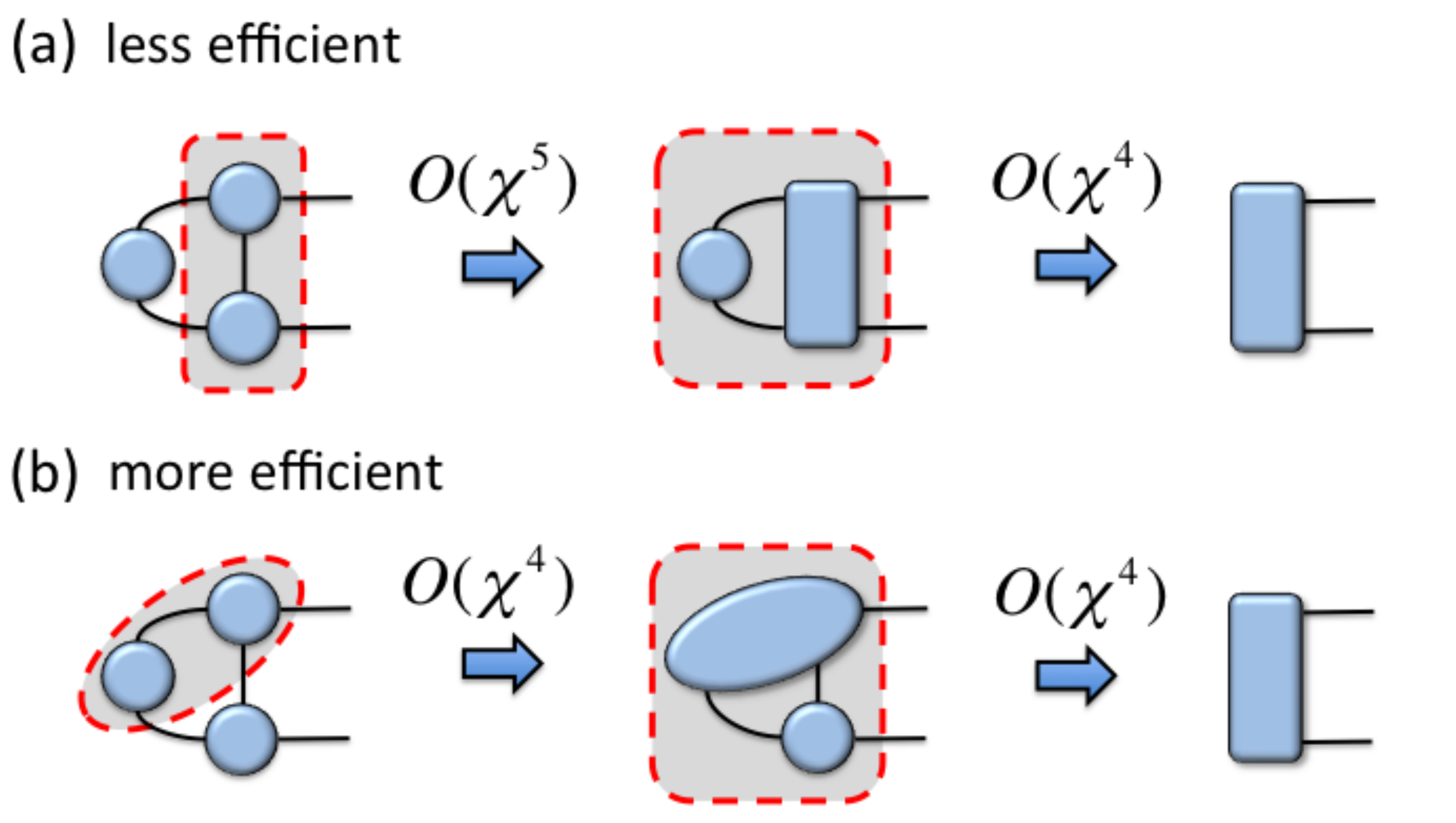}
\caption{(color online) Different orderings in the contraction lead to different number of operations. Assuming that all the indices of the tensors can take up to $\chi$ different values, then the number of operations is scheme (a) is $O(\chi^5)$, whereas it is $O(\chi^4)$ in scheme (b). Thus, scheme (b) is more efficient than scheme (a).}
\label{fig2}
\end{figure}

\section{Preliminary concepts}

The goal of this section is to introduce some preliminary concepts that we will need in the presentation of the algorithms of the forthcoming sections, as well as to provide some background on relevant topics. We start by quickly reminding some basic concepts on tensor networks and tensor network diagrams. Then, we review the basics of the CTM and its properties, as well as several aspects regarding possible gneralisations. After this, we quickly remind the basic definition of Matrix Product States (MPS) and Projected Entangled Pair States (PEPS), and then review the implementation from Ref. \cite{MPO} of time evolution operators (both in real and imaginary time) using Matrix Product Operators (MPO) and Projected Entangled Pair Operators (PEPO). 

\subsection{Tensor networks and diagrams}

For our purposes, a \emph{tensor} is a multidimensional array of complex numbers. A \emph{Tensor Network} (TN) is a network of tensors whose indices are connected according to some pattern. This connection of indices is done by summing over all the possible values of common indices between tensors. Summing over an index is also called \emph{contracting the index}. Summing over all the possible indices of a given TN is called \emph{contracting the TN}. 

Instead of using equations, tensors and TNs are more easily handled by using a diagrammatic notation in terms of \emph{tensor network diagrams}, see Fig.(\ref{fig1}). In these diagrams tensors are represented by shapes, and indices in the tensors are represented by lines emerging from the shapes, see Fig.(\ref{fig1}.a,b). A TN is thus represented by a set of shapes interconnected by lines. The lines connecting tensors between each other correspond to contracted indices, whereas lines that do not go from one tensor to another correspond to open indices in the TN, see Fig.(\ref{fig1}.c,d). As expected, the contraction of a TN with some open indices gives as a result another tensor, and in the case of not having any open indices the result is a scalar. Notice, though, that the total number of operations that must be done in order to obtain the final result of the contraction depends strongly on the order in which tensors in the TN are contracted, see Fig.(\ref{fig2}). To minimize the computational cost of a TN contraction, one must thus optimize over the different possible contraction orderings.  

\begin{figure}
\includegraphics[width=0.5\textwidth]{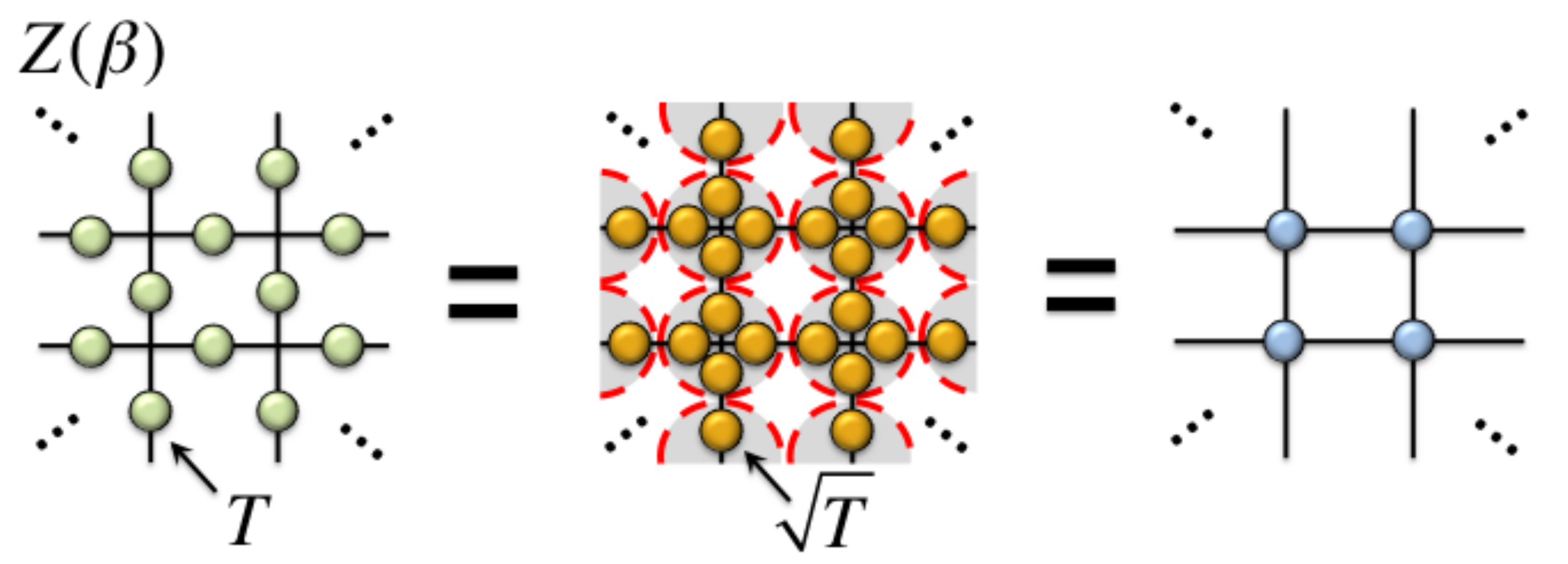}
\caption{(color online) Partition function $Z(\beta)$ at inverse temperature $\beta$ of some classical lattice model on the square lattice, defined by a symetric weight matrix $T$ between nearest neighbours (such as e.g. the classical Ising and Potts models). The partition function can be written as a TN with either one tensor $T$ per link, or one tensor per site resulting from the contraction of four $\sqrt{T}$ tensors (see e.g. Ref.\cite{iTEBD1, iTEBD2})}
\label{fig3}

\end{figure}
\begin{figure}
\includegraphics[width=0.45\textwidth]{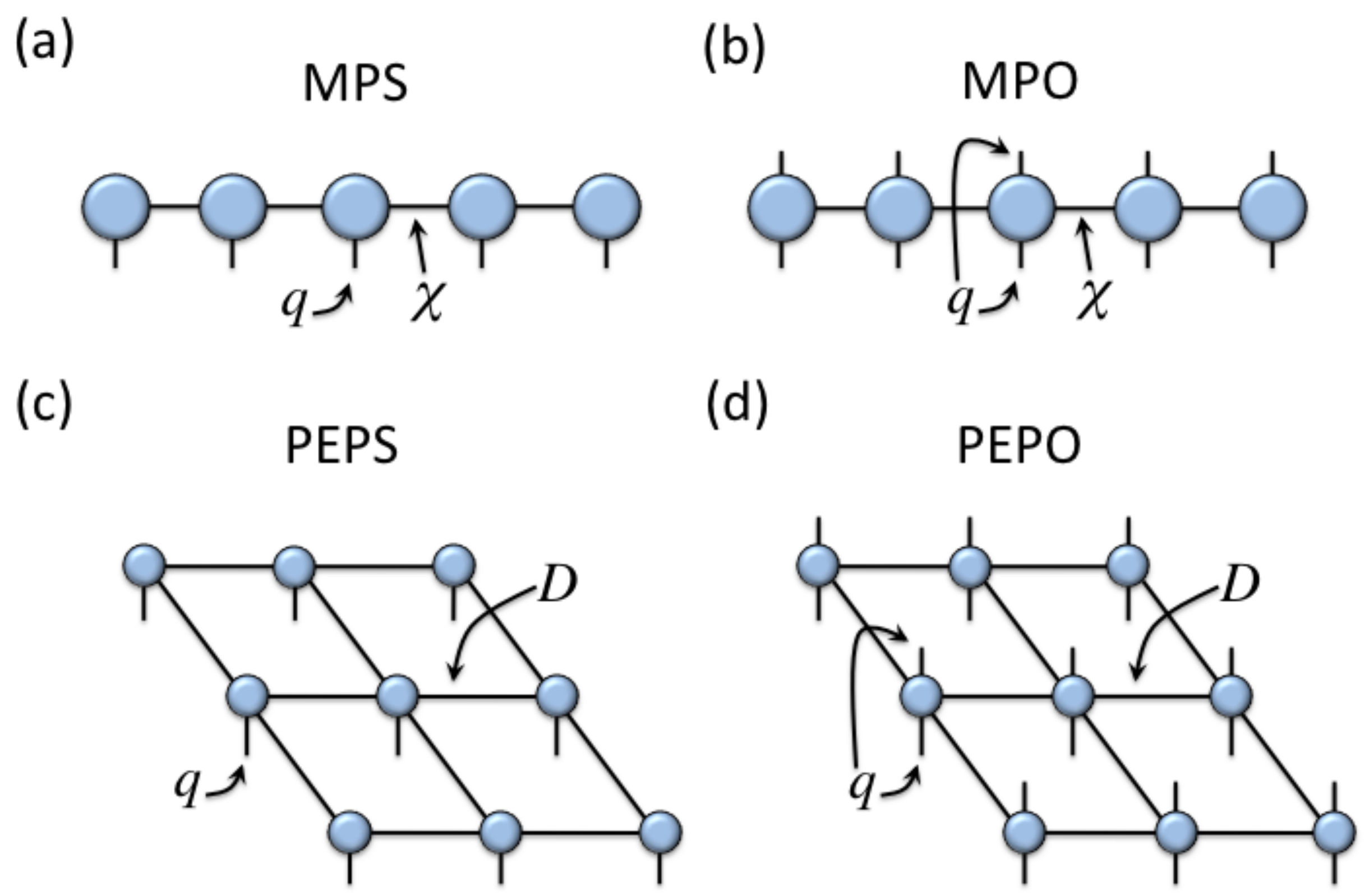}
\caption{(color online) (a) Matrix Product State (MPS). (b) Matrix Product Operator (MPO). (c) Projected Entangled Pair State (PEPS). (d) Projected Entangled Pair Operator (PEPO). The physical dimension is $q$ in all cases, whereas the bond dimension is $\chi$ for MPS and MPO, and $D$ for PEPS and PEPO.}
\label{fig4}
\end{figure}

There are famous examples of TN in the context of many-body physics. For instance, the partition function of a $d$-dimensional  classical lattice model with nearest-neighbour interactions is a TN in $d$ dimensions (see Fig.(\ref{fig3})). Also, for quantum lattice systems, the classes of MPS and PEPS are suitable to describe quantum states of $1d$ and $2d$ systems respectively (see Fig.(\ref{fig4}.a,c)). Other examples make use of an extra 'holographic' dimension accounting for some renormalization group scale \cite{holograph}, such as Tree Tensor Networks (TTN) \cite{TTN1, TTN2, TTN3, TTN4} (Fig.(\ref{fig5}.a)) and the so-called Multi Scale Entanglement Renormalization Ansatz (MERA) (Fig.(\ref{fig5}.b)), which is at the basis of Entanglement Renormalization \cite{ER1, ER2}. TNs can also be used to describe operators, such as MPOs in $1d$ and PEPOs in $2d$ (Fig.(\ref{fig4}.b,d)). 

\subsection{Corner Transfer Matrix: fundamentals}

\begin{figure}
\includegraphics[width=0.5\textwidth]{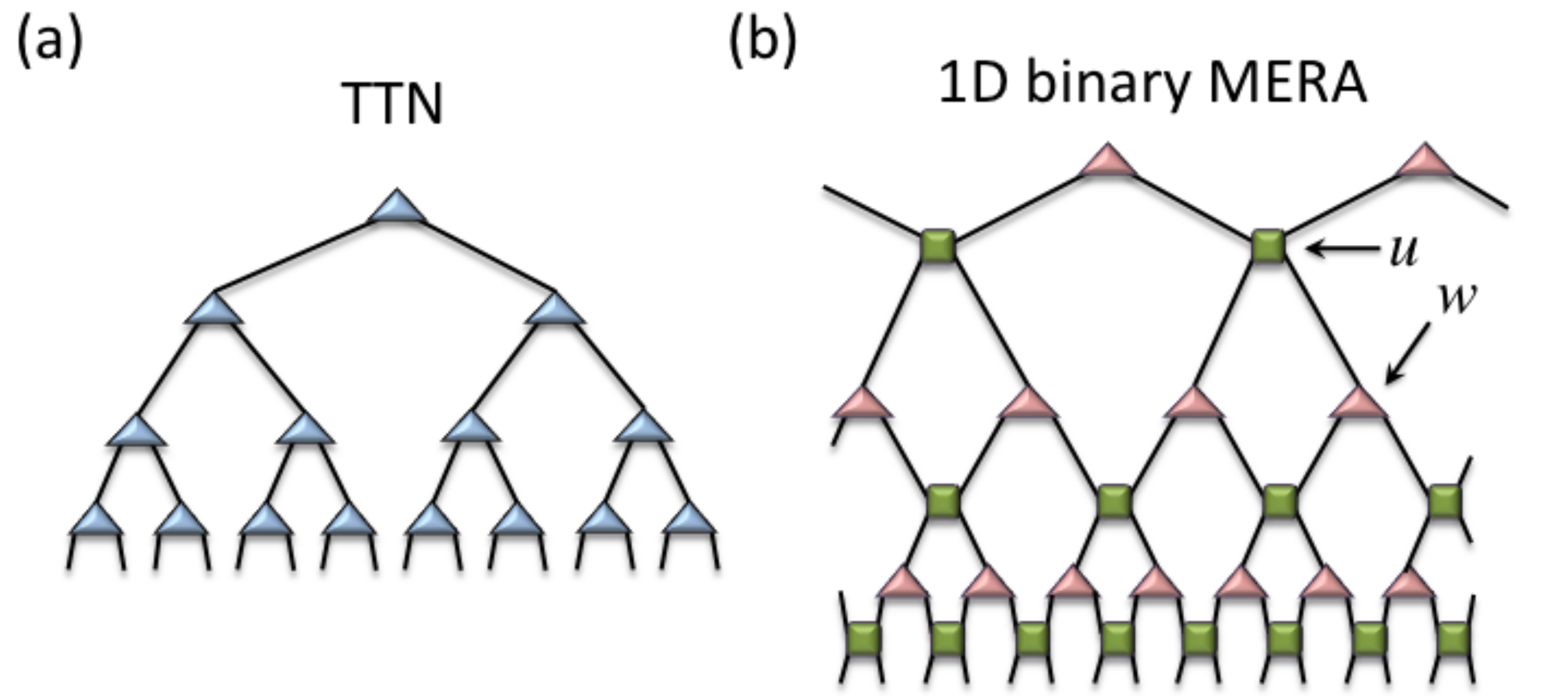}
\caption{(color online) (a) Tree Tensor Network (TTN). (b) $1d$ binary Multi Scale Entanglement Renormalization Ansatz (MERA), with unitaries $u$ and isommetries $w$. For details, see Ref. \cite{TTN1, TTN2, TTN3, TTN4, ER1, ER2}.}
\label{fig5}
\end{figure}

\begin{figure}
\includegraphics[width=0.45\textwidth]{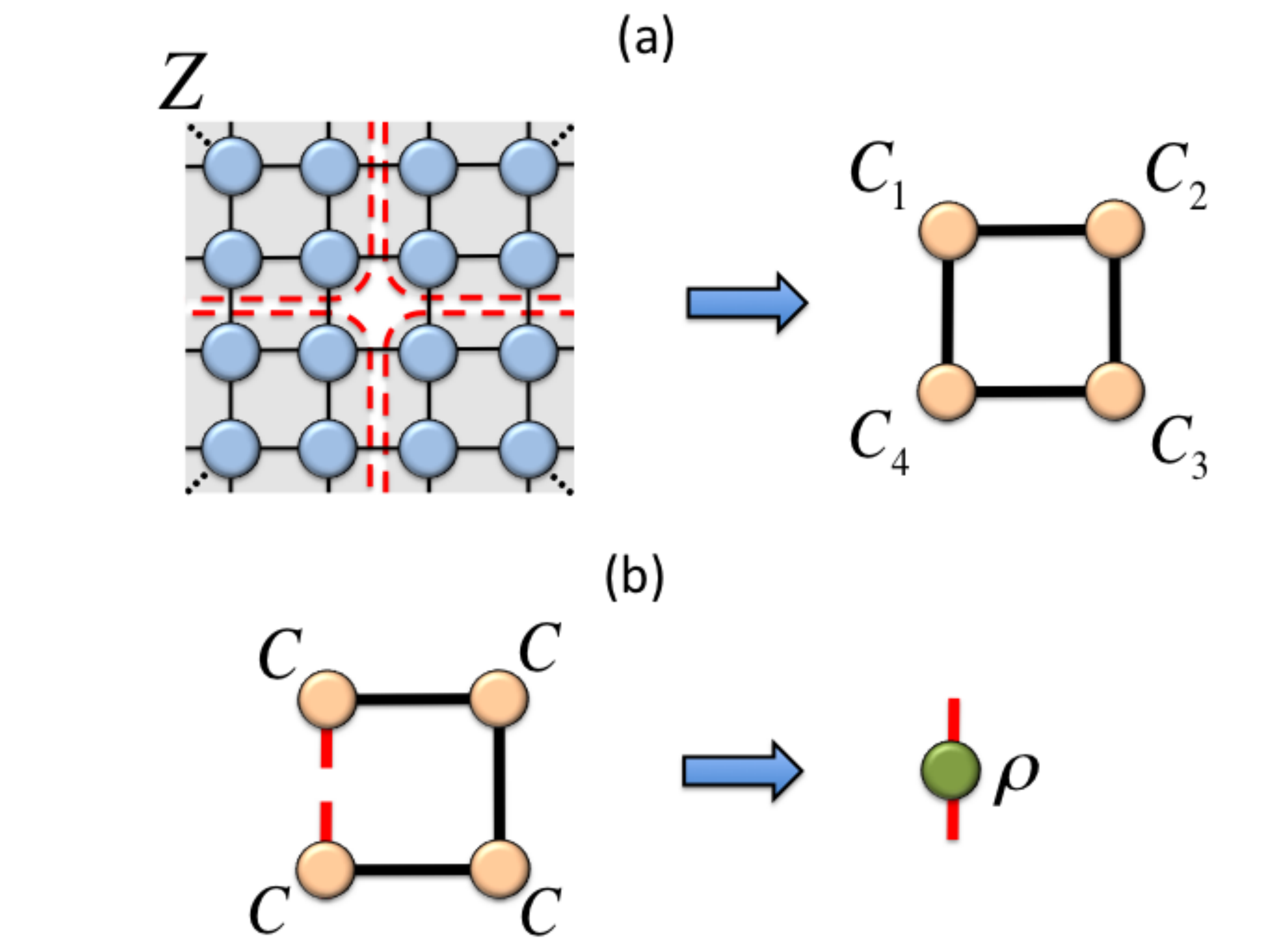}
\caption{(color online) (a) The contraction of a $2d$ square lattice of tensors results in a scalar $Z$. This contraction can be understood as the trace of the product of four CTMs, one for each corner. Notice that the exact CTMs will, in principle, have indices that can take up to exponentially many different values (in the size of the system). Thus, their indices are represented by a thick black line. The aim of CTM numerical methods is precisely to reduce the size of this index in some optimal or quasi-optimal way. (b) A possible reduced density matrix $\rho$ of a system with the same CTM at every corner. Open indices of $\rho$ are shown in red.}  
\label{fig6}
\end{figure}

The concept of Corner Transfer Matrix (CTM) was introduced by Baxter in the context of exactly solvable models in $2d$ classical statistical mechanics \cite{CTM681, CTM682}. These can be defined for any planar TN. However, for the sake of simplicity, here we shall assume the case of a $2d$ TN  on a square lattice as in Fig.(\ref{fig6}). This TN could represent, for instance, the partition function of some $2d$ classical lattice model as in Fig.(\ref{fig3}) or, as we will consider in Sec.III, the imaginary time evolution of a quantum $1d$ system. In order to define the CTM one makes the following observation: the contraction of the TN can be obtained by multplying four matrices $C_1,C_2,C_3$ and $C_4$, one for each corner (see Fig.(\ref{fig5}.a)). Thus, 

\begin{equation}
Z = {\rm tr} \left(C_1C_2C_3C_4 \right) \ , 
\label{ctmm}
\end{equation}
where $Z$ is the scalar resulting from the contraction. 

Matrices $C_1,C_2,C_3$ and $C_4$ are the \emph{Corner Transfer Matrices} of the system. They correspond to the contraction of all the tensors in each one of the four corners of the $2d$ lattice of tensors. The nomenclature 'transfer matrix' is appropriate, since the CTM 'transfers' a vector in the angular direction around the center of the lattice by an angle of $\pi/2$ in our case. To further simplify our explanation, let us assume that the four CTMs are equal, that is $C \equiv C_1 = C_2 = C_3 = C_4$. 

Sometimes it is convenient to define diagonal CTMs $C_d = P C P^{-1}$. Depending on the symmetries of the system (and thus of $C$), matrix $P$ may be arbitrary, unitary or orthogonal. Let us assume that there are $\chi$ different eigenvalues $\nu_{\alpha}$, with $\alpha = 1, 2, \ldots, \chi$. Then, the contraction of the full TN adopts the very simple expression 
\begin{equation}
Z = {\rm tr} \left(C_d^4 \right) = \sum_{\alpha=1}^\chi \nu_{\alpha}^4 . 
\label{ctm2}
\end{equation}
The fact that we choose to name the number of different eigenvalues as $\chi$ is made on purpose. There is in fact a direct relation between this parameter and the so-called \emph{bond dimension} of an MPS. The relation between these two seemingly different parameters will be made clearer later on in Sec.III.B, when considering $1d$ quantum lattice systems. 

CTMs are of paramount importance in the context of classical statistical mechanics. They have been used to solve the hard hexagon model, as well as many others \cite{CTM1, CTM2}. But they have also been useful in the context of quantum information, since it is known from long ago that their eigenvalue spectrum can be related to the entanglement spectrum of $1d$ quantum lattice systems, see e.g. Ref. \cite{peschel1, peschel2}. As we will see, this is a key property in the algorithms that will be further described in Sec.III.B. Also, from a numerical perspective,  a variational method to approximate the partition function per site of a $2d$ classical lattice model was developed by Baxter by truncating in the eigenvalue spectrum of the CTM \cite{CTM681, CTM682}. This idea was later on refined by Nishino and Okunishi, who developed the so-called Corner Transfer Matrix Renormalization Group (CTMRG) \cite{CTMRG1, CTMRG2}. As such, CTMRG is an algorithm to approximate properties of $2d$ classical lattice models with isotropic interactions (and thus a high degree of symmetry), and runs by truncating in the largest eigenvalues in magnitude of the spectrum of matrix $C^4$, which is a reduced density matrix of the system (see Fig.(\ref{fig6}.b)). This spectrum, in turn, is given by the numbers $\nu_\alpha^4$ in Eq.(\ref{ctm2}), and is thus in one-to-one correspondence with the spectrum of the CTM $C$. 

As we shall explain later on, from the point of view of quantum states of $1d$ quantum lattice systems the numbers $\nu_{\alpha}^4$ are, in fact,  the spectrum of eigevalues of the reduced density matrix of half an infinite chain. Or what is the same, the spectrum of Schmidt coefficients of half an infinite chain is given by $\lambda_{\alpha} = \nu_{\alpha}^2$. Thus, in a way, we can think of Baxter's variational method as sort of a precursor of the truncation scheme used in DMRG and TEBD, but in the context of $2d$ classical statistical mechanics \footnote{The aim of this analogy is simply to make manifest that all these algorithms somehow share the same spirit.}. Also, it is well known that these CTM methods work assymptotically in the limit of a system of infinite size. In fact, the closer the method is to convergence, the more faithful are the truncations in the eigenvalues of the corresponding reduced density matrix. The convergence and speed of CTMRG for classical $2d$ models is also remarkably better than other approaches such as the so-called Transfer Matrix Renormalization Group (TMRG) \cite{TMRG, Thermal1, Thermal2}. This is understandable, because in many situations away from criticality the largest eigenvalue of the CTM is non-degenerate, and there is a rather big gap to the next lower eigenvalues (as compared to the gap of the row-to-row transfer matrix)\footnote{In fact, the largest eigenvalue of the CTM could be degenerate. Yet, this does not spoil the performace of CTMRG if only a finite number of eigenvalues is degenerate. Nevertheless, at criticality all the eigenvalues tend to be degenerate, which translates into a critical slow-down of the method \cite{priv}.}. This ensures a fast numerical convergence of CTM methods. 

\begin{figure}
\includegraphics[width=0.45\textwidth]{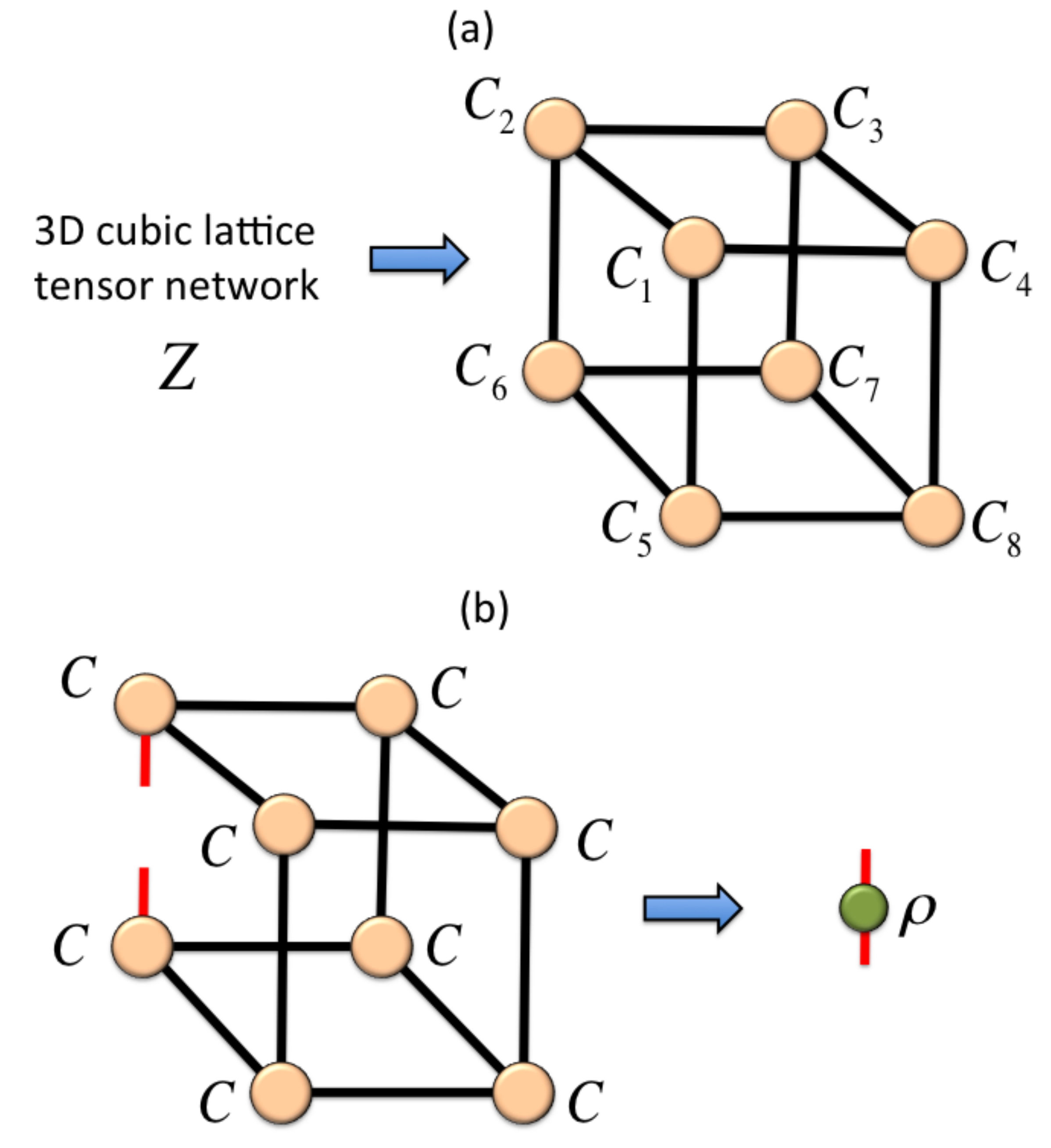}
\caption{(color online) Same as in Fig.(\ref{fig6}), but for a cubic lattice in $3d$. (a) A $3d$ cubic llatice of tensors results in a scalar Z. This contraction can be understood as the contraction of eight corner tensors. In the exact case, the corner tensors have 'fattened' indices, exactly as in the $2d$ case. We have chosen not to draw explicitely the $3d$ cubic lattice in order to keep the figure simple. (b) A possible reduced density matrix $\rho$ of a system with the same corner tensor at every corner. Open indices of $\rho$ are shown in red.}
\label{fig7}
\end{figure}

\subsection{Corner Transfer Matrix: generalizations}

Since the foundational works by Baxter, there have been several attempts to generalize the CTM in a number of different ways. Here we mention some basic possibilities. 

The first obvious generalization is for $2d$ lattices different from the square one. Of course, the CTM can be defined for any planar lattice as long as corners can be defined. This includes the usual lattices, but also more exotic constructions such as lattices with hyperbolic geometry (or negative curvature), e.g. lattice discretizations of Anti de Sitter (AdS) manifolds \cite{Hyper}. 

An attempt to improve the efficiency of the truncations involved in CTMRG was also proposed in Ref. \cite{iPEPS2}, where a directional version of the method was put forward in the context of  $2d$ iPEPS calculations, and where tensors were no longer real and broke rotational symmetry. In Sec.III.B of this paper we will also propose another plausible way of doing this generalization to non-symmetric tensors in the context of the algorithms that we explain here. 

Another natural generalization is the case of periodic boundary conditions. Even if this sounds counterintuitive (since a periodic system does not have any corners), it turns out that the ideas from CTMRG can be generalized to deal with this type of systems as well \cite{perCTM}. We will briefly comment on how one can perform this generalization in Sec.III.D. 

Moreover, the CTMRG technique can also be generalized to deal with stochastic models, see Ref. \cite{stochastic} for details. Also, a quantum counterpart of the CTM is the so-called \emph{Corner Hamiltonian}, which essentially is the logarithm of the CTM. Numerical techniques for Corner Hamiltonians have also been developed, see e.g. Ref.\cite{CH}. 

Finally, the CTM can also be generalized to higher-dimensional systems. The natural corner object now is not a matrix, but a tensor with three indices (for a cubic lattice) which we call the \emph{corner tensor}\footnote{We adopt here the same notation as in Ref. \cite{CTM3d}.}, see Fig.(\ref{fig7}). In terms of corner tensors $C_1, C_2, \ldots, C_8$, the contraction of a $3d$ lattice of tensors reads

\begin{equation}
Z = f \left(C_1,C_2,C_3,C_4,C_5,C_6,C_7,C_8 \right) \ ,
\end{equation}
where $f$ is a function that performs the corresponding TN contraction (see Fig.(\ref{fig7}.a)). This higher-dimensional generalization is quite obvious but, yet, it has some non-trivial consequences since one is dealing now with tensors instead of matrices. In particular, the corner tensors can no longer be diagonalized (or more precisely, there is no unique eigenvalue decomposition since it depends on how one chooses to split the indices of the tensor). Thus, no expression like the one in Eq.(\ref{ctm2}) for the $2d$ case can be obtained in general for $3d$. Nevertheless, corner tensors still have interesting spectral properties with respect to eigenvalue/singular value decomposition of bipartitions of their indices. The behavior of these singular values will be the key to define simplified numerical approaches for 
higher-dimensional systems, as we shall do in Sec.III.C. 

\subsection{Matrix Product States and Projected Entangled Pair States}

Let us now revise briefly the concepts of Matrix Product States (MPS) and Projected Entangled Pair States (PEPS). There is a vast amount of literature on these two families of states, and we refer the interested reader to it for further details (see e.g. Ref.\cite{tn} and references therein). 

Consider a quantum many-body system of $N$ particles. The quantum state of the system is $\ket{\Psi} \in \mathcal{H}$, where $\mathcal{H} = \otimes_{r=1}^N \mathcal{H}^{[r]}$ is the total Hilbert space of the system and $\mathcal{H}^{[r]}$ is the individual Hilbert space of particle $r$. Let us assume that each particle is modelled by a $q$-level system, so that ${\rm dim}(\mathcal{H}^{[r]}) = q$. Given a local basis  $\ket{i_r}$ for each site $r$, with $i_r = 1, 2, \ldots, q$, the quantum state of the system reads
\begin{equation}
\ket{\psi} = \sum_{i_1 i_2 \cdots i_N} c_{i_1 i_2 \cdots i_N} \ket{i_1 i_2 \cdots i_N}. 
\eeq
The coefficient $c_{i_1 i_2 \cdots i_N}$ can be understood as a tensor of $N$ indices with $O(q^N)$ complex coefficients. Thus, this is clearly an inneficient representation of the quantum state because the required number of parameters scales exponentially with the size of the system. In order to find an efficient description, one can consider a decomposition of the above tensor into a MPS for $1d$, or a PEPS in $2d$. These decompositions are shown in the tensor network diagrams of Fig.(\ref{fig4}.a.c), where open boundary conditions are assumed. 

Both MPS and PEPS offer an efficient description of the quantum state $\ket{\Psi}$ for $1d$ and $2d$ respectively. Moreover, MPS and PEPS are known to have many interesting (and important) properties. For instance, both of them satisfy the so-called \emph{area law} scaling of the entanglement entropy \cite{entangent1, entangent2, entangent3, entangent4}, which is a key property of low-energy states of most quantum many-body systems (albeit with notable exceptions, see Refs. \cite{ferment,spinbose}). Moreover, it is known that ground states of $1d$ gapped local Hamiltonians can be efficiently approximated with exponential accuracy by an MPS \cite{hastings1}, and the same holds for thermal states in $2d$ with PEPS \cite{hastings2}. From the numerical point of view, MPS is the relevant class of states at the heart of efficient methods for $1d$ systems such as DMRG and TEBD. PEPS is also at the basis of simulation methods for $2d$ systems such as the finite and infinite PEPS methods \cite{PEPS,iPEPS}, as well as TERG \cite{TERG}.  

We wish to remark here a couple of properties of MPS and PEPS. First, for a system of size $N$, the number of parameters in both families of states scales linearly with $N$, and polynomially in the \emph{bond dimension} of the tensors (that is, the number of different values for the connecting indices in the MPS or PEPS tensor network, see Fig.(\ref{fig4})). We call this bond dimension $\chi$ in the case of MPS, and $D$ in the case of PEPS. Importantly, this bond dimension can be regarded as a measure of the number of parameters in the TN, but also as a measure of the maximum amount of entanglement that can be handled by the wavefunction (see Ref. \cite{PEPS}). Second, for systems invariant under translations, it is possible to take the thermodynamic limit $N \rightarrow \infty$ and consider a MPS or PEPS for a system of infinite size. This is done by repeating the same unit cell of tensors across the whole lattice \cite{iTEBD1, iTEBD2, iDMRG1, iDMRG2, iPEPS}. This trick is at the basis of methods to study infinite-size systems such as iDMRG, iTEBD and iPEPS. The methods that we shall propose in this paper will be for infinite-size systems, and thus rely on this idea as well. 

\subsection{Time evolution with Matrix Product Operators and Projected Entangled Pair Operators}

Let us now consider the problem of time evolution. We assume that this evolution is generated by a Hamiltonian $H$ that is the sum of local interaction terms on lattice. For simplicity of the explanation, let us also assume a time-independent Hamiltonian with nearest-neighbour interactions,  namely
\beq
H = \sum_{\langle r, r' \rangle} h^{[r,r']} \ ,
\eeq
although time-dependent cases and more generic types of intereactions could also be considered (including long-range ones \cite{iDMRG1, iDMRG2}). The real time evolution of a given state $\ket{\Psi(0)}$ reads

\beq
\ket{\Psi(t)} = e^{-i H t} \ket{\Psi(0)}. 
\label{real}
\eeq
It is equally possible to consider the evolution in imaginary time $\tau$ in order to get the ground state $\ket{\Psi_{gs}}$ of $H$, namely 

\beq
\ket{\Psi_{gs}} = \lim_{\tau \rightarrow \infty} \frac{e^{-H \tau} \ket{\Psi(0)}}{|| e^{-H \tau} \ket{\Psi(0)} ||} \ ,
\label{imag}
\eeq
where we have assumed that the initial state $\ket{\Psi(0)}$ has a non-zero overlap with the ground state $\ket{\Psi_{gs}}$, and the appropriate normalization of the state has been included. 

As is well known, these two different types of evolutions can be approximated using a TN approach for $1d$ and $2d$ systems, both for finite and infinite systems. What is important for us, though, is that the evolution operators $e^{-i H t}$ and $e^{-H \tau}$ can usually be approximated by a sequence of well-behaved MPOs\footnote{MPOs were first introduced in Ref.\cite{Thermal1, Thermal2}.} in $1d$ or PEPOs in $2d$, as explained in Ref. \cite{MPO}. This MPO and PEPO approximation will turn out to be a key step in the algorithms that will be proposed in Sec.III. 

Let us now remind the basic steps in obtaining MPO and PEPO descriptions for the evolutions generated by $H$. Here we simply review some of the results from Ref. \cite{MPO}, but which are important for our purposes. For concreteness of the explanation, let us imagine that $H$ corresponds to the spin-1/2 ferromagnetic quantum Ising model in a homogeneous transverse field, 

\beq
H = - \sum_{\langle r,r' \rangle} \sigma_z^{[r]} \sigma_z^{[r']} - h\sum_r \sigma_x^{[r]} \ ,
\label{ham}
\eeq
where $\sigma_z$ and $\sigma_x$ are the usual Pauli matrices, and the sum in the first term is over nearest-neighbours. Let us now assume, for simplicity, the case of a $1d$ system of $N$ particles with periodic boundary conditions. Without loss of generality, we also consider the evolution in imaginary time under this Hamiltonian for a total time $T$. The first thing to do, and as also done also in other approaches \cite{iTEBD1, iTEBD2, iPEPS,iPEPS2}, is to approximate the whole evolution by breaking the total evolution time $T$ into smaller steps of size $\delta \tau \ll 1$. Then, we have that the evolution operator $U \equiv \exp(-H T)$ can be written as $U = [U(\delta \tau)]^m = [\exp(- H \delta \tau)]^m$, with $m \equiv T/\delta \tau$. Next, we would like to describe the term $U(\delta \tau) = \exp (-H \delta \tau)$ by an MPO. We wish to make it in such a way that the resultant MPO is invariant under translations and symmetric under space inversion, since these are symmetries of the original Hamiltonian, and also favors stabiltity in numerical manipulations. As explained in Ref. \cite{MPO}, this can be achieved in the following way: first, split $H = H_z + H_x$, so that we can perform a Suzuki-Trotter decomsposition \cite{suztrot1, suztrot2} $\exp(-H \delta \tau) \sim \exp(-H_x \delta \tau) \exp(-H_z \delta \tau) + O(\delta \tau^2)$\footnote{Here we have used a first-order Suzuki-Trotter decomposition, but it is possible to consider higher-order decompositions to reduce the error, e.g. $\exp(-H \delta \tau) \sim \exp(-H_x \delta \tau/2) \exp(-H_z \delta \tau) \exp(-H_x \delta \tau/2) + O(\delta \tau^4)$.}. In this splitting, $H_z$ contains all the terms with $\sigma_z$ operators and $H_x$ the terms with $\sigma_x$ operators. The term $ \exp(-H_x \delta \tau)$  is just a tensor product of one-body operators, 
\beq
e^{-H_x \delta \tau} = \otimes_{r=1}^N e^{h \sigma_x^{[r]} \delta \tau} \equiv \otimes_{r=1}^N L^{[r]} \ ,
\eeq
and its MPO representation is trivial (see Fig.(\ref{fig8}.a)). However, the term  $\exp(-H_z \delta \tau)$ requires a little bit more consideration. In order to evaluate the MPO for this term, we first remember that 

\beq
e^{\omega  \sigma_z^{[r]} \sigma_z^{[r']}}= \cosh{(\omega)} \mathbb{I}^{[r]}  \mathbb{I}^{[r']} + \sinh{(\omega)} \sigma_z^{[r]} \sigma_z^{[r']} 
\eeq
for any $\omega$. Using this property, and a little bit of algebra, it is easy to arrive at an expression for the MPO. If the operator is expressed as
\beq
e^{\delta \tau \sum_{r=1}^N  \sigma_z^{[r]} \sigma_z^{[r+1]}} =  \sum_{i's, j's} c^{i_1 i_2 \cdots i_N}_{j_1 j_2 \cdots j_N} \ket{i_1 i_2 \cdots i_N} \bra{j_1 j_2 \cdots j_N} \ ,
\eeq
where $\ket{i_r}$ is the basis of eigenstates of $\sigma_z$ at site $r$ (and same for the $j$'s), then the coefficients $c^{i_1 i_2 \cdots i_N}_{j_1 j_2 \cdots j_N}$ are given by the MPO
\beq
c^{i_1 i_2 \cdots i_N}_{j_1 j_2 \cdots j_N} = {\rm tr} \left((M^{i_1}_{j_1})(M^{i_2}_{j_2}) \cdots (M^{i_N}_{j_N}) \right), 
\label{mpo1}
\eeq
where the non-zero components $(M^i_j)_{\alpha \beta}$ of tensor $M$ are 
\beqa
(M^i_j)_{1 1} = \cosh{(\delta \tau)} \mathbb{I}^i_j, ~~ (M^i_j)_{2 2} = \sinh{(\delta \tau)} \mathbb{I}^i_j, \nonumber \\
(M^i_j)_{1 2} = (M^i_j)_{2 1} = \sqrt{\sinh{(\delta \tau)} \cosh{(\delta \tau)}} (\sigma_z)^i_j \ , 
\label{mpo2}
\eeqa
as explained in Ref. \cite{MPO}. This MPO is represented in Fig.(\ref{fig8}.b). Then, in order to obtain a complete MPO combining the $H_x$ and $H_z$ terms, we just need to combine the tensors for the two pieces as shown in Fig.(\ref{fig8}.b). This MPO description is particularly convenient for numerics, since it is symmetric with respect to space inversion, the tensors are real numbers, and it is also invariant under translations. Notice also that one can easily take the thermodynamic limit $N \rightarrow \infty$.  

It is a good idea to make the tensors in the MPO description real and as symmetric as possible. This can usually be achieved by a variety of tricks depending on the system considered. For instance, as explained in Ref. \cite{MPO}, for the antiferromagnetic Heisenberg models on bipartite lattices it may be convenient to perform a sublattice rotation prior to finding the MPO representation in order to make the MPO tensors real. It is a good idea to apply tricks like this also to other models, whenever this is possible.  

\begin{figure}
\includegraphics[width=0.5\textwidth]{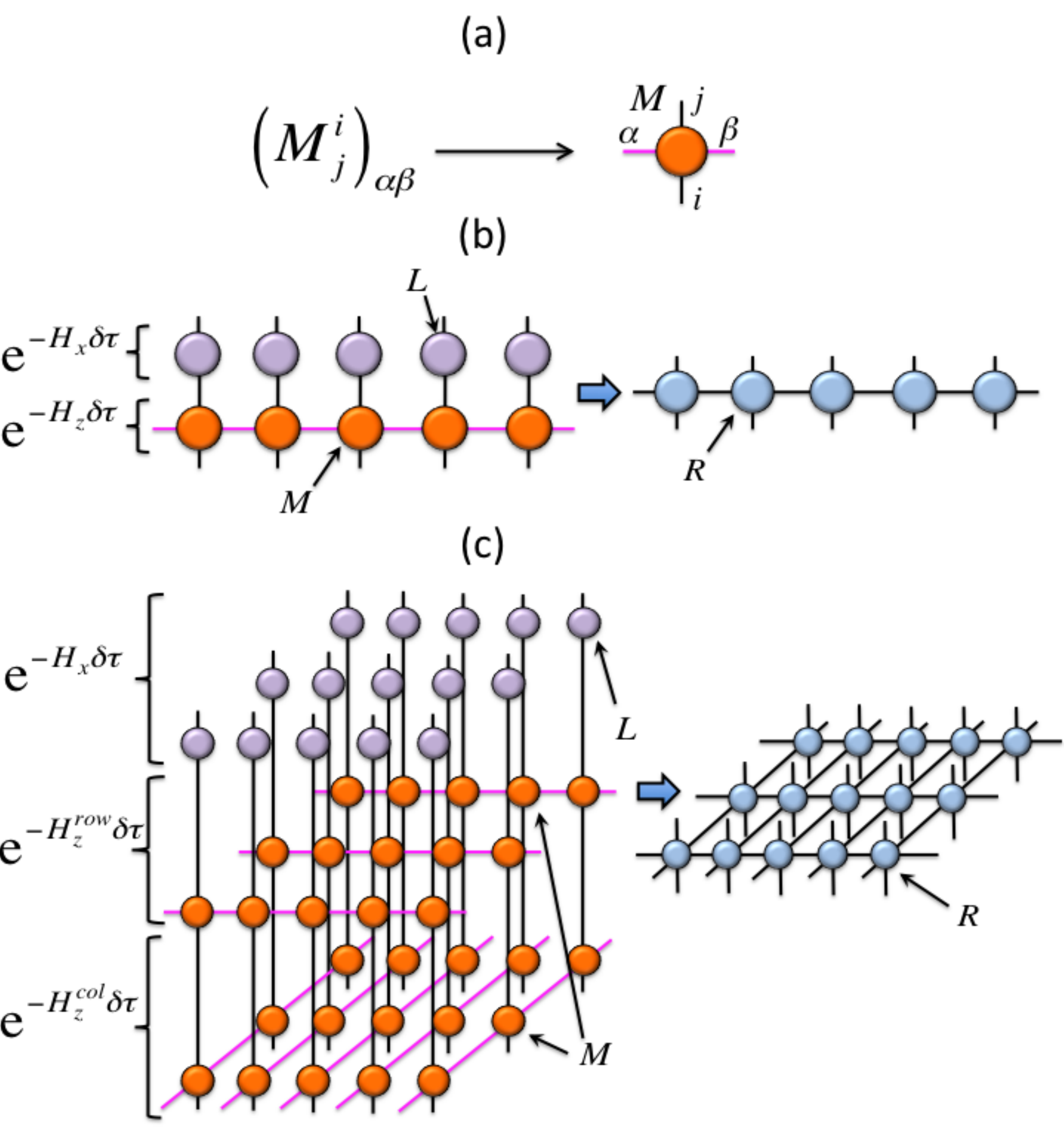}
\caption{(color online) (a) Diagrammatic representation of the tensor $M$ in Eqs.(\ref{mpo1},\ref{mpo2}). Bond indices are highlighted in pink. (b) Construction of the $1d$ MPO. The contraction of tensors $L$ and $M$ results in the MPO tensor $R$. (c) Construction of the $2d$ PEPO. The contraction of one tensor $L$ and two tensors $M$ (one along each direction of the lattice) results in the PEPO tensor $R$.} 
\label{fig8}
\end{figure}

In $2d$ it is equally simple to find a PEPO description for the evolution operator. The construction is based on that of the MPO for $1d$. Let us imagine that this time we have the same Hamiltonian as in Eq.(\ref{ham}) but in a square lattice. Then, we can decompose the evolution term generated by $H_z$ into a sum of two contributions: one for interactions along the rows, and one for interactions along the columns, $H = H^{row} + H^{col}$. Then, we have that $\exp(-H_z \delta \tau) = \exp(-H^{row}_z \delta \tau) \exp(-H^{col}_z \delta \tau)$, where $H_z^{row/col}$ contains the interactions alog the rows/columns. It is clear that, for each individual row or column, one can find MPO descriptions as the ones described before for the $1d$ case. Therefore, it is just a matter of combining together all these MPOs in a appropriate way in order to obtain a PEPO for the desired evolution. This is represented in the diagram of Fig.(\ref{fig8}.c). This PEPO has again nice properties: it is real, and also is symmetric with respect to space inversions in the two lattice directions.

\begin{figure}
\includegraphics[width=0.45\textwidth]{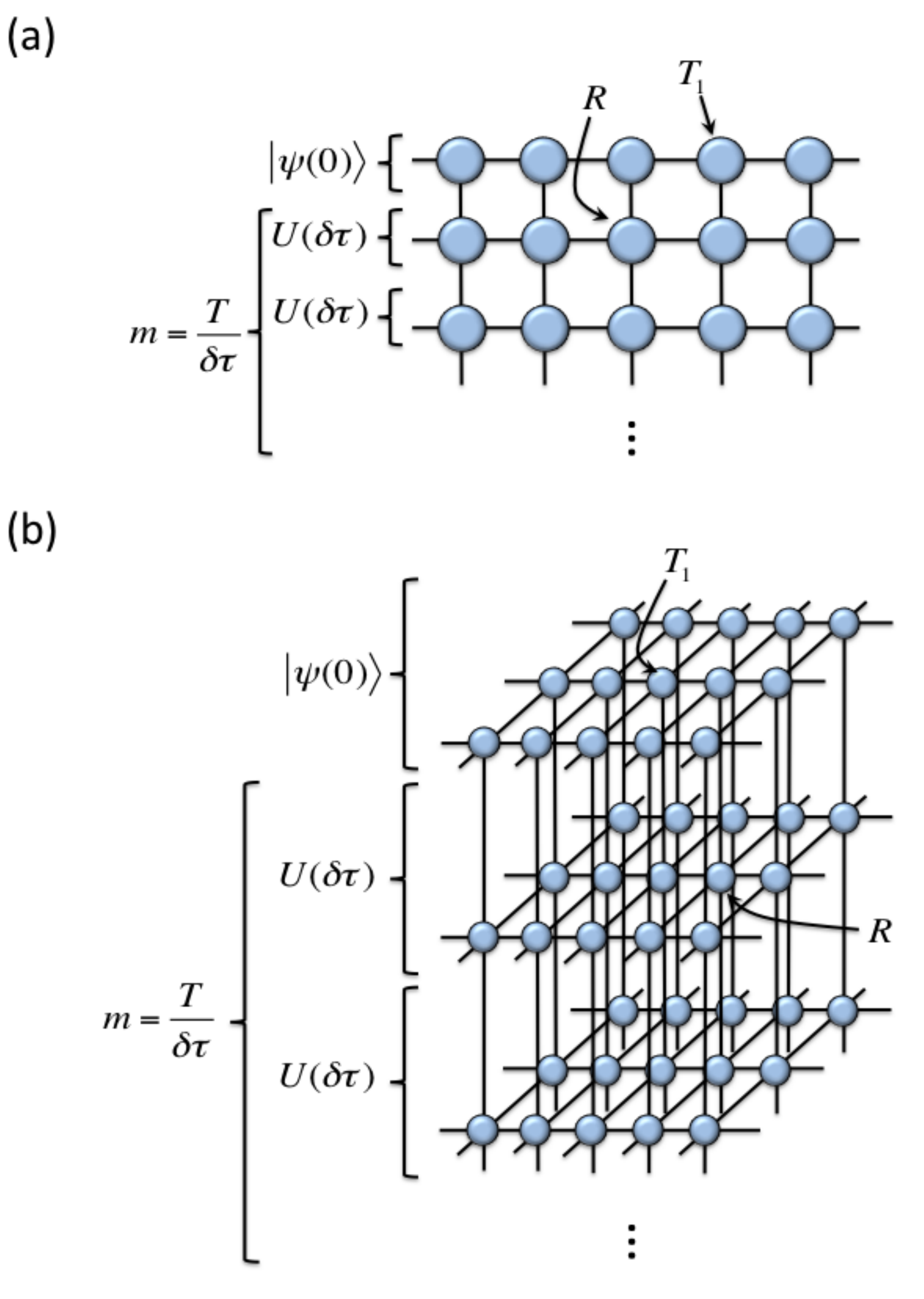}
\caption{(color online) (a) Time evolution of an MPS for $\ket{\Psi(0)}$ as driven by an MPO for $U(\delta \tau)$. (b) Time evolution of a PEPS for $\ket{\Psi(0)}$ as driven by a PEPO for $U(\delta \tau)$.}
\label{fig9}
\end{figure}

The construction of MPOs and PEPOs that we have reviewed here can be further generalized to a variety of other models and interactions. Notice that it is of course possible to use other alternative approaches to build MPOs and PEPOs for evolution operators. However, these constructions may not always guarantee the symmetry conditions of the obtained TN such as space inversion and translational invariance. As we shall see, it is important for our purposes that the obtained MPOs and PEPOs have these nice symmetries, since this improves the efficiency of the algorithms that we shall propose in Sec.III. 

\section{Algorithms}

In this section we propose a number of numerical simulation algorithms that allow to compute ground state properties of quantum lattice systems of infinite size. Our main focus is on $1d$ and $2d$, but we also discuss briefly the possibility of $3d$ systems, periodic boundary conditions, and real time evolution. Our algorithms are valid to deal with systems with a different amount of spatial symmetry. As we shall see, more spatial symmetry means simpler and more efficient simulation methods. In $1d$ this may not be too relevant, since the number of operations in the proposed methods differ only in subleading and multiplicative constant terms. But in $2d$ this is crucial since the difference in the leading number of operations turns out to be huge, in fact several orders of magnitude, depending on the amount of symmetry. The algorithms that we present here allow to compute ground state properties. As we shall discuss more quantitatively in Sec.IV,  these methods are a possible alternative to other methods for systems of infinite size such as iDMRG, iTEBD, iPEPS and TERG \cite{iDMRG1, iDMRG2, iTEBD1, iTEBD2, iPEPS,iPEPS2,TERG}. Importantly, all the algorithms of this section preserve at every step the invariance under translations. 

First we discuss the general approach, and then we describe the details of each algorithm. The cases of $3d$ and periodic boundary conditions are briefly discussed in Sec.III.D.  For an overall view of the complexity of the different the methods, one can jump directly to Sec.III.E. 

\begin{figure}
\includegraphics[width=0.45\textwidth]{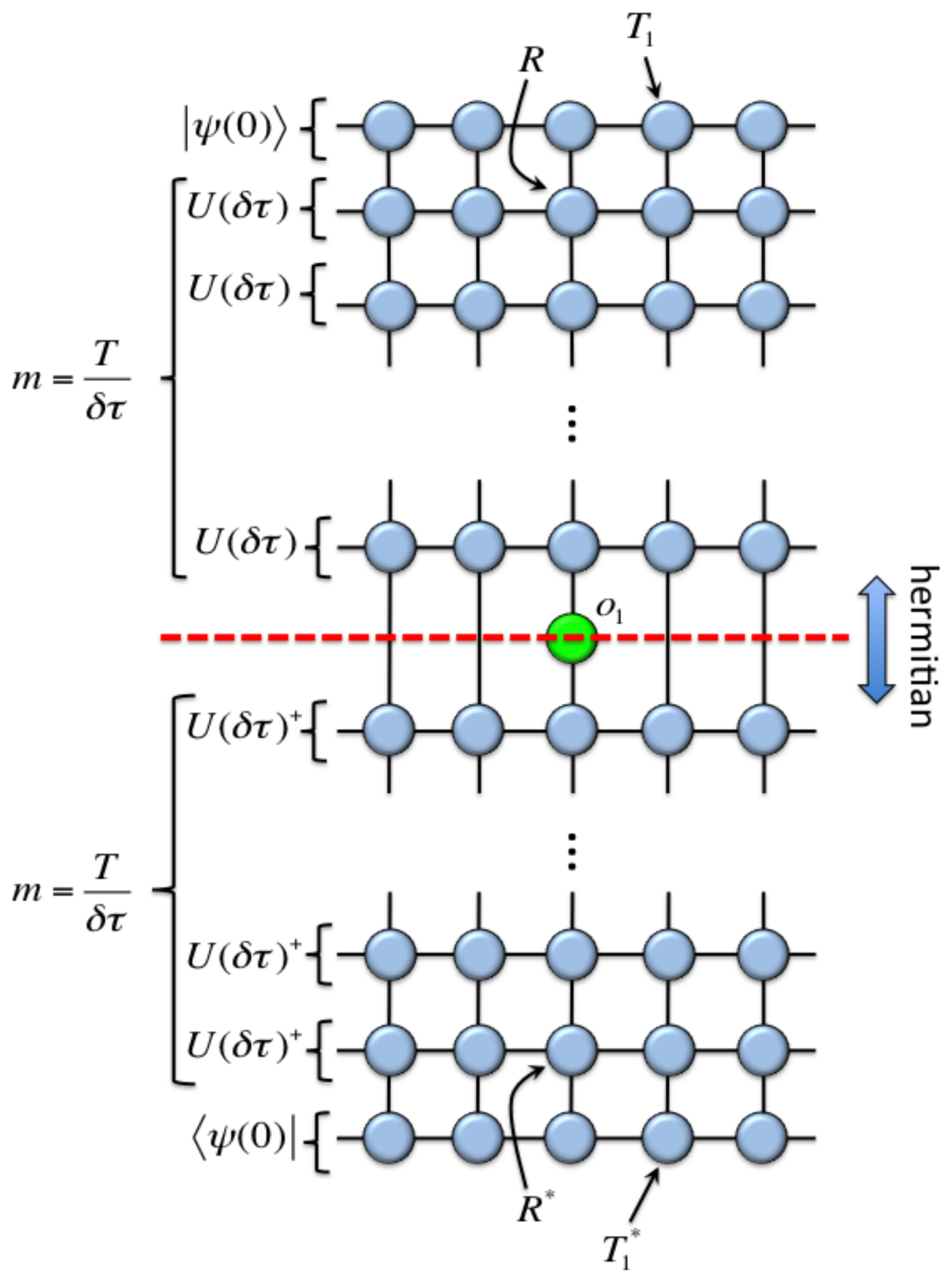}
\caption{(color online) Diagram for the numerator in Eq.(\ref{expec}). The expression is hermitian with respect to the vertical indices, hence the TN is symmetric with respect to a specular reflection across the red dashed line plus complex conjugation. This means that the individual tensors are symmetric with respect to transposition of the vertical indices plus complex conjugation.} 
\label{fig10}
\end{figure}

\subsection{General approach}

The general idea is quite simple, and can be understood from the diagrams in Fig.(\ref{fig9}) and Fig.(\ref{fig10}). The goal is to compute the evolution of some quantum state as driven by some Hamiltonian with local interactions for a total time $T$. As we have explained in the previous section the time evolution operator, both in real and imaginary time, can be decomposed as the action of MPOs or PEPOs for a given time interval $\delta \tau$ on a given quantum state described by an MPS or PEPS. Thus, the whole evolution of the system can be represented by some initial MPS or PEPS, to which one applies the MPO or PEPO driving the evolution for as many steps $m = T/\delta \tau$ as needed, see Fig.(\ref{fig9}). In the case of a ground state calculation the number of steps $m$ is infinite or, in practice, very large until convergence of some relevant quantity is achieved. 

The main goal of our algorithms is the efficient approximation of expectation values of local obsevables.  For instance, let us consider a one-body operator $o_1$. Let us also consider the evolved state $\ket{\Psi(T)}$ at some given time $T$, as represented in Fig.(\ref{fig10}) for an MPS (the case of a PEPS is a straightforward generalization). The expectation value of operator $o_1$ in the evolved state is given by 
\beq
\langle o_1 \rangle = \frac{\bra{\Psi(T)}o_1 \ket{\Psi(T)}}{\braket{\Psi(T)}{\Psi(T)}} =   \frac{\bra{\Psi(0)}U^{\dagger}o_1 U \ket{\Psi(0)}}{\bra{\Psi(0)}U^{\dagger} U \ket{\Psi(0)}} \ , 
\label{expec}
\eeq
where $U$ is the corresponding evolution operator that can be decomposed as $U = [U(\delta \tau)]^m$, with $U(\delta \tau)$ being approximated by an MPO or PEPO. Thus, e.g. the numerator in Eq.(\ref{expec}) can be represented diagramatically as in Fig.(\ref{fig10}). The actual expectation value is the ratio of two contracted TNs like the one in the figure, one with the obserble $o_1$ for the numerator, and one without the observable for the denominator. For a quantum lattice system in $d$ spatial dimensions, these are $(d+1)$-dimensional TNs, where the extra vertical dimension is \emph{time}. Therefore, for an MPS one has to deal with the contraction of a $(1+1)d$ TN, whereas in $2d$ one has the contraction of a $(2+1)d$ TN.

All the existing TN methods that compute real and imaginary-time evolutions approximate, in one way or another, a contraction like the one described above or similar. For instance, iTEBD and iPEPS methods compute an MPS or PEPS approximation to the dominant eigenvector of some transfer matrix operator defined as the MPO or PEPO driving the evolution \cite{iTEBD1, iTEBD2, iPEPS,iPEPS2}. But there are more ways of contracting these TNs. For instance, one could think of 'folding' the TN across the red dashed line in Fig.(\ref{fig10}) accompanied by a transversal contraction, as has been done in $1d$ in Ref. \cite{folding}. 

Our approach in this paper is to use CTMs and corner tensors to approximately contract TNs like the one in Fig.(\ref{fig10}) for different types of systems. This is actually a direct generalization of the ideas of CTMRG in Ref. \cite{CTMRG1, CTMRG2}. In our case, though, it is worth stressing a peculiarity inherent to the case of quantum lattice systems: any expectation value is by construction a hermitian expresison. This means that the coresponding TN is \emph{symmetric with respect to transposition of the vertical indices and complex conjugation}, see Fig.(\ref{fig10}). Even if quite obvious, this property needs to be explicitely taken into account in all our algorithms.  

The methods that we propose are different ways of approximating these expectation values by using CTMs and corner tensors. This is done in two steps. First, one finds a set of \emph{renormalized tensors} (e.g. renormalized CTMs) accounting effectivly for the most important correlations in the TN. Second, expectation values are approximated by using these renormalized tensors. In what follows we show how to do this in $1d$ and $2d$. 

\begin{figure}
\includegraphics[width=0.5\textwidth]{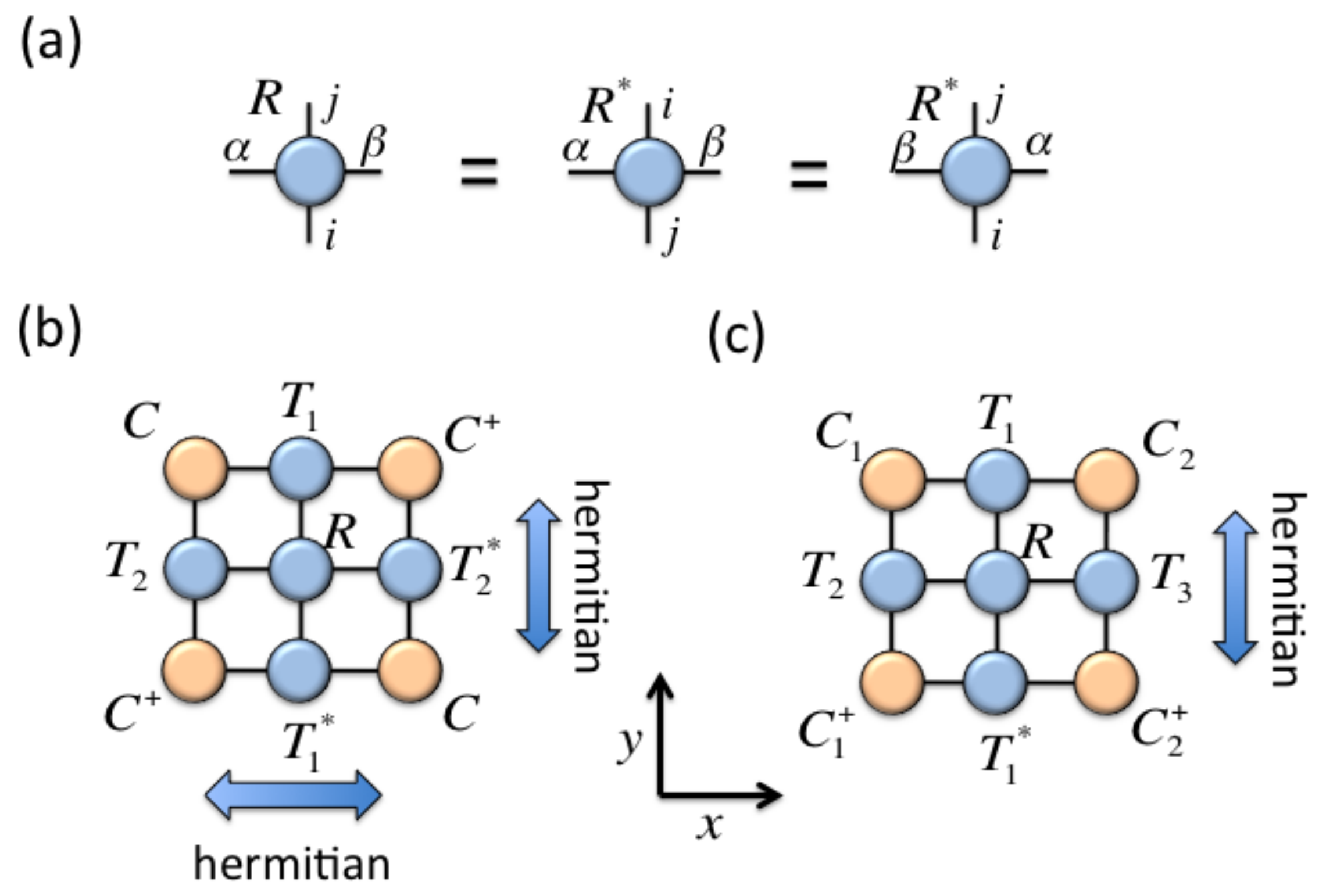}
\caption{(color online) (a) Symmetries of the MPO tensor $R$ for the simplfied one-directional $1d$ method. In the general case (full one- and two-directional $1d$ methods), only the first equality holds. (b) TN structure for the simplified one-directional $1d$ method. (c) General TN structure for the full one- and two-directional $1d$ methods.}
\label{fig11}
\end{figure}

\subsection{$1d$ quantum lattice systems}

In this paper we consider three algorithms using CTMs to approximate expectation values as in Fig.(\ref{fig10}). The first algorithm is called 'simplified one-directional $1d$ method', and is very efficient. This approach is designed for systems such that the MPO is also hermitian with respect to the horizontal indices and is explained here. The other two algorithms are considered in Appendix A.a,b. The second algortithm is the 'full one-directional $1d$ method', valid for an MPO without the previous symmetry requirement. The third algorithm is the 'full two-directional $1d$ method', which is equivalent to the CTMRG algorithm generalized to deal with the MPO considered here without extra symmetry requirements. As we shall see, the notion of 'one-directional' or 'two-directional' for each approach refers to the number of directions in which the lattice is simultaneously expanded at every step. 

The relevant parameter in these methods will be $\chi$, which is the rank of the renormalized CTMs at every step. From Eqs.(\ref{ctmm},\ref{ctm2}) and Figs.(\ref{fig12}.d,\ref{fig13}.d,\ref{fig14}.c,\ref{fig15}.c,\ref{fig17}), one can see that this is also the rank of the the renormalized reduced density matrices of the system. Thus, some of the truncations performed by the algorithms in this section are really truncations in the \emph{entanglement spectrum} (or, equivalently, the spectrum of Schmidt coefficients) of half an infinite chain \footnote{This analogy is formally valid for the truncations of the indices along the spatial directions.} 

The leading scaling of the running time (or \emph{complexity}) of the three methods is the same, namely $O(\chi^3)$. This is exactly the same complexity as iDMRG and iTEBD methods. However, the multiplicative corrections to this overall complexity are very different for each case, which implies that the \emph{total} number of operations is also quite different. The simplfied one-directional $1d$ method is the less time-consuming, whereas the full two-directional $1d$ method is the one that takes more time.  

The procedure in these algorithms is similar to the directional CTM algorithm from Ref. \cite{iPEPS2}. Namely, we (i)  \emph{insert} rows and columns in the TN in order to expand its structure, (ii) \emph{absorb} the inserted tensors by contractions towards the horizontal ($x$) and vertical ($y$) directions, or towards the corners, and (iii) \emph{renormalize} the resultant tensors in some proper way.

\subsubsection{Simplified one-directional $1d$ method}

Here we assume that the MPO is invariant under transposition of the horizontal indices plus complex conjugation, see Fig.(\ref{fig11}.a). This symmetry is taken into account in order to produce a very efficient algorithm. Thus, the TN must be invariant under the index transposition plus complex conjugation both in the horizontal ($x$) and vertical ($y$) directions, see Fig.(\ref{fig11}.b). This requirement must be satisfied at all steps in the algorithm. In order to achieve this, we consider a renormalized TN that is specified at every step by one CTM $C$, a half-column transfer matrix $T_1$, a half-row transfer matrix $T_2$, and the MPO  tensor $R$, see Fig.(\ref{fig11}.b). Notice that tensors $T_1$ and $T_2$ can also be interpreted as the tensors of two MPSs respectively in the horizontal and vertical directions (see e.g. Fig.(\ref{fig19})). The algorithm implements what we call $x$- and $y$-moves. These are as follows:

\begin{figure}
\includegraphics[width=0.5\textwidth]{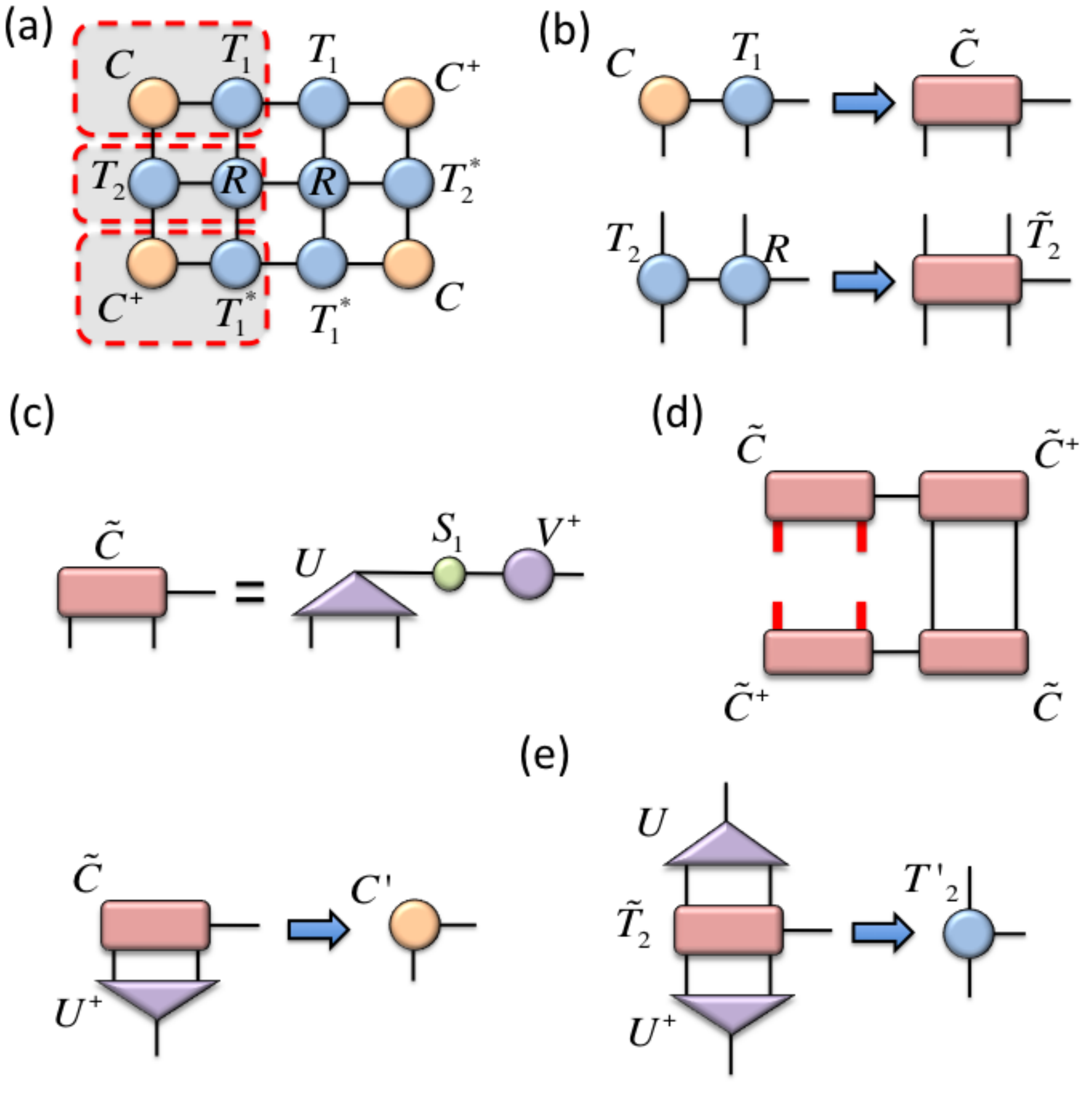}
\caption{(color online) $x$-move for the simplified one-directional $1d$ method, see text. Open indices in reduced density matrices are shown in red.} 
\label{fig12}
\end{figure}

\begin{enumerate}
\item{{\bf $x$-move:}}
\begin{enumerate}
\item{\underline{Insertion}. We insert one new column in the TN, as shown in Fig.(\ref{fig12}.a).}
\item{\underline{Absorption}. We absorb the new column towards the left, as indicated in Fig.(\ref{fig12}.a,b). At this step we produce the new (unrenormalized) tensors $\widetilde{C}$ and $\widetilde{T}_2$.}
\item{\underline{Renormalization}. This is done by means of the isommetry $U$ as shown in Fig.(\ref{fig12}.e). Importantly, this isommetry is obtained from the \emph{singular value decomposition} ot the CTM $\widetilde{C}$ as in Fig.(\ref{fig12}.c). As a result of this, we obtain the renormalized CTM $C'$ and half-row transfer matrix $T_2'$, see Fig.(\ref{fig12}.e). Notice that $U$ is, in fact, the unitary matrix that diagonalizes the reduced density matrix from Fig.(\ref{fig12}.d). Finally, since the system is symmetric with respect to horizontal transposition plus complex conjugation, we can just use the new tensors to obtain the complete TN as shown in the diagram of Fig.(\ref{fig11}.b).} 
\end{enumerate}

\item{{\bf $y$-move:}} 
\begin{enumerate}
\item{\underline{Insertion}. We insert one new row in the TN, as shown in Fig.(\ref{fig13}.a).}
\item{\underline{Absorption}. We absorb the new row towards up, as indicated in Fig.(\ref{fig13}.a,b). At this tep we produce the new (unrenormalized) tensors $\widetilde{C}$ and $\widetilde{T}_1$.}
\item{\underline{Renormalization}. This is done by means of the isommetry $V$ as shown in Fig.(\ref{fig13}.e). This isommetry is obtained again from the \emph{singular value decomposition} of the CTM $\widetilde{C}$, see Fig.(\ref{fig13}.c). As in the previous case, $V$ is the unitary operator that diagonalizes the reduced density matrix in Fig.(\ref{fig13}.d). Finally, we use the new tensors to obtain once again the complete TN as in Fig.(\ref{fig11}.b).}
\end{enumerate}
\end{enumerate}

\begin{figure}
\includegraphics[width=0.5\textwidth]{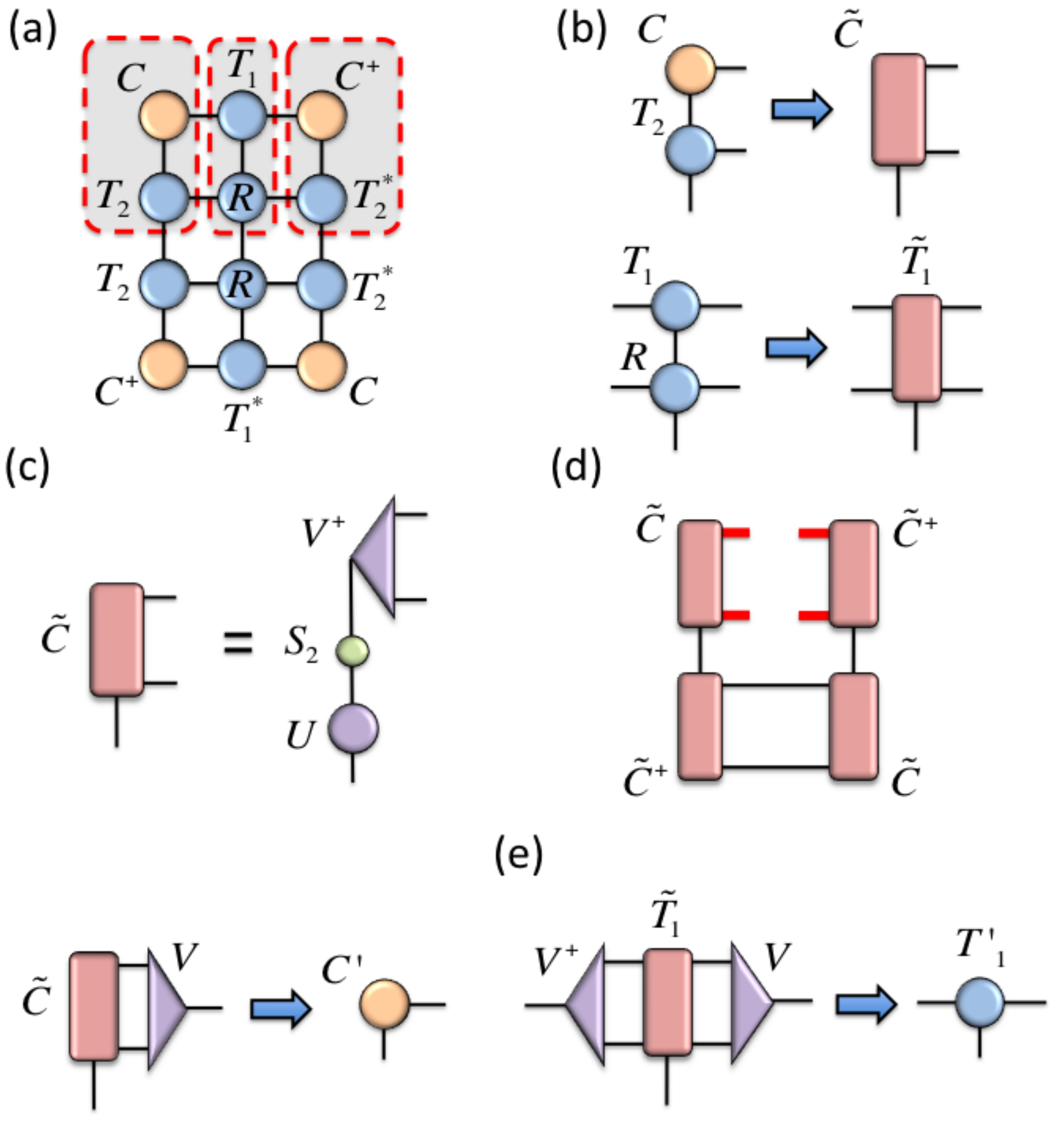}
\caption{(color online) $y$-move for the simplified one-directional $1d$ method, see text. Open indices in reduced density matrices are shown in red.} 
\label{fig13}
\end{figure}

The whole algorithm now proceeds by iterating these two moves until convergence of some relevant quantity (e.g. the spectrum of singular values of the CTM). This approach is valid for computing imaginary time evolution (and thus ground states) under an MPO with the required symmetry properties, e.g. the quantum Ising or antiferromagnetic Heisenberg and XX models described in Sec.II.E. The leading number of operations of the metod is $O(\chi^3)$, where $\chi$ is the rank of the CTM $C$. This approach has a number of advantages: it is extremely efficient in finding ground states (the \emph{total} number of operations is very low), and also preserves the spatial symmetries of the evolution operator at every step. Let us also stress the fact that the net result of the $y$-move is, in fact, equivalent to (i) a rotation of the $2d$ TN by $\pi/2$, followed by (ii) an $x$-move, and then followed by (iii) a rotation of the TN by $-\pi/2$. This relation is possible because of the symmetries of the TN in Fig.(\ref{fig11}.b). 

Let us stress another important and intriguing fact about this algorithm, which is that \emph{there are no explicit truncations}. That is, the rank $\chi$ of the CTM is specified from the very beggining in the initial TN, and \emph{does not grow at all throughout the evolution}. This is a key difference with other methods such as CTMRG and iTEBD, where the analogous to this rank grows at every step in the algorithm and thus needs to be truncated at every step. Here, though, the situation is more subtle. One can think of the truncation begin implemented \emph{implicitely}. The initial choice of rank $\chi$ for the CTM defines already this implicit truncation, and then $\chi$ is preserved all along the algorithm. Even if this is conceptually strange, it is really not a problem for the method: every time we insert a column or row, the number of indices of the CTM proliferates in one direction while it is kept constant in the other, which means that the rank of the CTM can not grow. It is this property of the algorithm what implements an automatic implicit truncation and, thus, we do not need to truncate explicitely at any step. However, in Sec.III.D we will see that this property no longer holds in $2d$ since the indices of the corresponding corner tensors always proliferate in, at least, two directions, so explicit truncations will always be needed in that case. 

\subsubsection{Expectation values}

Once the algorithm has converged, one is in position of computing expectation values of local operators. For this, one needs to consider the contraction of the TN together with the operator for the observable that one wishes to calculate. The CTM method explained before allows to approximate such a calculation. We show this for the general case in Fig.(\ref{fig19}) (up to normalization of the wavefunction \footnote{Again, we remark that the actual expectation value is the ratio between these expressions, and the same ones but where observable operators are replaced by identities.}) for one-site operators, two-site operators between nearest-neighbours, and also for two-point correlation functions. As is well known from MPS methods, all these calculations can also be done in $O(\chi^3)$ operations. 

\begin{figure}
\includegraphics[width=0.4\textwidth]{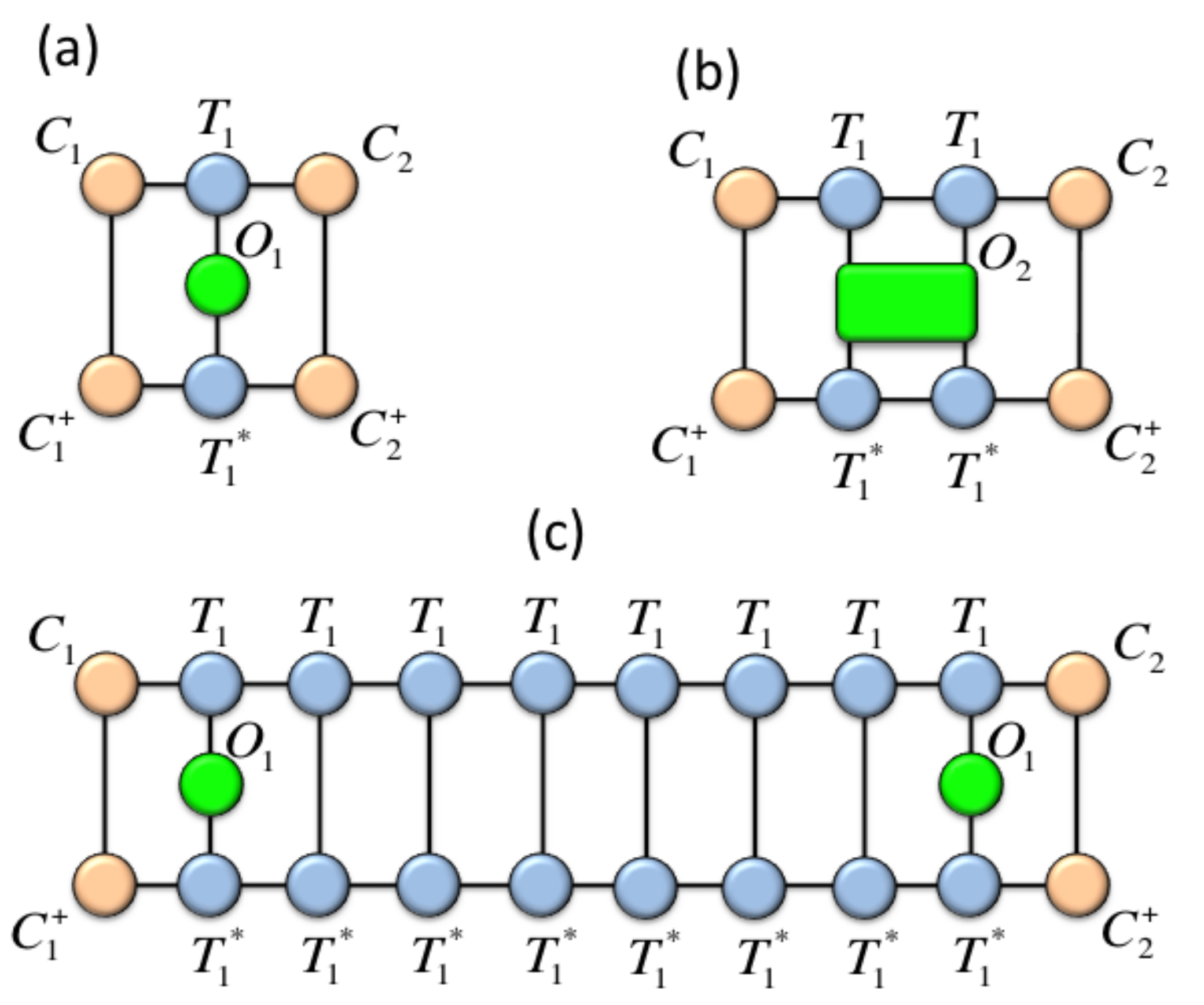}
\caption{(color online) Different expectation values, up to normalization, for a $1d$ system: (a) one-body operator $o_1$, (b) two-body operator $o_2$, and (c) two-point correlation function with a separation of seven lattice sites. The normalization is done by dividing each one of these expressions by the same expression where observables are replaced by identity operators.} 
\label{fig19}
\end{figure}

\subsection{$2d$ quantum lattice systems}

In this section we discuss how to generalize the ideas from the previous section to the $2d$ case. This generalization is conceptually straightforward, but there are a number of significant differences with respect to the $1d$ case that are worth stressing. Let us summarize these differences: 
\begin{enumerate}[(i)]
\item{Since more spatial dimensions come into play, we have more possibilities to effectively expand the relevant TN. Thus, we discuss \emph{four} possible corner algorithms, as opposed to the \emph{three} discussed in $1d$.}
\item{In any of the approaches the indices in the different tensors proliferate at every step in, at least, two different directions. Thus, this growth needs to be always truncated explicitely, regardless of the approach.}
\item{The spectrums of the corner tensors are no longer directly associated to the spectrums of the reduced density matrices of the system. Nevertheless, these spectrums still carry important information about the correlations in the system.}
\item{The complexity of each approach depends dramatically on the level of symmetry and chosen renormalization scheme. Unlike in the $1d$ case, where all the methods had the same complexity (namely $O(\chi^3)$), in $2d$ there are differences in several orders of magnitude (see Table (\ref{tabu}) in Sec.III.E).}
\item{Our algorithms produce separate tensors for the 'bra' and 'ket' parts of the TN, see Fig.(\ref{fig20}). Thus, the TN is always a positive-definite object by construction. This was also the case in $1d$, but was not found to be of special relevance there. However, in $2d$ this is important since the truncation approach differs from the one used in e.g iPEPS and TERG algorithms. In those algorithms,  the 'bra' and 'ket' parts of the TN are approximated simultaneously by some common tensors, which in practice breaks the positivity requirement and may lead to numerical instabilities (see e.g. the analysis in Ref. \cite{Jordan}). In a way, the approach in this paper is similar to the single-layer methods in Ref. \cite{single}.} 
\end{enumerate}

\begin{figure}
\includegraphics[width=0.5\textwidth]{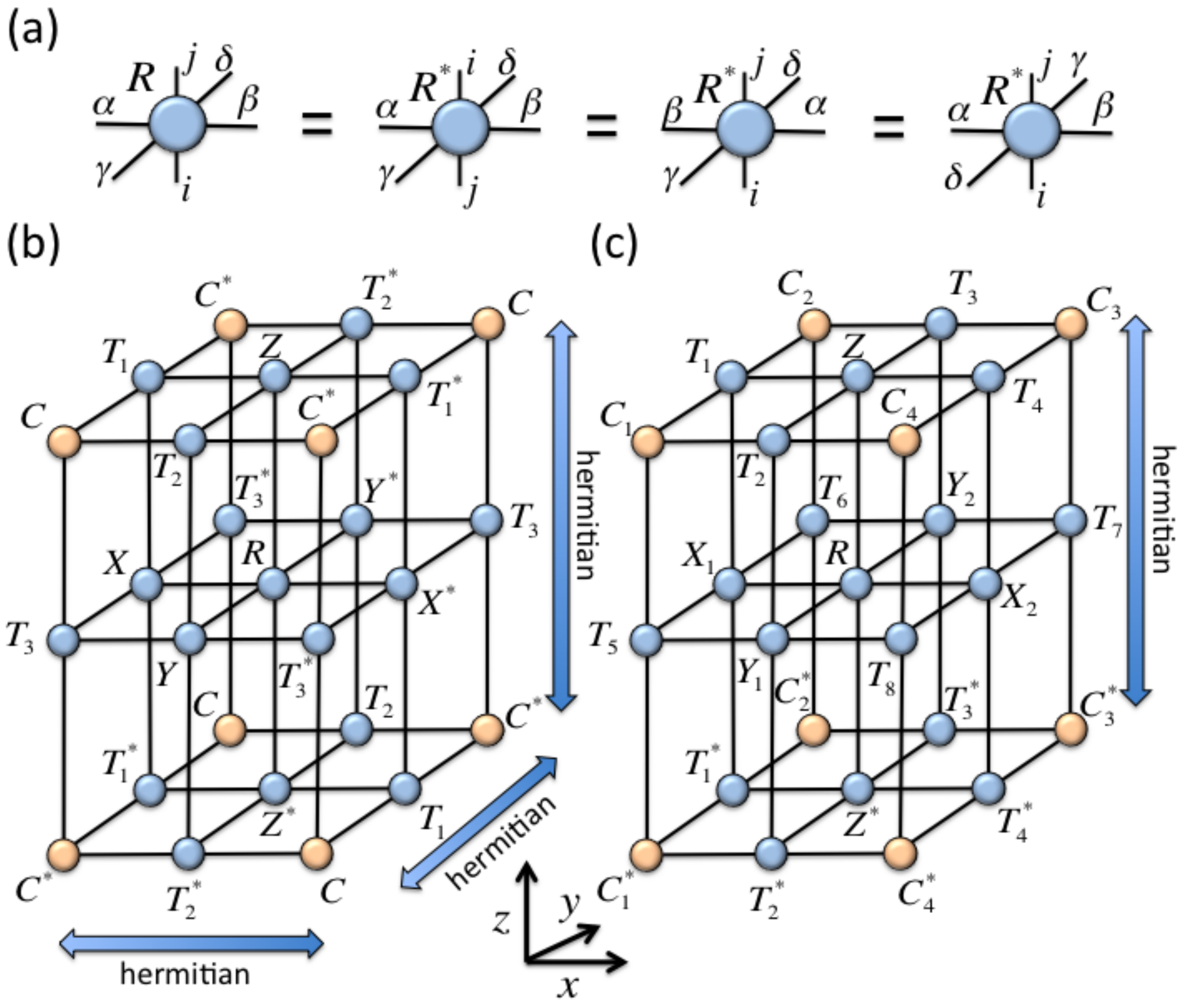}
\caption{(color online) (a) Symmetries of the PEPO tensor $R$ for the simplified one-directional $2d$ method. In the general case (full one-, two- and three-directional $2d$ methods) only the first equality holds. (b) TN structure for the simplified one-directional $2d$ method. (c) General TN structure for the full one-, two- and three-directional $2d$ methods. } 
\label{fig20}
\end{figure}

Before entering into details of the methods, let us mention that another alternative for a $2d$ algorithm based on CTMs is actually possible. Namely, one could use some $1d$ method with CTMs to compute effective environments of a $2d$ iPEPS, and use it in the context of some tensor update to simulate time evolution (e.g. 'simplified' or 'full' updates \cite{Xiang,iPEPS}). This approach is also valid, and has already been explored \cite{iPEPS2}. However, the algorithms that we explain in this section are based on a completely different approach. 

Here, we first present a 'simplified one-directional $2d$ method'. As in the $1d$ case, this simplified method is designed for systems such that the tensor defining the PEPO is hermitian with respect to the two spatial directions independently. The implementation of this algorithm is quite efficient and  the steps will be explained in detail. In Appendix A.c, we discuss three different 'full' $2d$ methods which are valid for PEPO tensors without the previous symmetry requirement. In increasing level of complexity, these approaches are called 'full one-directional, full two-directional, and full three-directional $2d$ methods'. We only sketch briefly the main idea behind them, and from here the interested reader can infer very easily their step-by-step implementation. As we shall see in the Appendix, these algorithms are less efficient than the simplified directional $2d$ method, and the complexity for each one of them is also very different. In fact, sometimes the complexity may be higher than the calculation of expectation values itself, as we will see in Sec.III.E. Still, we believe that it is important (at least from a conceptual perspective) to be aware of the existence of all these possibilities. 

For $2d$ quantum lattice systems, the CTMs are generalized to corner tensors (see Fig.(\ref{fig7})), which are tensors with at least three indices, one for each direction of the lattice. Thus, different singular value decompositions of corner tenors may have different ranks. In practice, one can work with an effective rank $\chi$ for all the decompositions, and this is the relevant parameter for the $2d$ methods. However, it is also possible to work with separate ranks, e.g. $\chi$ for the vertical direction and $D$ for the horizontal directions. This choice of ranks would be analogous to the iPEPS algorithm, where we have an iPEPS of bond dimension $D$ for which an effective environment with tensors of bond dimension $\chi'$ is computed (in our approach, we have that $\chi' \approx \chi^2$ very roughly, since we produce 'bra' and 'ket' tensors independently). For simplicity, in this paper we choose always the same rank when explaining the details of the methods (including complexity issues). However, we sometimes choose different ranks for the different directions in the numerical calculations of Sec.IV. Let us also remark that, again and as in the $1d$ case, the truncations in the vertical indices for the full methods are in fact in the entanglement spectrum of a corner of the $2d$ quantum lattice system, e.g. see again Fig.(\ref{fig7}).  

As in the $1d$ case, the procedure for these methods is always the same: (i) \emph{insertion} of tensors, (ii) \emph{absorption} of the inserted tensors towards some direction or towards some corner, and (iii) \emph{renormalization} by means of some isommetry or rectangular matrix. Also, in this section we use the following notation: the spatial (horizontal) directions are called $x$ and $y$, whereas the temporal (vertical) direction is called $z$ (see Fig.(\ref{fig20})). Also, \emph{$x$-indices} refer to indices that connect the tensors in the TN in the $x$ direction (and similar definitions apply for $y$ and \emph{$z$-indices}). 

\begin{figure}
\includegraphics[width=0.5\textwidth]{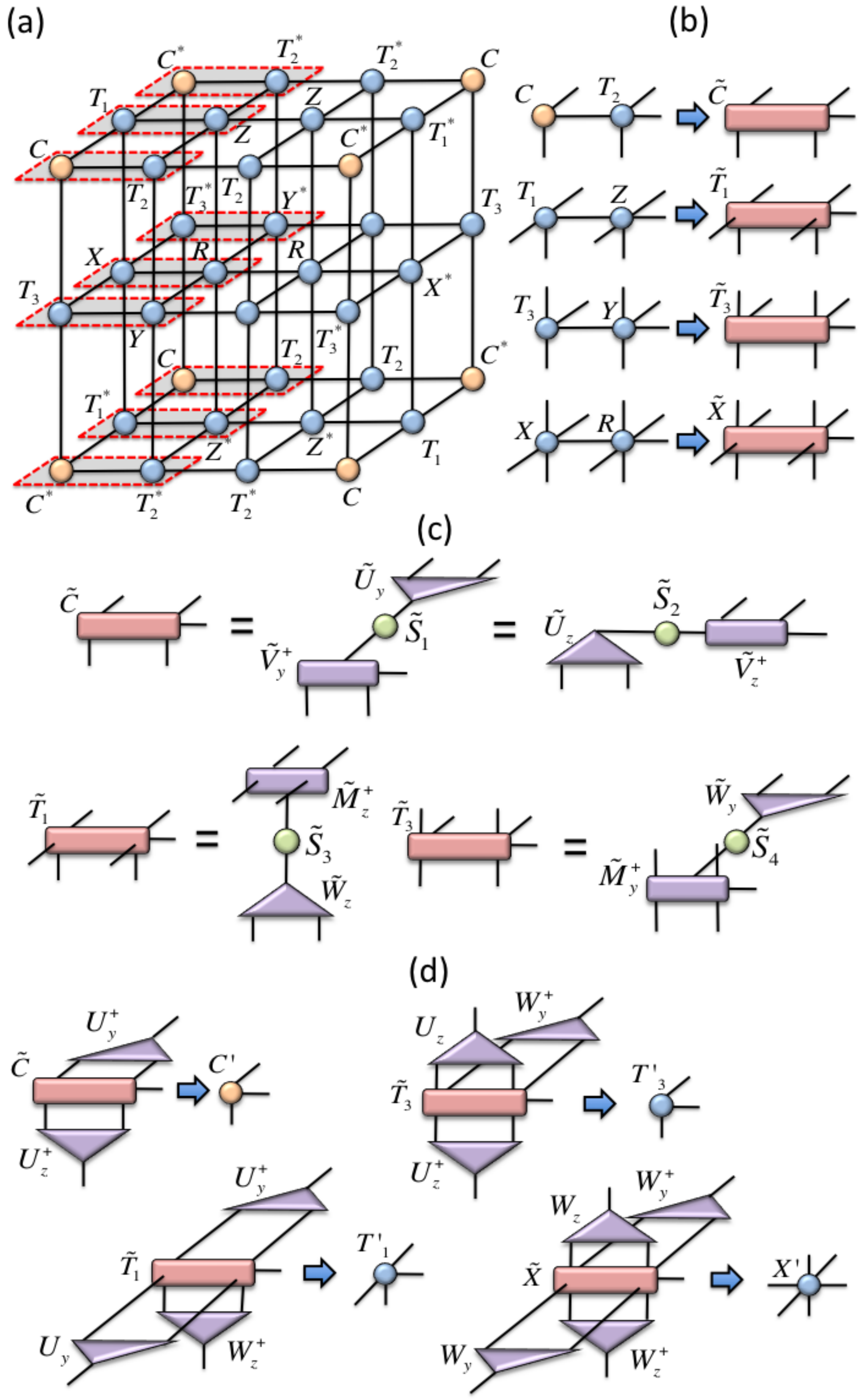}
\caption{(color online) $x$-move for the simplfied one-directional $2d$ method, see text.} 
\label{fig21}
\end{figure}

\subsubsection{Simplified one-directional $2d$ method}

Here we assume that the PEPO is invariant under transposition of the $x$ and $y$ indices independently plus complex conjugation, see Fig.(\ref{fig20}.a). As in the $1d$ case, this extra symmetry is taken into account in order to produce a very efficient algorithm. Thus, the TN must be hermitian in the three directions $x, y$ and $z$, see Fig.(\ref{fig20}.b). This requirement will be satisfied at all steps in the algorithm. Therefore, the renormalized TN is specified at every step by one corner tensor $C$ (analogue to the CTM in $1d$), three tensors $T_1, T_2$ and $T_3$ (analogue to the half-row and half-column transfer matrices in $1d$), and three tensors $X,Y$ and $Z$ (coresponding to some renormalized iPEPS for each one of the three planes $yz,zx$ and $xy$), see Fig.(\ref{fig20}.b) The algorithm now implements what we call $x$-, $y$- and $z$-moves. As in $1d$, these moves are equivalent up to rotations of the whole lattice by $\pi/2$. Thus, for simplicity we explain in detail e.g. the $x$-move, and then explain how the $y$- and $z$-moves can be related to the $x$-move by rotations. 

\begin{enumerate} 
\item{{\bf $x$-move:}}
\begin{enumerate} 
\item{\underline{Insertion}. We insert one new plane of tensors in $y$ and $z$ directions of the TN, as shown in Fig.(\ref{fig21}.a).}
\item{\underline{Absorption}. We absorb the tensors from the new plane towards the left hand side in the $x$ direction, as indicated in Fig.(\ref{fig21}.a). At this step we produce the new (unrenormalized) tensors $\widetilde{C}, \widetilde{T}_1, \widetilde{T}_3$ and $\widetilde{X}$, see Fig.(\ref{fig21}.b).}
\item{\underline{Renormalization}. This is done by the isommetries $U_y, U_z, W_y$ and $W_z$ as shown in Fig.(\ref{fig21}.d). These are found as follows: first, we find the isommetries $\widetilde{U}_y, \widetilde{U}_z, \widetilde{W}_y$ and $\widetilde{W}_z$ from the \emph{singular value decompositions} of tensors $\widetilde{C}, \widetilde{T}_1$ and $\widetilde{T}_3$ that are shown in Fig.(\ref{fig21}.b). Then, we perform an explicit truncation in the $\chi$ largest singular values respectively of all these decompositions, and find the isommetries $U_y, U_z, W_y$ and $W_z$. With this, we obtain the renormalized tensors $C', T'_1, T'_3$ and $X'$, see Fig.(\ref{fig21}.d).}
\end{enumerate}

\item{{\bf $y$-move:} rotate the TN by $\pi/2$ in the $xy$ plane as shown in Fig.(\ref{fig22}), and do an $x$-move. Then, rotate the TN back to the original position.}

\item{{\bf $z$-move:} rotate the TN by $\pi/2$ in the $zx$ plane as shown in Fig.(\ref{fig22}), and do an $x$-move. Then, rotate the TN back to the original position.}

\end{enumerate}

\begin{figure}
\includegraphics[width=0.5\textwidth]{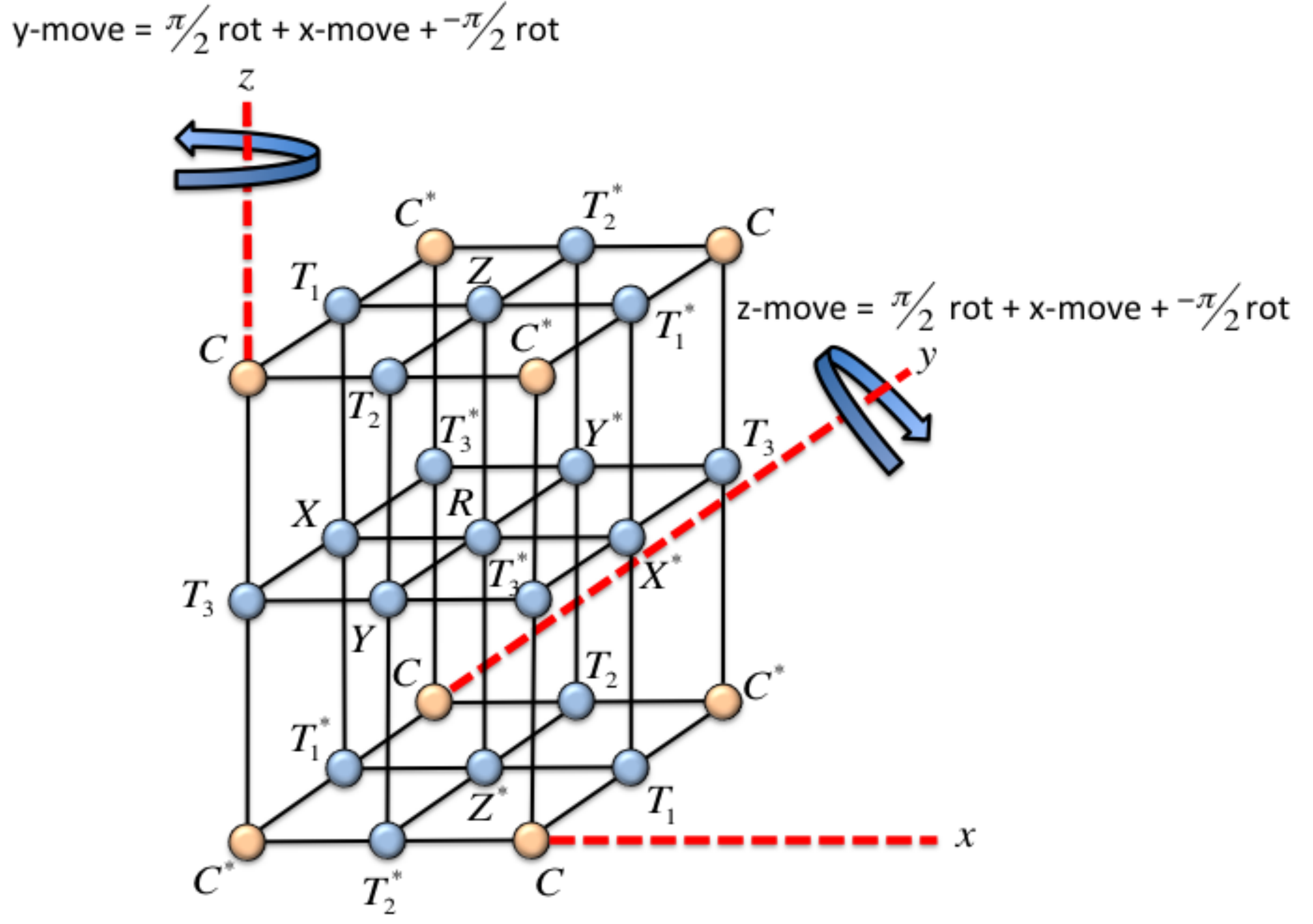}
\caption{(color online) Relation between the $x$-, $y$- and $z$-moves in the simplified one-directional $2d$ method.} 
\label{fig22}
\end{figure}

The algorithm works again by iteration of the above three steps until convergence of e.g. the singular value spectrums of the corner tensors. As in the $1d$ case, this approach is valid for computing ground states by doing imaginary time evolution driven by a PEPO with the required symmetry properties, e.g. the quantum Ising, Heisenberg and XX models described in Sec.II.E. The leading number of operations in this method is $O(\chi^7)$ if the different contractions are done by making use of the geometric structure of the TN (remember the example from Fig.(\ref{fig2})). As in $1d$, this approach has the advantage of being quite efficient, and also keeps the translational invariance of the evolution operator at every step. 

Several remarks are in order. First notice that,  as hinted previously, the indices in the tensors proliferate in two different directions at each move, and thus one needs explicit truncations in $\chi$. But second, and also unlike in $1d$, this time the different truncations \emph{are not associated to the truncations in any reduced density matrix of the system}. We thus can think of this method as an over-simplified algorithm that implements some 'acceptable' truncation scheme in a very efficient way. We believe that the isommetries found in this way, despite not being the best possible, are still good choices as long as the amount of entanglement in the system is not too large (in a way similar to the simplified update \cite{Xiang}). In the end, whether this approach is useful or not can only be assessed by numerical simulations, and this is what we will do in Sec.IV. Notice also that these truncations do not correspond to truncations in any entanglement spectrum of the system. For this, we should construct explicitely  the reduced density matrices of the system, thus leading to less efficient but probably more accurate algorithms. This is precisely what we discuss in the algorithms of Appendix A.c. 

\subsubsection{Expectation values}

As in $1d$, once the algorithm has converged it is possible to compute expectation values of local operators. This is shown for the general case in Fig.(\ref{fig24}) (again up to normalization of the wavefunction) for one-site and two-site operators between nearest neighbours as well as for a two-point function along the $x$ direction. This time the required calculations can all be done in $O(\chi^{11})$ operations by choosing the appropriate order in the TN contraction of the TN. The fact that the complexity is $O(\chi^{11})$ means that, for the simplfied one-directional $2d$ method, this is actually the bottleneck of the calculation, whereas for the rest of the methods explained in Appendix A.c the bottleneck is the calculation of the reduced density matrices in Fig.(\ref{fig23}). 

\begin{figure}
\includegraphics[width=0.5\textwidth]{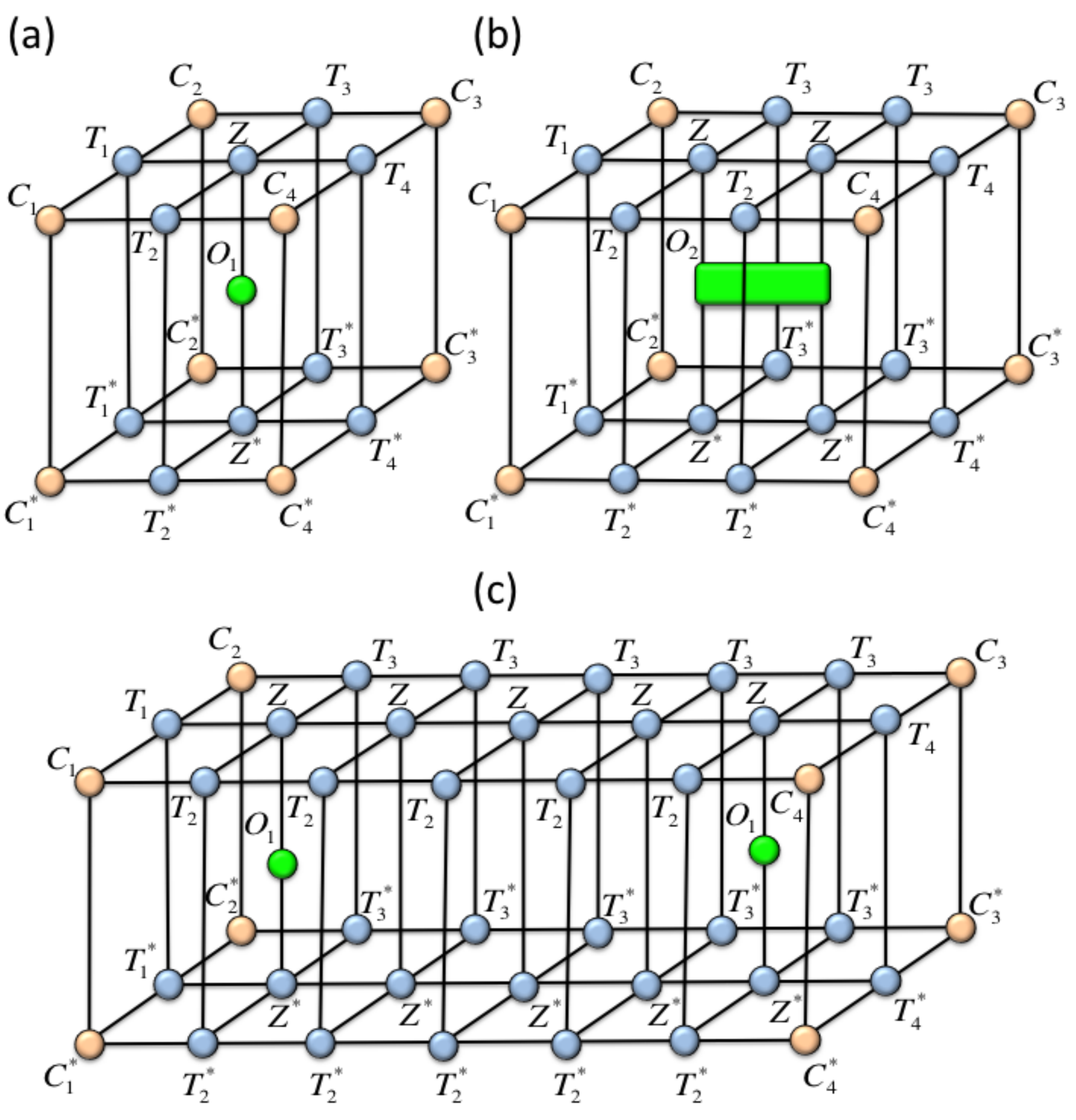}
\caption{(color online) Different expectation values, up to normalization, for a $2d$ system: (a) one-body operator $o_1$, (b) two-body operator $o_2$, and (c) two-point correlation function with a separation of four lattice sites. The normalization is done by dividing each one of these expressions by the same expression where observables are replaced by identity operators.}
\label{fig24}
\end{figure}

\subsection{Summary of methods and complexities}

In the previous sections we have discussed two different approaches to simulate quantum lattice systems with corner transfer matrices and corner tensors. The first approach is the one from Ref. \cite{iPEPS2}: use a $(d-1)$-dimensional corner method to approximate effective environents in the context of an iPEPS algorithm in $d$ dimensions. This is the approach that we discussed in Sec.III.D for $3d$ quantum lattice systems. The other approach, which is the one that we presented in detail here, is to implement directly a corner method in $(d+1)$ dimensions to approximate the contraction involved in the calculation of expectation values of local observables for quantum lattice systems in $d$ dimensions, assuming an evolution driven by some suitable MPO or PEPO. This has been done in Sec.III.B and Sec.III.C, as well as Appendix A.  

The complexity in the algorithms presented in Sec.III.B, Sec.III.C and Appendix A is summarized in Table \ref{tabu}. In $1d$, all the methods that we have studied here have the same complexity, which is also the same as in iTEBD and iDMRG. However, in $2d$ the complexity depends on the chosen method. As expected, $2d$ methods are harder to implement numerically than $1d$ methods. Nevertheless, these methods can also be implemented in practice within some limitations\footnote{See e.g. Ref. \cite{CTM3d}.}. 

\begin{table}[h]
\begin{center}
\begin{tabular}{||c||c||c||c||}
\hline
~~~~~~~{\bf Method}~~~~~~~ & ~~~~~{\bf $1d$}~~~~~ & ~~~~~{\bf $2d$}~~~~~  \\ 
\hline
\hline
~~~~~~Simplified one-directional~~~~~~ & $O(\chi^3)$ & $O(\chi^7)$ \\
\hline
Full one-directional & $O(\chi^3)$ & $O(\chi^{11})$ \\
\hline
Full two-directional & $O(\chi^3)$ & $O(\chi^{13})$ \\
\hline
Full three-directional & - & $O(\chi^{17})$ \\
\hline
\hline
Expectation values & $O(\chi^3)$ & $O(\chi^{11})$ \\ 

\hline 
\end{tabular} 
\end{center}
\caption{Complexity of the methods in Sec.III.B, Sec.III.C and Appendix A, defined as the leding term in the number of operations. The calculation of expectation values is also added for comparison. In $1d$ the three complexities are the same, but the total running time for each algorithm is different because of subleading corrections and constant multiplicative terms.}
\label{tabu}
\end{table}

\section{Benchmark: $1d$ and $2d$}

In what follows we present preliminary numerical results for some of the algorithms discussed previously. Specifically, we have considered the calculation of some of the ground state properties of the ferromagnetic quantum Ising model in transverse magnetic field $h$ from Eq.(\ref{ham}), both in $1d$ and $2d$. The properties of this model are well known \cite{McCoy,IsingMC}, and hence it is useful as a first benchmark for our methods. 

\begin{figure}
\includegraphics[width=0.48\textwidth]{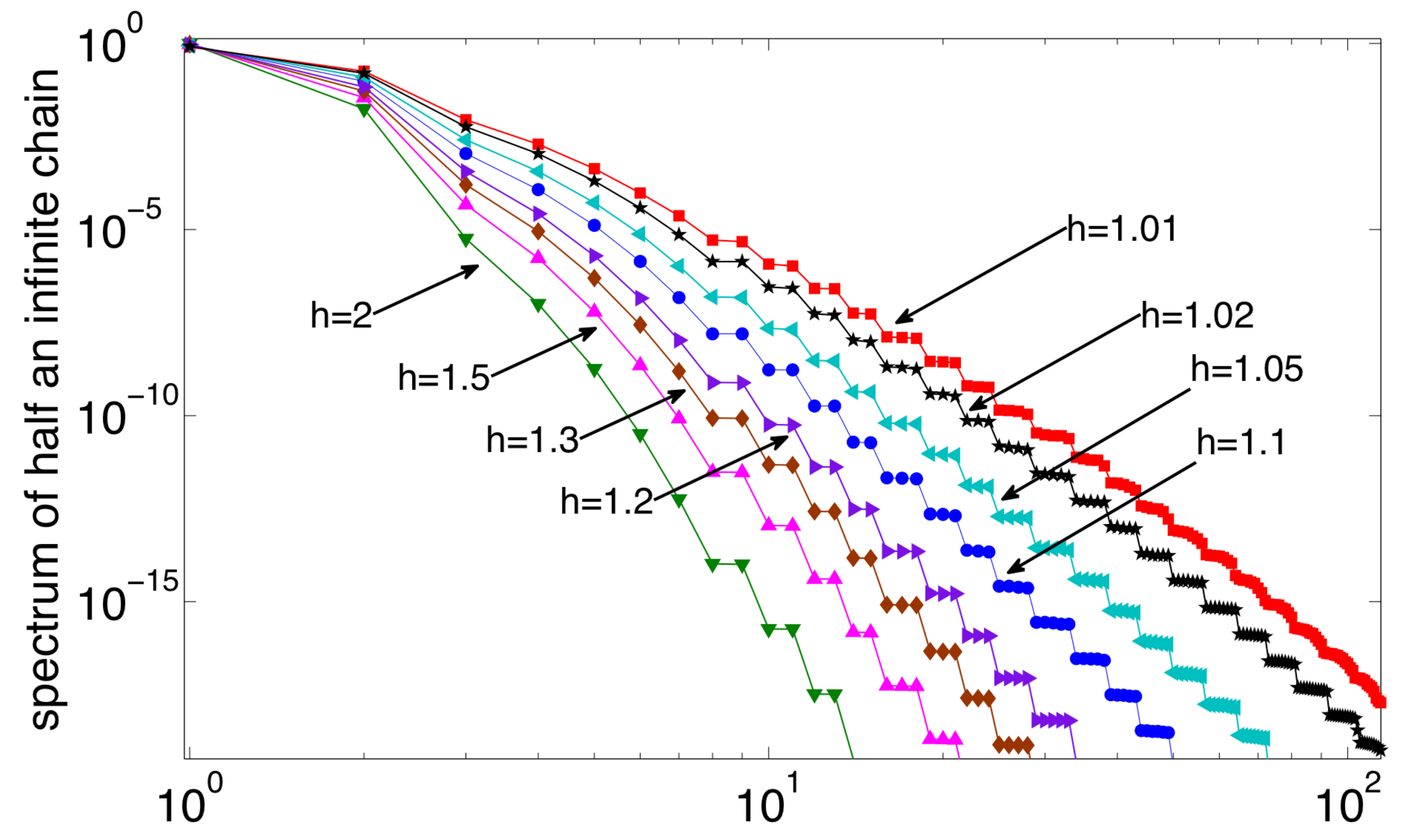}
\caption{(color online) Entanglement spectrum of half an infinite chain for the $1d$ ferromagnetic quantum Ising model in transverse field, for different field values $h$.} 
\label{plot1}
\end{figure}
\begin{figure}
\includegraphics[width=0.45\textwidth]{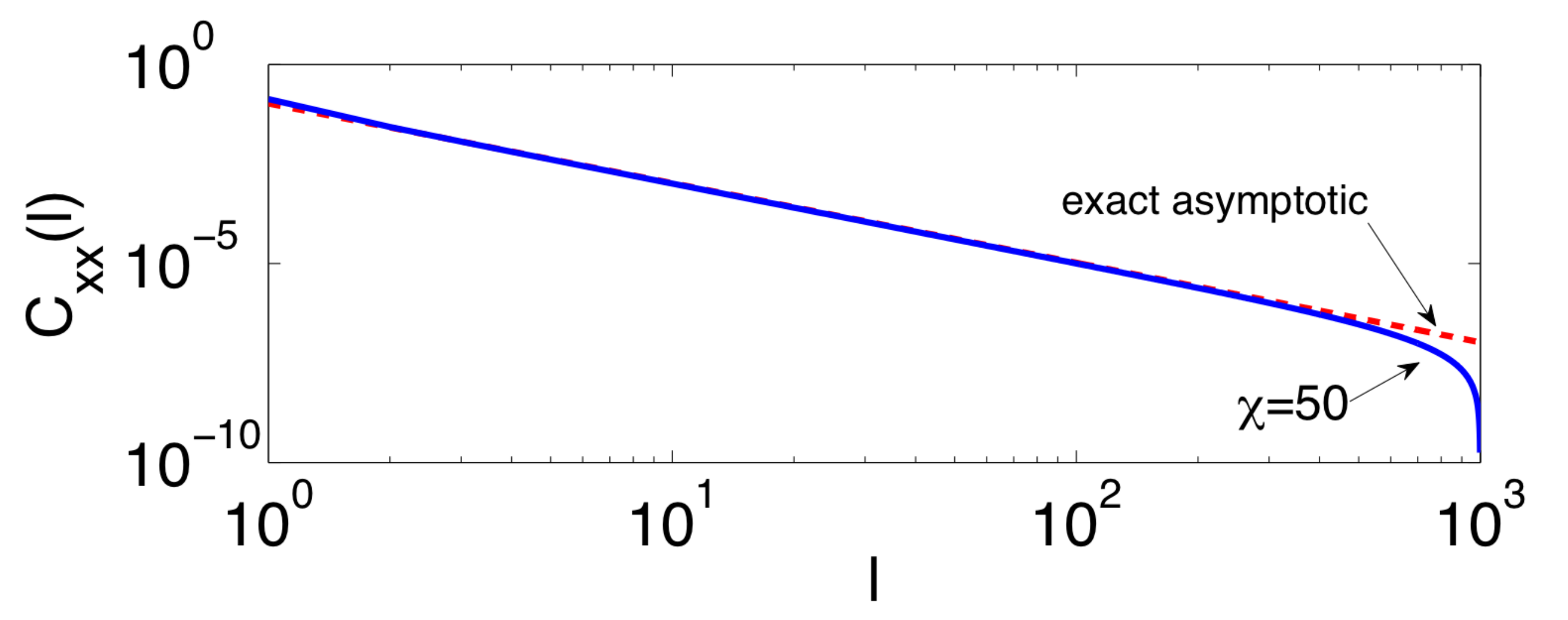}
\caption{(color online) Critical correlation function $C_{xx}(l)$ as a funtion of the spin separation $l$ for the $1d$ ferromagnetic quantum Ising model in transverse field at criticality. The dashed dotted line is the asymptotic behavior in Eq.(\ref{asy}).} 
\label{corr}
\end{figure}

\subsection{1d}

In $1d$ the algorithm in Sec.III.B and the ones in Appendix A.a,b seem to produce similar results for this model. In Fig.(\ref{plot1}) we show a calculation of the entanglement spectrum of half an infinite chain for different values of $h$ (focusing on the first $130$ spectral values). This is computed from the converged spectrum of eigenvalues of the CTM as in Eq.(\ref{ctm2}). The calculated spectrums coincide with remarkable accuracy with the ones in the literature (see e.g. Ref. \cite{iTEBD1, iTEBD2}), and were obtained in just a few minutes with very modest computational resources.

We have also computed the two-point correlation function $C_{xx}(l)$, defined as 
\beq
C_{xx}(l) \equiv \frac{\bra{\Psi_{gs}}\sigma_x^{[r]} \sigma_x^{[r+l]} \ket{\Psi_{gs}}}{\braket{\Psi_{gs}}{\Psi_{gs}}} - m_x^2 \ ,
\eeq
where we have substracted the long-distance term $m_x^2$, with $m_x$ the expectation value of $\sigma_x$ at one site. This correlation function can be computed exactly \cite{McCoy}, and at criticality ($h=1$) it tends to decay algebraically with the separation distance $l$ as
\beq
C_{xx}(l) \sim \frac{a}{l^2} + O\left( \frac{1}{l^4} \right) 
\label{asy}
\eeq
for some constant $a$ and large $l$. In Fig.(\ref{corr}) we plot the calculated value of this correlator for $\chi=50$ and $h=1$. We see that our simulation is indeed able to reproduce the assymptotic behavior in Eq.(\ref{asy}) for a large number of sites even for this small value of $\chi$. 

\subsection{2d}

In $2d$ we have computed ground state properties of the Hamitonian in Eq.(\ref{ham}) by using the simplified one-directional $2d$ method from Sec.III.C.1. In particular, we have computed the ground state energy per site $e_0$, as well as the magnetizations per site $m_x$ and $m_z$, defined respectively as the expectation values of the $\sigma_x$ and $\sigma_z$ operators at a given site. For completeness and comparison to the $1d$ case, we have also computed the entanglement spectrum of a corner of the $2d$ lattice, as well as of half an infinite plane. In our numerical calculations we employed different truncation parameters depending on the direction. Hence, we used a truncation parameter $D$ for the indices of tensor $Z$ in Fig.(\ref{fig20}.b) in the $xy$ plane, and a parameter $\chi$ everywhere else.

\begin{figure}
\includegraphics[width=0.48\textwidth]{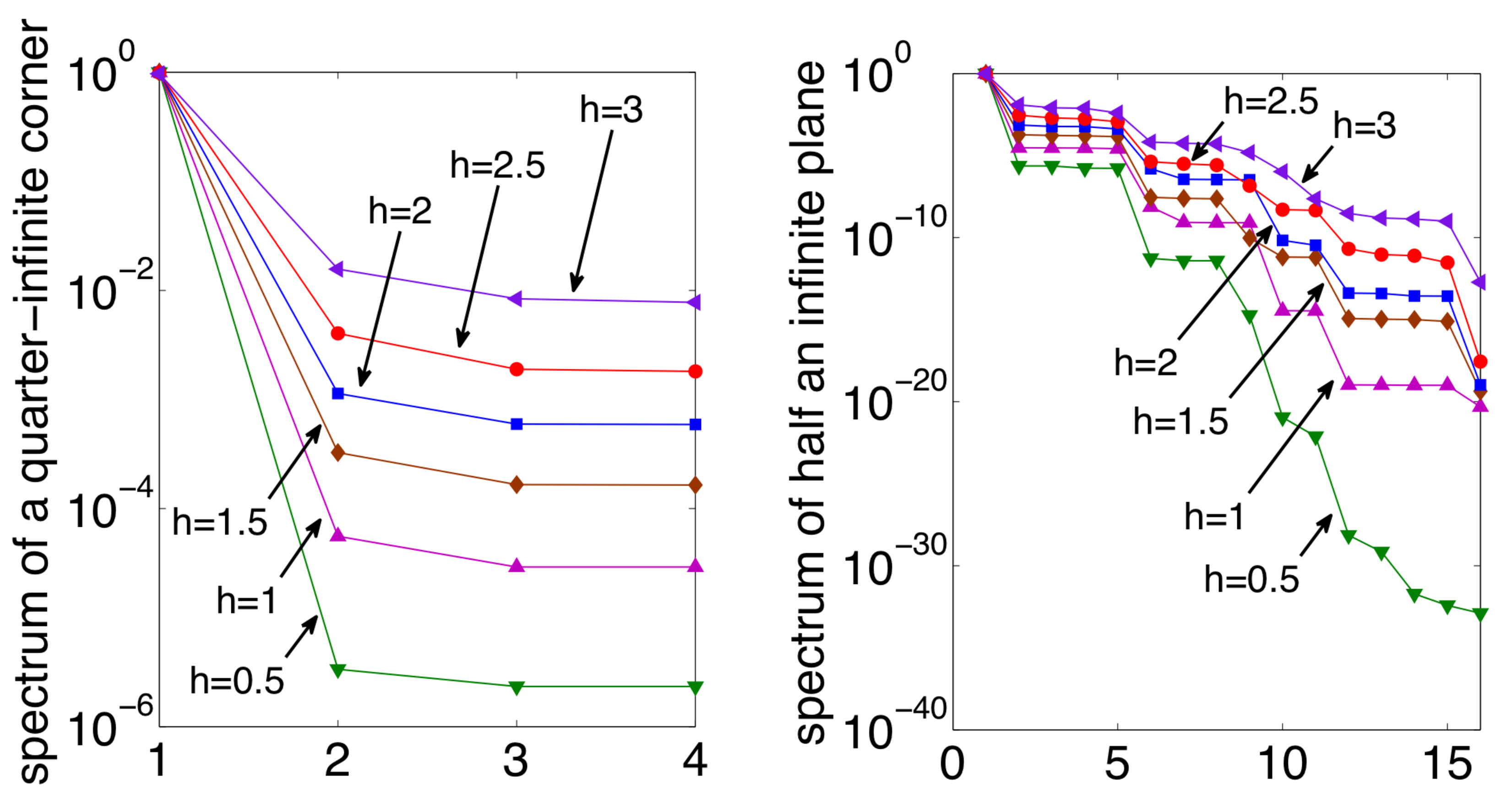}
\caption{(color online) Two different entanglement spectrums for the $2d$ ferromagnetic quantum Ising model in a transverse field for different values of $h$: for a corner (left), and for a half-infinite plane (right).}
\label{NewFig}
\end{figure}
\begin{figure}
\includegraphics[width=0.38\textwidth]{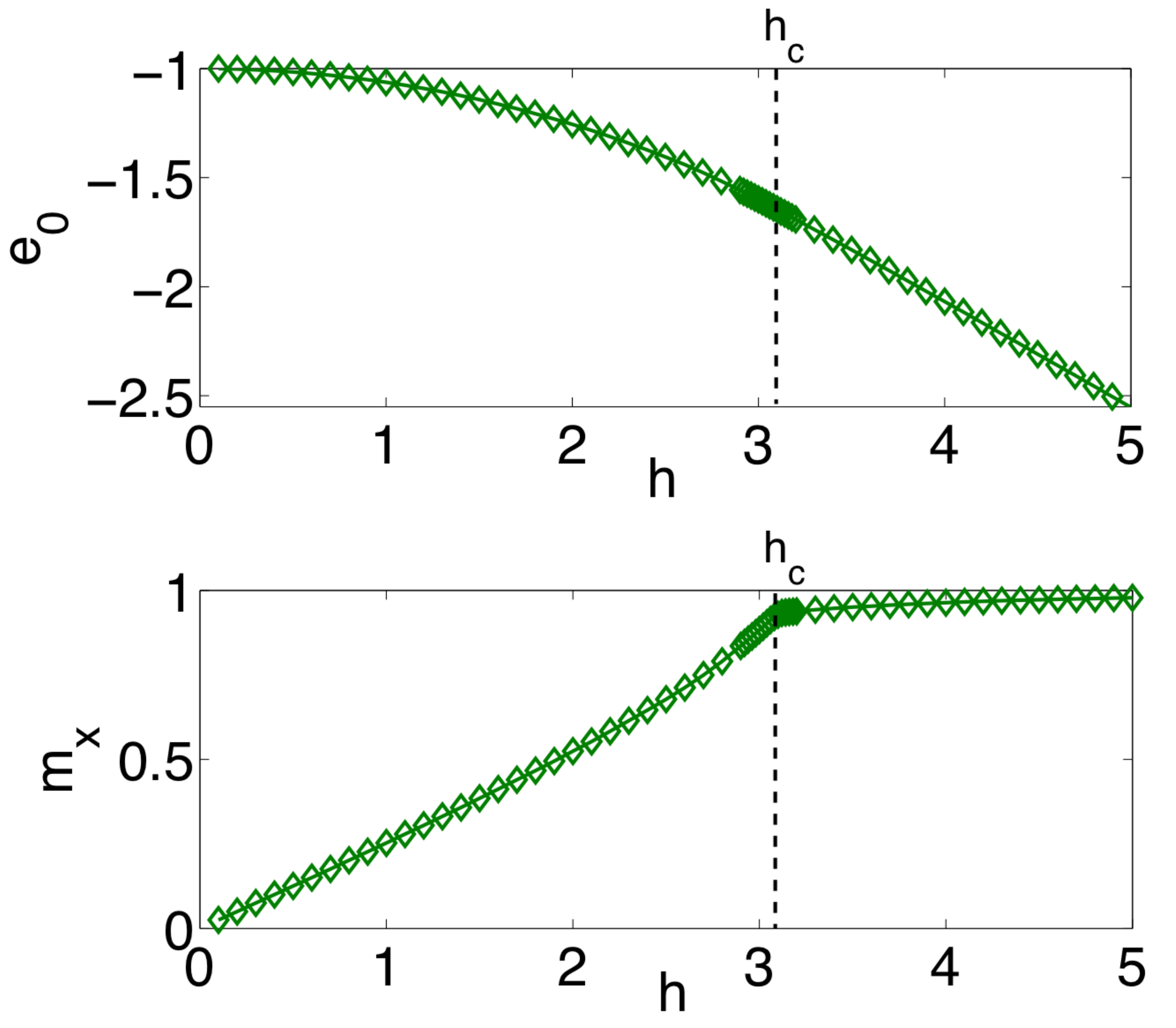}
\caption{(color online) Energy per site (left) and magnetization $m_x$ per site (right) for the $2d$ ferromagnetic quantum Ising model in transverse field. The calculation is for the simplified one-directional $2d$ method with truncation parameters $(D,\chi) = (4,4)$} 
\label{plot2}
\end{figure}

In Fig.(\ref{NewFig}) we plot the entanglement spectrum for different values of $h$ both for a corner of the infinite plane, as well as for half an infinite plane, as computed with $(D,\chi) = (4,4)$. These have been calculated from the corner tensors as in e.g. Fig.(\ref{fig7}). The finite value of $\chi$ used in our simulations truncates these spectrums in $4$ and $16$ values respectively. Our calculations show that the entanglement spectrum flattens as the critical point is approached, thus involving a much larger amount of entanglement. Interestingly, we also see that long tails in the spectrums tend to be produced close to criticality. When moving slightly away from criticality, we find a crossover region to a regime wehere the spectrums tend to decay very quickly. 

In Fig.(\ref{plot2}) we show the energy per site $e_0$ and the magnetization $m_x$ as a function of the magnetic field $h$ for $(D,\chi)$ = $(4,4)$. They follow the usual behavior for this $2d$ quantum system found by other methods (see e.g. Fig.(4) in Ref.\cite{iPEPS}) with good accuracy. In Fig.(\ref{plot3}) we show the behavior of the magnetization $m_z$, which is the order parameter, for $(D,\chi) = (4,2), (4,4)$ and $(4,6)$. As expected, close to criticality the order parameter goes to zero according to a critical exponent $\beta$ as 
\beq
m_z \sim b(h_c - h)^{\beta}
\eeq
for some constant $b$. The estimated values of $\beta$ as well as our estimations of the critical point for these values of the truncation parameters are shown in Fig.(\ref{plot4}). Notice that for $(D,\chi) = (4,2)$ the value of the critical exponent is very close to the mean field one $\beta_{MF} = 1/2$, but this gets closer to the known Montecarlo result $\beta_{MC} \sim 0.327$ \cite{IsingMC} as higher $\chi$ is considered. A comparison of the critical exponents and critical points as computed by different methods is provided in Table \ref{tabd}. We see that this approach, even if quite simple, already obtains a value for the critical exponent that is compatible with the iPEPS approach using CTMs from Ref. \cite{iPEPS2}, and a value for the critical point already better than the Vertical Density Matrix Approach (VDMA) from Ref. \cite{VDMA1, VDMA2}.

\begin{figure}
\includegraphics[width=0.48\textwidth]{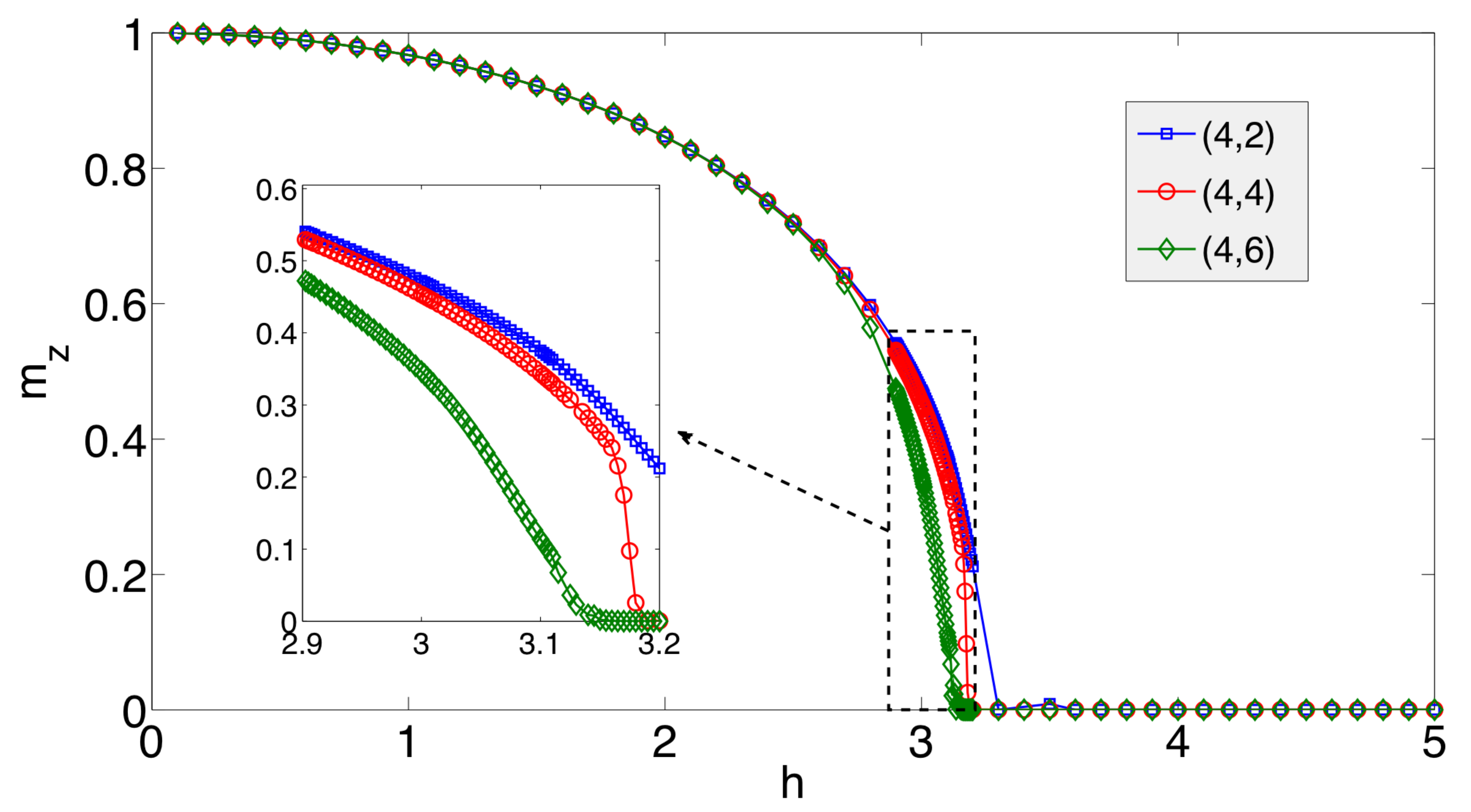}
\caption{(color online) Magnetization $m_z$ per site for the $2d$ ferromagnetic quantum Ising model in transverse field. The calculation is for the simplified one-directional $2d$ method with truncation parameters $(D,\chi) = (4,2), (4,4)$ and $(4,6)$. The inset shows the behavior around the critical region.} 
\label{plot3}
\end{figure}
\begin{figure}
\includegraphics[width=0.48\textwidth]{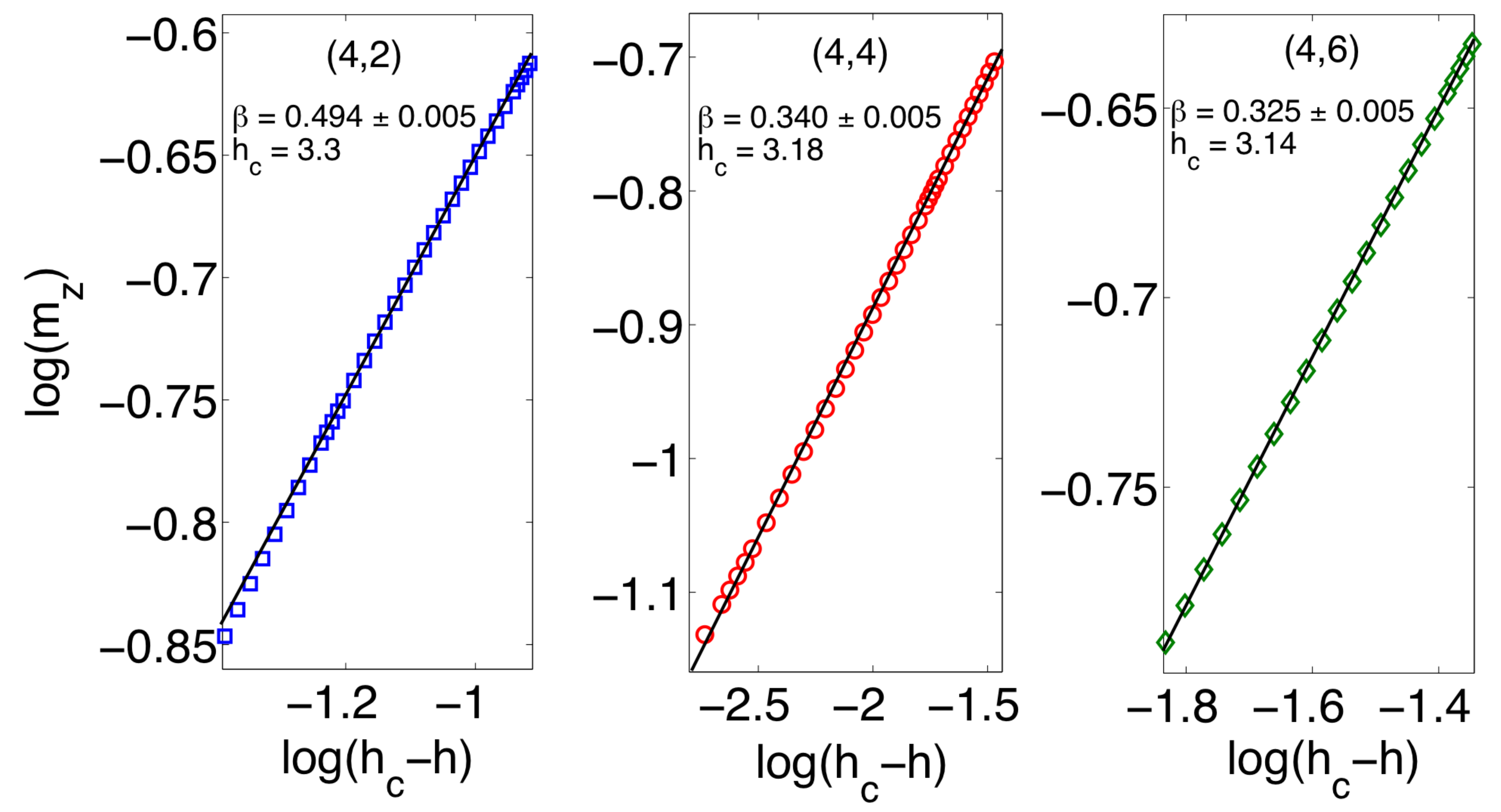}
\caption{(color online) Linear fits to the logarithm of $m_z$ around the critical region of Fig.(\ref{plot3}) to extract the critical exponent $\beta$. The different truncation parameters as well as the estimated values of $\beta$ and the critical point $h_c$ are indicated inside the plots. These fits are done at a region sufficiently close to criticality, but also slightly far from it, in such a way that the inherent mean-field effective behavior of the TN ansatz for too large correlation lengths is not observed \cite{anders}.} 
\label{plot4}
\end{figure}

Let us stress that our $2d$ results are based on the simplest possible $2d$ algorithm from this paper. At this point,  we wish to remind that this algorithm is sort of an over-simplified method because the truncating isommetries are not computed from reduced density matrices at all. Thus, much better accuracies in the critical properties of the system are expected if instead one implements some of the 'full' methods from Appendix A.c. Therefore, the numerical results in this paper should be considered only as indicative, and not as representative of the best possible performance achievable by corner methods for $2d$ problems. Nevertheless, we feel that it is quite encouraging that such an over-simplified approach is already able to capture the essential properties of the system within some accuracy. The numerical performance of some of the full $2d$ methods from Appendix A.c will be considered in a future work \cite{fut}. 

\begin{table}
\begin{center}
\begin{tabular}{||c||c||c||}
\hline
{\bf Method} & $\beta$ & $h_c$ \\
 \hline
 \hline
 Mean Field Theory& 0.5 & 4 \\ 
 \hline
Quantum Montecarlo\cite{IsingMC} & ~~0.327~~ & ~~3.044~~ \\
\hline
$D=3$ VDMA \cite{VDMA1, VDMA2} & - & 3.2 \\
\hline
$D=3$ MPS + iPEPS \cite{iPEPS} & 0.332 & 3.06 \\
\hline
$D=3$ CTM + iPEPS \cite{iPEPS2} & 0.328 & 3.04 \\ 
\hline
$D=2$ TERG \cite{TERG} & 0.333 & 3.08 \\
\hline
~~~$D=4,\chi=4$ simp. one-dir. $2d$~~~& 0.325 & 3.14   \\ 
\hline
\end{tabular} 
\end{center}
\caption{Critical exponent $\beta$ and critical point $h_c$ of the $2d$ quantum Ising model in a transverse magnetic field, as computed by different methods. In our case, $h_c$ has been estimated as the point at which $m_z$ becomes roughly $10^{-2}$.}
\label{tabd}
\end{table}

\section{Conclusions and final remarks}

In this paper we have explored the practical use of CTMs and corner tensors to develop tensor network algorithms for the classical simulation of quantum lattice systems of infinite size. At every renormalization step, these methods try to minimize the trucation error by either (i) keeping the largest singular values of a CTM or corner tensor, or (ii) keeping the largest magnitude eigenvalues of some reduced density matrix. The second scheme maximizes the fidelity between the unrenormalized and renormalized reduced density matrices. In some cases, e.g. $1d$ systems with some extra symmetries, we have seen that the truncation in (i) is equivalent to that in (ii). 

We have focused mainly on ground states of $1d$ and $2d$ systems, although we have also discussed briefly other possibilities ($3d$ systems, periodic boundary conditions, and real time evolution). The methods that have been proposed preserve the spatial symmetries of the system, including invariance under translations. We have benchmarked some of this methods by numerically computing several ground state properties of the ferromagnetic spin-$1/2$ quantum Ising model in $1d$ and $2d$. 
These numerics, which should be regarded as preliminary, are already quite encouraging. The algorithms of this paper could be a possible alternative to other well-established ways to compute ground state properties of quantum lattice systems in the thermodynamic limit, such as iDMRG and iTEBD in $1d$, and iPEPS and TERG in $2d$. The computational complexity of the proposed algorithms has also been analized, and we have seen that in $2d$ it depends strongly on the simulation scheme, leading to differences in several orders of magnitude. 

From a broader perspective, it would be interesting to understand the differences between explicit and implicit truncations in the rank of the CTMs. In particular, it would be good to know whether one type of truncation is, by construction, more accurate than the other. In a way, this could be similar to the difference between conventional DMRG and single-site DMRG \cite{single2}, where single-site DMRG seems to produce more accurate data with the same computational resources. 

The methods explored in this paper can also be generalized easily to deal with invariance under translations every two (or more) lattice sites. In such a case, one just needs to choose the tensors accordingly in such a way that everything is compatible with the unit cell of the lattice. The different renormalizations need also to be implemented in accordance with the lattice periodicity. Finally, let us mention that these algorithms can also be used in the context of further tensor network generalizations in order to study e.g. systems with internal symetries \cite{sym1, sym2, sym3, sym4, sym5, sym6, sym7, sym8, sym9} and fermionic quantum lattice systems \cite{ferm1, ferm2, ferm3, ferm4, ferm5, ferm6, ferm7, ferm8}. We believe that the methods of this paper will be useful for the practical implementation and development of further tensor network algorithms in the future.

{\bf Acknowledgements}
R. O.  acknowledges T. Nishino for a critical reading of this manuscript as well as for providing many insightful comments and important references. EU is also acknowledged for funding through a Marie Curie IIF.

\appendix

\section{Full $1d$ and $2d$ methods}

\subsubsection{Full one-directional $1d$ method}

\begin{figure}
\includegraphics[width=0.5\textwidth]{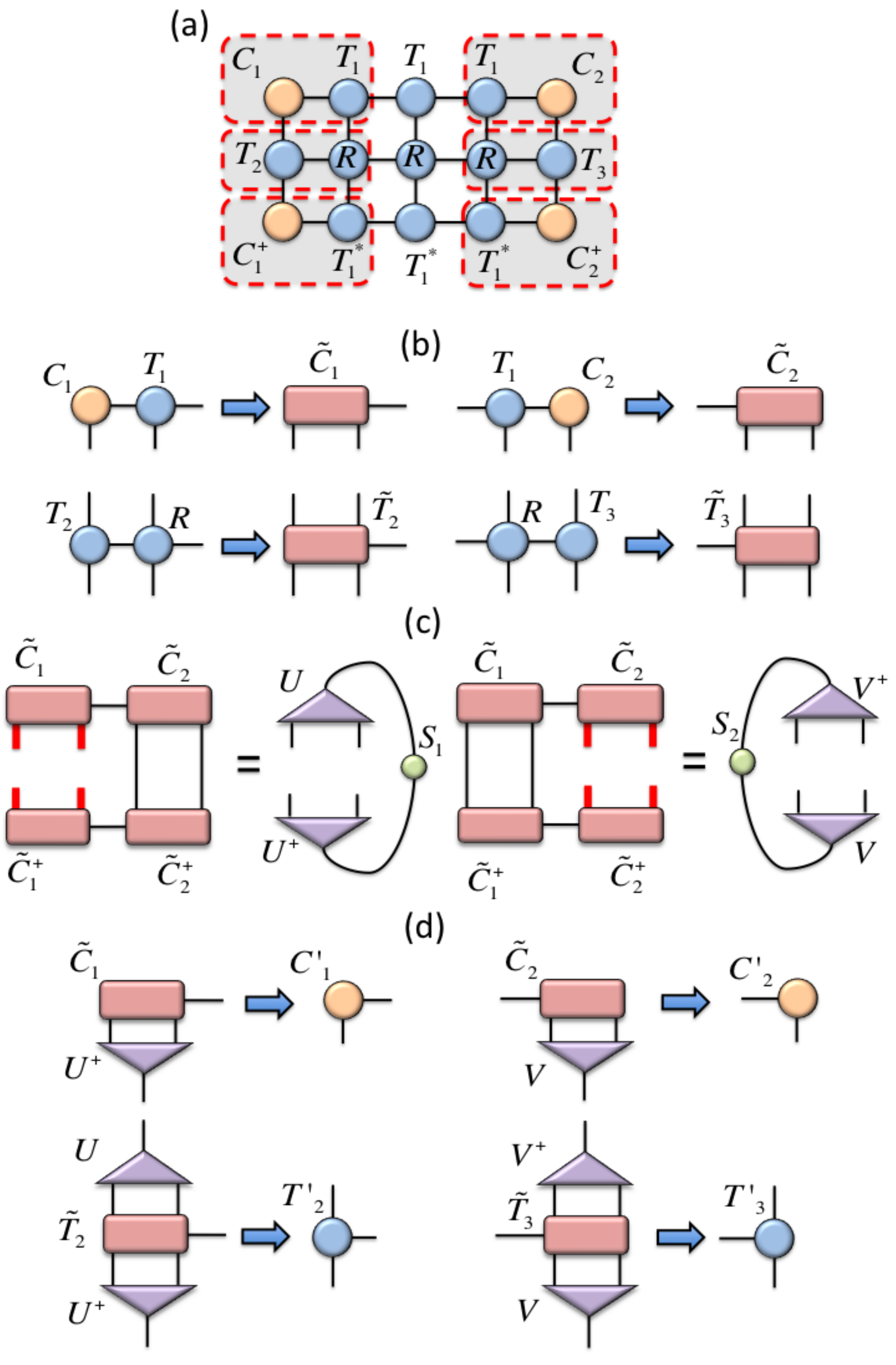}
\caption{(color online) $x$-move for the full one-directional $1d$ method, see text. Open indices in reduced density matrices are shown in red.} 
\label{fig14}
\end{figure}

\begin{figure}
\includegraphics[width=0.5\textwidth]{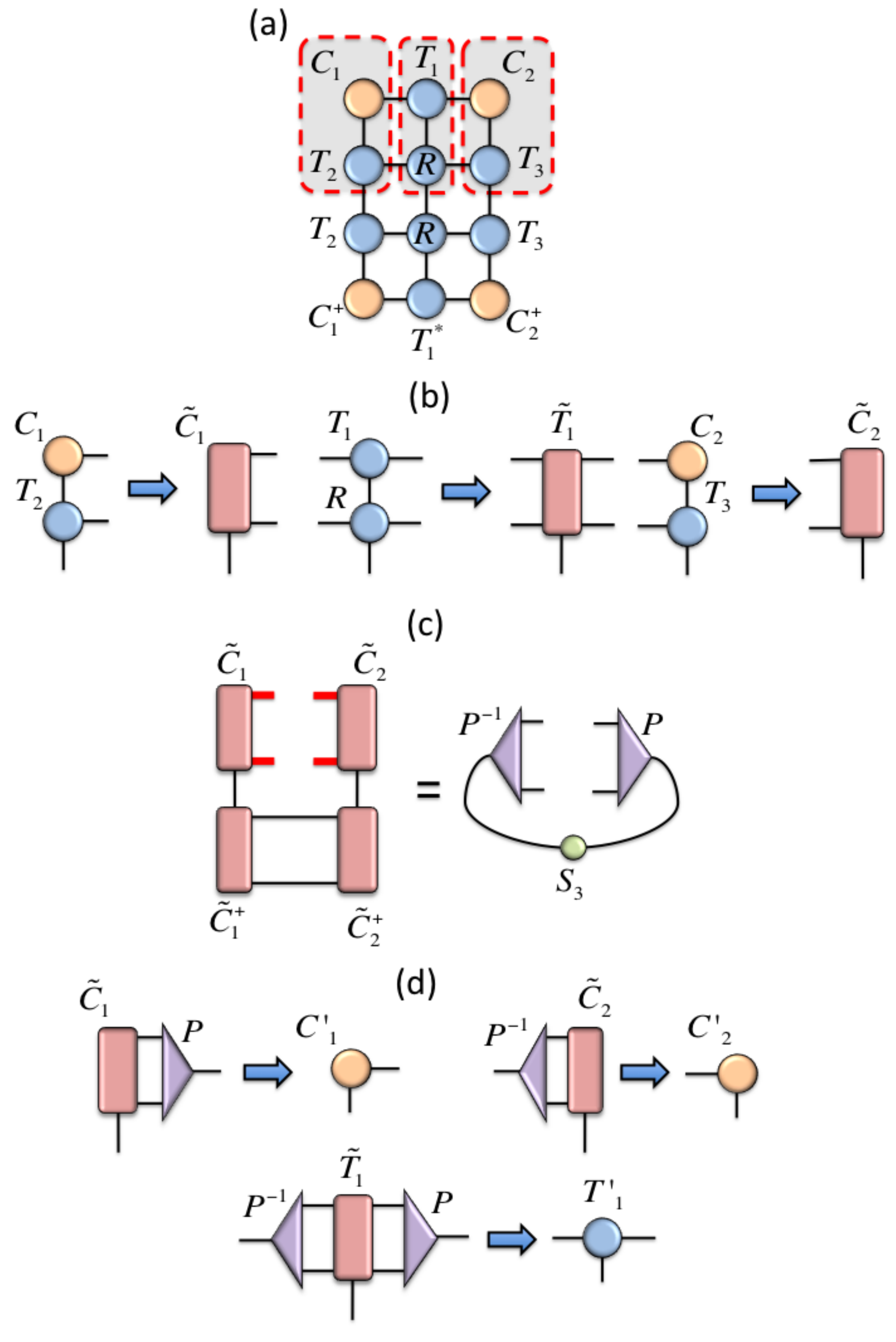}
\caption{(color online) $y$-move for the full one-directional $1d$ method, see text. Open indices in reduced density matrices are shown in red.} 
\label{fig15}
\end{figure}

In this algorithm the only existing symmetry in the TN is the vertical transposition of the tensors plus complex conjugate, and also translation invariance. Thus, this time the relevant TN is represented by two CTMs $C_1$ and $C_2$, two half-row transfer matrices $T_2$ and $T_3$, and one half-column transfer matrix $T_1$, see Fig.(\ref{fig11}.c). Notice that tensors $T_1, T_2$ and $T_3$ can again be interpreted as the tensors of three MPS respectively in the horizontal and vertical directions (in the vertical direction there is one for the left hand side, and one for the right hand side). The general idea now is the same as in the previous algorithm: insert, absorb and renormalize. However, since there are less symmetries than in the previous method, the renormalization needs to be done in a different way. Here are the main steps of the algorithm: 

\begin{enumerate}
\item{{\bf $x$-move:}}
\begin{enumerate}
\item{\underline{Insertion}. We insert two new columns in the TN, as shown in Fig.(\ref{fig14}.a).}
\item{\underline{Absorption}. We absorb one new column towards the left, and one new column towards the right, as shown in Fig.(\ref{fig14}.a,b). At this step we produce four new (unrenormalized) tensors $\widetilde{C}_1, \widetilde{C}_2, \widetilde{T}_2$ and $\widetilde{T}_3$.}
\item{\underline{Renormalization}. This is done by means of two isommetries $U$ and $V$ as shown in Fig.(\ref{fig14}.d). This time, these isommetries are computed by calculating the \emph{eigenvalue decomposition} of the reduced density matrices in Fig.(\ref{fig14}.c). We have that $U$ renormalizes tensors $\widetilde{C}_1$ and $\widetilde{T}_2$, whereas $V$ renormalizes $\widetilde{C}_2$ and $\widetilde{T}_3$, see Fig.(\ref{fig14}.d). It is important that $U$ and $V$ are different, since the TN does not have extra symmetries in the horizontal direction.} 
\end{enumerate}
\item{{\bf $y$-move:}}
\begin{enumerate} 
\item{\underline{Insertion}. We insert one new row, as shown in Fig.(\ref{fig15}.a).}
\item{\underline{Absorption}. We absorb the row towards up, as in Fig.(\ref{fig15}.a,b). At this tep we produce three new (unrenormalized) tensors $\widetilde{C}_1, \widetilde{C}_2$ and $\widetilde{T}_1$.}
\item{\underline{Renormalization}. This is done as in Fig.(\ref{fig15}.d) by means of a matrix $P$. This matrix is computed by finding the \emph{eigenvalue decomposition} of the reduced density matrix in Fig.(\ref{fig15}.c). Importantly, this time $PP^{\dagger}$ is not equal to the $\chi \times \chi$ identity matrix, so we must do the renormalization using $P$ and its inverse $P^{-1}$. Notice, however, that the rank of the reduced density matrix in Fig.(\ref{fig15}.c) is always $\chi$. Thus, $P$ is a rectangular matrix, and $P^{-1}$ is the \emph{pseudoinverse} of $P$. That is, if the singular value decomposition of $P$ is given by $P = FSG^{\dagger}$, then its pseudoinverse is defined as $P^{-1} \equiv GS^{-1}F^{\dagger}$.}
\end{enumerate}
\end{enumerate}

The method proceeds again by iterating the above $x$- and $y$-moves for as long as required until e.g. convergence of the eigenvalue spectrums of the reduced density matrices. Importantly, we can apply this algorithm to compute evolutions by MPOs that are not symmetric. Regarding complexity, this algoithm is again $O(\chi^3)$, with $\chi$ the rank of the reduced density matrices. Also, similar conclusions as in the simplified one-directional $1d$ method apply for parameter $\chi$ in this algorithm: it does not grow throughout the evolution and is fixed from the beginning of the algorithm. Finally, in the case of having an MPO invariant under horizontal transposition plus complex conjugation, this algorithm is just equivalent to the simplfied directional method, although with a less efficient implementation. 

\subsubsection{Full two-directional $1d$ method: CTMRG revisited}

\begin{figure}
\includegraphics[width=0.5\textwidth]{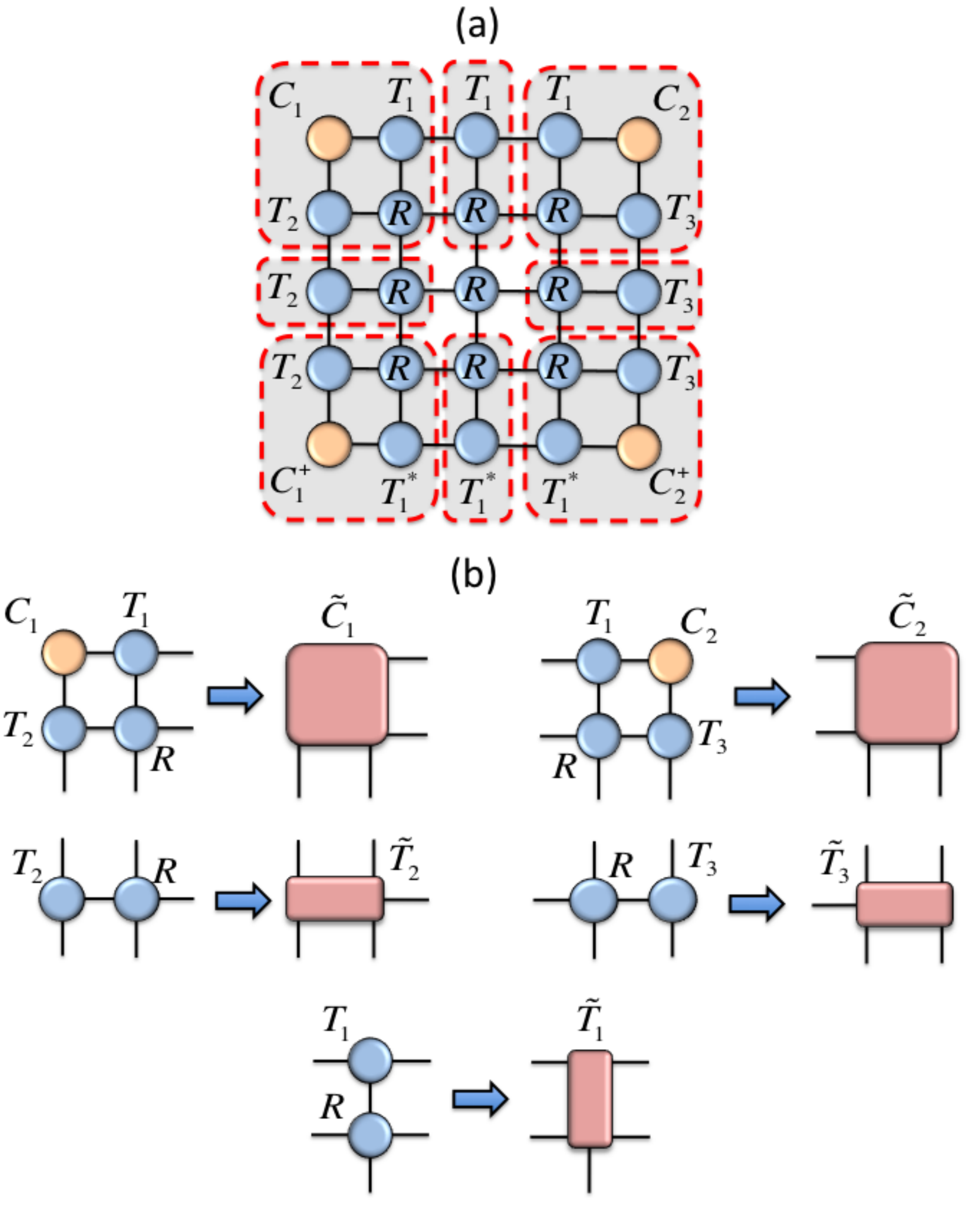}
\caption{(color online) Full two-directional $1d$ method: absorptions, see text. Open indices in reduced density matrices are shown in red.} 
\label{fig16}
\end{figure}

The following algorithm does no longer use a directional approach with $x$- and $y$-moves. Instead, it uses a complete 'radial'-move, expanding both directions of the TN at the same time. It is in fact the CTMRG method \cite{CTMRG1, CTMRG2} but adapted to our case (time evolution of a $1d$ quantum lattice system driven by an MPO). As in the previous section, the system is assumed not to have any spatial symmetry in the horizontal direction (apart from translation invariance). Thus, the relevant TN is described again by two CTMs $C_1$ and $C_2$, two half-row TM $T_2$ and $T_3$, and one half-column TM $T_1$, as in Fig.(\ref{fig11}.c). Again, the half-row and half-column TMs can be interpreted as defining three different MPSs. The details of the method are as follows:

\begin{figure}
\includegraphics[width=0.4\textwidth]{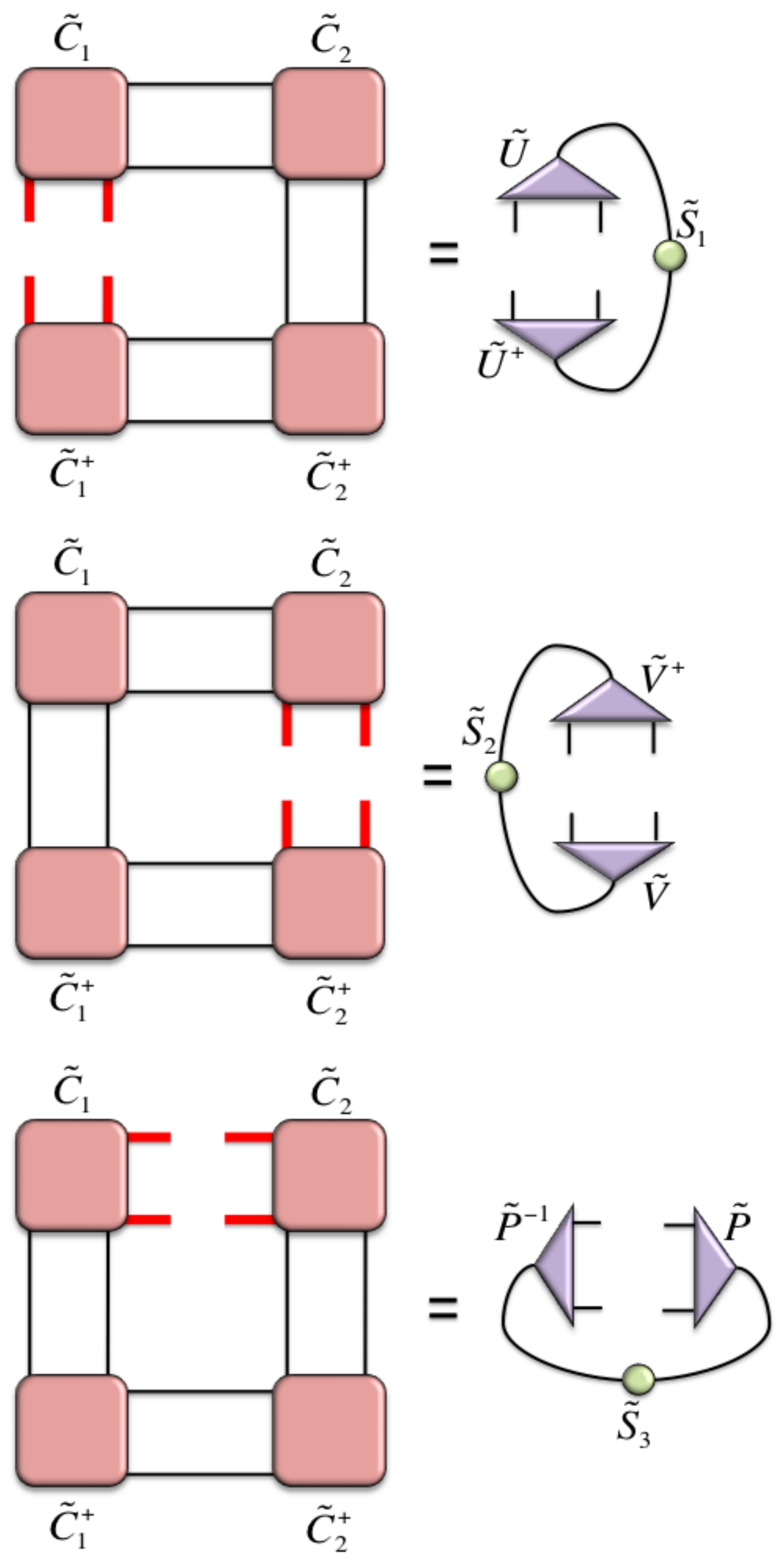}
\caption{(color online) Full two-directional $1d$ method: reduced density matrices, see text. Open indices in reduced density matrices are shown in red.} 
\label{fig17}
\end{figure}
\begin{figure}
\includegraphics[width=0.5\textwidth]{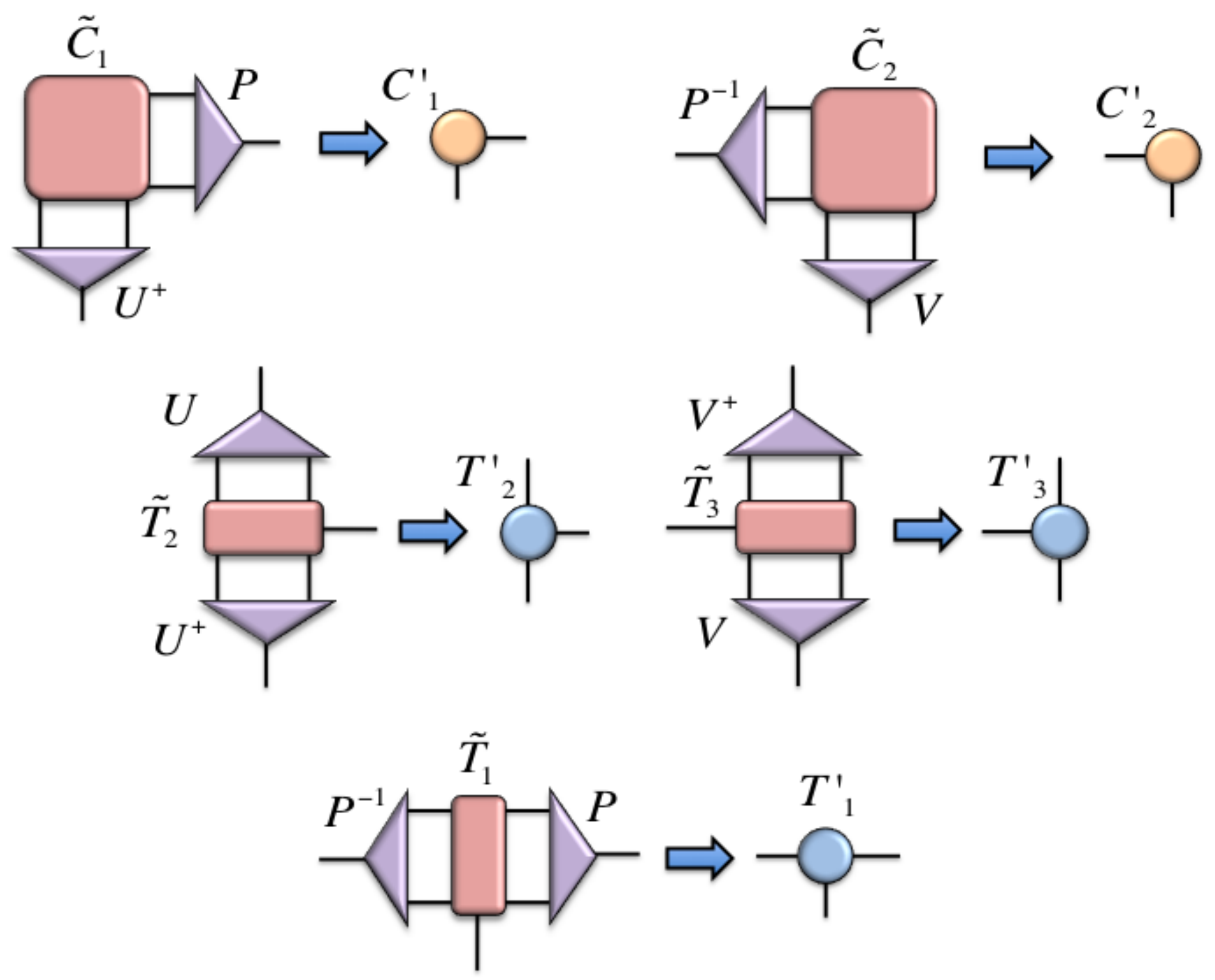}
\caption{(color online) Full two-directional $1d$ method: renormalizations, see text.} 
\label{fig18}
\end{figure}

\begin{enumerate}[(a)]
\item{\underline{Insertion}. We insert two new rows and two new columns, as shown in Fig.(\ref{fig16}.a)}. 
\item{\underline{Absorption}. We absorb the new tensors towards the corners by computing the two (unrenormalized) corner transfer matrices $\widetilde{C}_1$ and $\widetilde{C}_2$, half-row transfer matrices $\widetilde{T}_2$ and $\widetilde{T}_3$ and half-column transfer matrix $\widetilde{T}_1$ as in Fig.(\ref{fig16}.b).}
\item{\underline{Renormalization}. Compute the reduced density matrices from Fig.(\ref{fig17}), and perform their \emph{eigenvalue decomposition}. From these decompositions, obtain the two unitary matrices $\widetilde{U},\widetilde{V}$ and the matrix $\widetilde{P}$ as well as their inverses $\widetilde{U}^{\dagger}, \widetilde{V}^{\dagger}$ and $\widetilde{P}^{-1}$. Truncate explicitely in the $\chi$ largest eigenvalues in magnitude for each one of these decompositions, and obtain the matrices $U,V,P$ and their pseudoinverses. Then, renormalize the CTMs, half-row transfer matrices and half-column transfer matrix as shown in Fig.(\ref{fig18}).}
\end{enumerate}

These steps are again repeated until convergence of the spectrum of the reduced density matrices. Notice that this time the rank of the CTM grows at each iteration of the method and hence needs to be truncated explicitely in $\chi$ at every step. The leading number of operations for this algorithm is again $O(\chi^3)$. However, the subleading and multiplicative corrections are higher than in the previous two methods, and hence the total running time is also higher.

\subsubsection{Full one-, two- and three-directional $2d$ methods}

From a generic perspective, the 'full' methods in $2d$ have several things in common. First, the PEPO does not need to be hermitian in the spatial directions, so we will deal with TNs like the one in Fig.(\ref{fig20}.c). Second,  the renormalizations are implemented by tensors found from the \emph{eigenvalue decomposition} of the relevant reduced density matrix of the indices that one wishes to truncate. Third, truncations in the $z$ direction are implemented by isommetries, whereas in the $x$ and $y$ directions one uses rectangular matrices and their pseudoinverses. And fourth, and unlike in $1d$, now explicit truncations are always needed.

\begin{figure}
\includegraphics[width=0.5\textwidth]{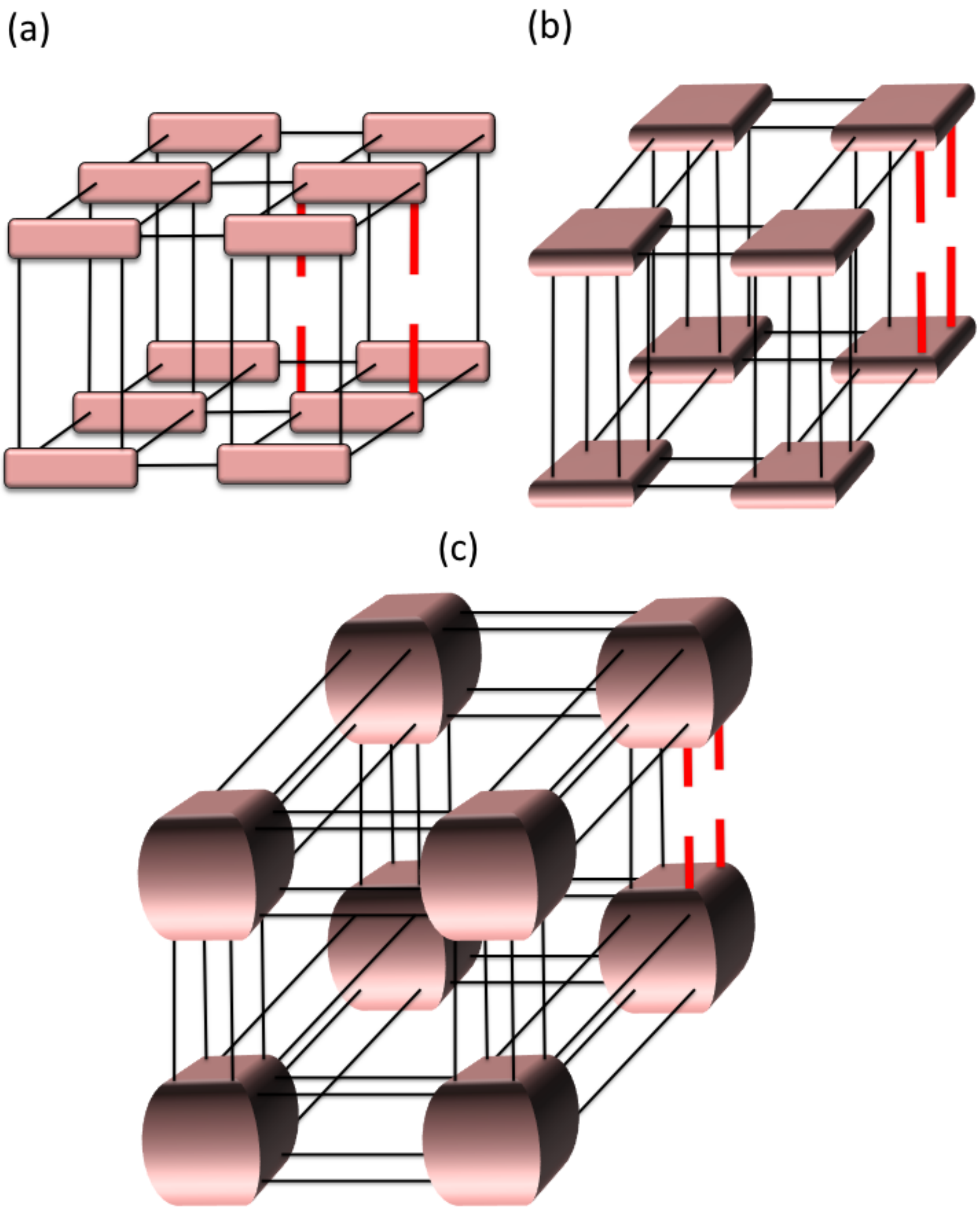}
\caption{(color online) Different reduced density matrices for two indices that appear at some step in the (a) full one-directional, (b) full two-directional and (c) full three-directional $2d$ methods. Reduced density matrices for four neighbouring indices  from the corners may also be considered in the cases (b) and (c), see text. Open indices are shown in red.} 
\label{fig23}
\end{figure}

There are many different ways of puting the above ideas in practice, see Fig.(\ref{fig23}). For instance, one could think of a full one-directional approach, similar to the simplified one, but where the tensors that implement the truncations are computed from the reduced density matrices as in Fig.(\ref{fig23}.a). Another approach is a full two-directional approach, which keeps the idea of the full two-directional method in $1d$ but being restricted to the planes in the $(2+1)d$ lattice. In this approach, the relevant reduced density matrices are as in Fig.(\ref{fig23}.b). In fact, one could even think of combining the one-directional approach in one direction with the two-directional approach in the other two directions, giving rise to another possible way of expanding the system. Finally, one can think of a full three-directional approach, where the reduced density matrices look as in Fig.(\ref{fig23}.c). This last approach is the most direct generalization of the CTMRG algorithm in Ref. \cite{CTMRG1, CTMRG2} to higher dimensions. However, this is also the most inneficient alternative. Notice also that in Fig.(\ref{fig23}) we consider only reduced density matrices of two indices. However, it would also be possible to consider reduced density matrices of four neighbouring indices in the same direction within a corner in the full two- and three-directional 2d methods (not shown). Such an approach would of course be less efficient, but also more accurate. 

We expect stability and numerical accuracy of these methods to increase as the reduced density matrix captures more relevant correlations. But the price that one has to pay is the growth in complexity which, in practice, makes it very difficult to implement numerical simulations in some cases. Assuming calculations of reduced density matrices for two indices (and thus two-index truncations), we see that the full one-directional approach has complexity $O(\chi^{11})$, the full two-directional approach $O(\chi^{13})$, and the full three-directional approach $O(\chi^{17})$. This is to be compared with the simplified one-directional method from the previous section, which has complexity $O(\chi^7)$. 

These approaches are expected to produce more accurate results than the simplified one-directional approach. However, in this paper we have just implemented numerically the simplfied one-directional $2d$ approach and seen that this already produces sensible results, see Sec.IV. Thus, we expect these full methods to work even better, despite one has to pay a price in the number of operations required for the different contractions. The numerical exploration of some of these methods will be presented elsewhere \cite{fut}. 

\section{Further posibilities}

We now discuss briefly some other possibilities for corner-inspired methods, namely, $3d$ quantum lattice systems, periodic boundary conditions, and real time evolution.  

\subsubsection{$3d$ quantum lattice systems}

$3d$ quantum lattice systems (or equivalently, $4d$ classical lattice systems) are also of relevance for the study of important physical phenomena, e.g. the emergence of fermions and gauge bosons in $3d$ string-net models \cite{stringnet}, or confinement of quarks in Quantum Chromodynamics. In principle, it should be possible to generalize all the approaches discussed so far in this paper to the $3d$ case. However, from our experience with the $2d$ case, we expect that the complexity of the algorithms will be quite large and thus they may become unpractical. Yet, another possibility to deal with a $3d$ quantum lattice system would be to follow the same idea as in Ref. \cite{iPEPS2} for the $2d$ case, namely, to use some $2d$ method with corner tensors such as the one described in Appendix A.c to approximate the environment of a $3d$ iPEPS in the context of a $3d$ iPEPS algorithm. In fact, this would possibly lead to a quite efficient algorithm for studying $3d$ quantum lattice systems if combined with a simplified update of the tensors \cite{Xiang}. For instance, if the full one-directional $2d$ method is used, then we would have a $3d$ method of complexity $O(\chi^{11})$, which is certainly doable. As in Ref. \cite{iPEPS2}, such an approach may break the traslational invariance by one lattice site of the lattice, but it would be straightforward to generalize our methods accordingly. The numerical exploration of this approach will also be presented elsewhere \cite{fut}. 

\begin{figure}
\includegraphics[width=0.5\textwidth]{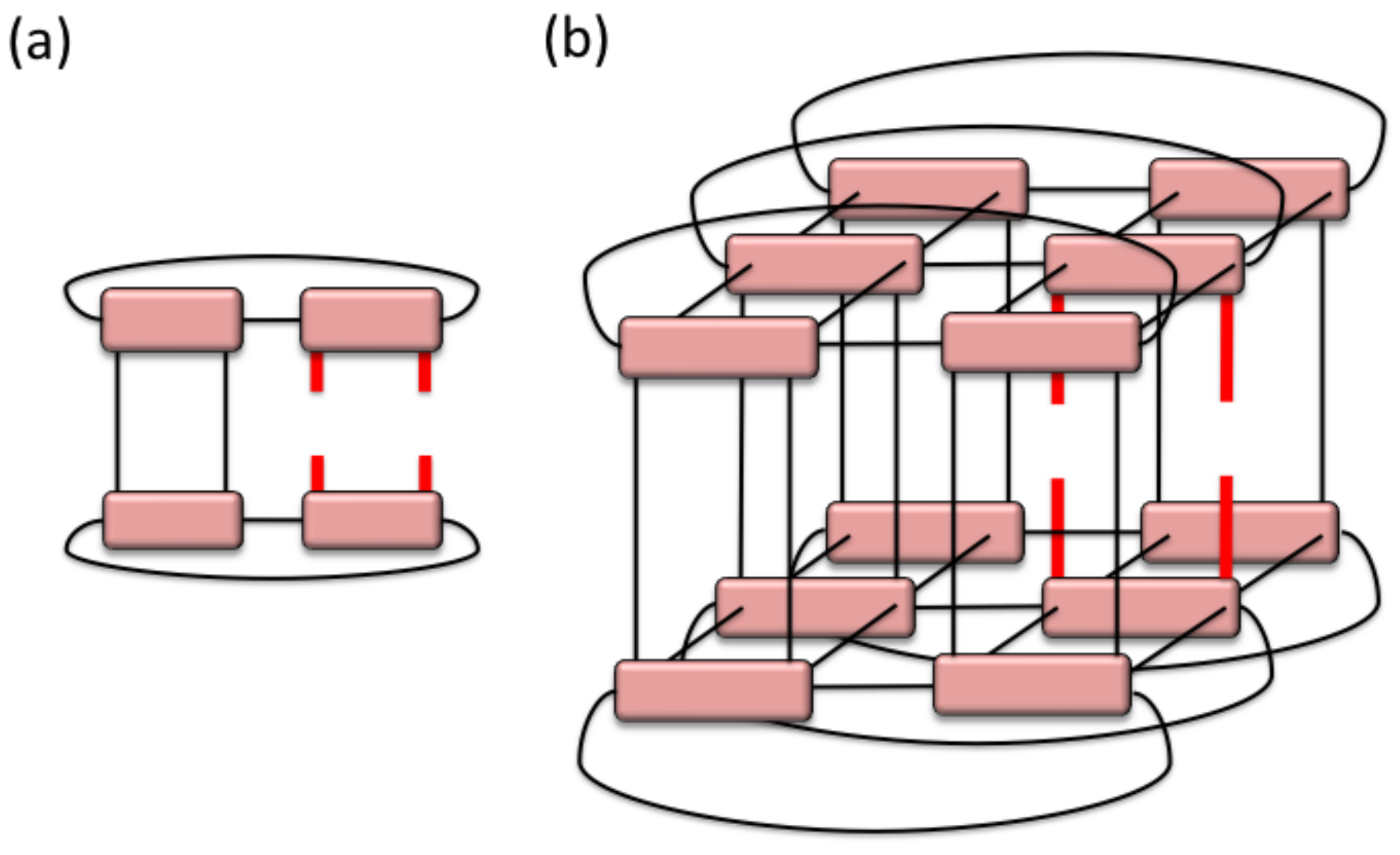}
\caption{(color online) Examples reduced density matrices for e.g. full one-directional methods with periodic boundary conditions in (a) $1d$ and (b) $2d$. In the $2d$ case periodic boundary conditions are assumed in one spatial direction only, but these could also be in both spatial directions. If necessary, further periodic boundary conditions in the temporal (vertical) direction could also be considered. Open indices are shown in red.} 
\label{fig25}
\end{figure}

\subsubsection{Periodic boundary conditions}

Indeed, the case of periodic boundary conditions can be considered as well in the context of full algorithms, where the renormalizing matrices are computed by eigenvalue decompositions of the relevant reduced density matrices. This is so since, in the end, periodic boundary conditions just add some extra indices to the TN diagram that needs to be contracted, see e.g. Fig.(\ref{fig25}). This means that the complexity of the corresponding algorithm will increase (such an increase in complexity is well-known in the DMRG community, see e.g. Ref.\cite{dmrg3, dmrg4} and also Ref.\cite{pbc1, pbc2}). From the point of view of the contraction of a $2d$ TN this provides an alternative to the method in Ref. \cite{levin}. Let us also remark that the case of periodic boundary conditions has already been considered in the literature in the context of CTMRG, see e.g. Ref. \cite{perCTM}. Periodic boundary conditions in the temporal direction may also be useful to compute thermal properties, see e.g. Ref.\cite{Thermal1, Thermal2, Thermal3, Thermal4, Thermal5}. 

\subsubsection{Real time evolution: iTEBD and iPEPS reinterpreted}

\begin{figure}
\includegraphics[width=0.5\textwidth]{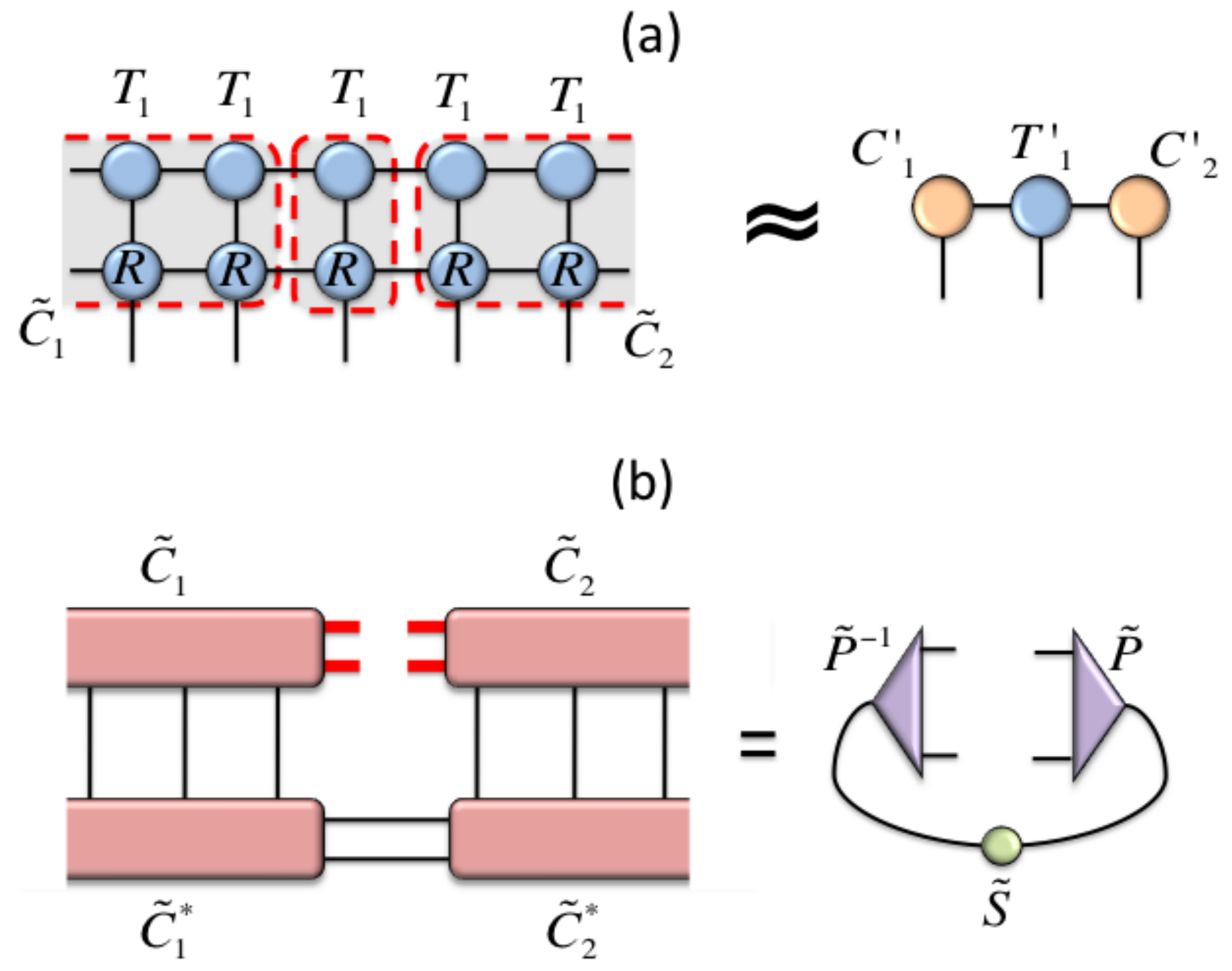}
\caption{(color online) (a) The action of an MPO over an MPS can be approximated by e.g. three renormalized tensors.  (b) Reduced density matrix of the bond index of the MPS obtained after applying the MPO, in terms of four CTMs. Open indices in the reduced density matrix are shown in red.} 
\label{fig26}
\end{figure}

So far we have discussed the problem of finding ground states, which is done by imaginary time evolution. However, one may wonder whether the techniques and methods explained here are of any use to compute also the real time evolution of a quantum lattice system, and hence its dynamical properties. We discuss this in what follows. 

In a real time evolution, the MPO or PEPO driving the evolution is usually only hermitian with respect to the vertical (time) indices. Thus, if one wished to implement a time-evolution method based on CTMs and corner tensors, 'full' methods (or variations thereof) should be considered. Moreover, in a real time evolution the total running time is finite (unlike in an imaginary time evolution for finding ground states). Therefore, we need to deal with a 'full' approach such that the time direction can grow \emph{independently} of the spatial directions. The natural options for algorithms are thus the full one-directional approaches from Appendix A.a,b, or in $2d$ also a combination of the full one-directional $2D$ approach in the time direction and the full two-directional $2d$ approach in the space directions (Appendix A.c). 

For concreteness let us focus on the $1d$ case, for which some variation of the full one-directional approach from Appendix A.a may be used. Based on the above considerations, one may be tempted to propose the followiing real-time evolution algorithm: at a given step, (i) repeat \emph{many} $x$-moves until convergence (effectively making the size of the system infinite), and (ii) perform \emph{one} $y$-move. This can be understood as approximating the action of an MPO over an MPS by means of three tensors, see Fig.(\ref{fig26}.a). Then, just repeat these operations for as many (finite) number of steps as required. 

The above approach is indeed a possibility. However, numerical simulations show that its accuracy at long times is not as good as the one that can be obtained by using e.g. the standard iTEBD method (results not shown). This is understandable, since the truncations implemented in this approach are more stringent than the ones in iTEBD. 
 
Let us now think of a different approach: we could think again on the action of the MPO over the MPS in terms of CTMs. Then, the aim is to truncate in the horizontal indices by means of some tensor. Following the spirit of the full algorithms described earlier, this can done by computing the reduced density matrix of the indices by means of the CTMs $\widetilde{C}_1$ and $\widetilde{C}_2$, see Fig.(\ref{fig26}.b). Then, we just find its eigenvalue decomposition. With this procedure we obtain a matrix $\widetilde{P}$ and its inverse $\widetilde{P}^{-1}$. A further truncation in the $\chi$ largest eigenvalues in magnitude produces the rectangular matrices $P$ and $P^{-1}$, which we can use to renormalize the horizontal indices of the tensors. 

Remarkably, this procedure is \emph{nothing but a reformulation of the the standard iTEBD algorithm for non-unitary evolutions} \cite{iTEBD1, iTEBD2}. To see this, notice that  iTEBD proceeds by (i) finding the canonical form of the resultant MPS after the MPO has been applied, and (ii) truncating its bond index in the largest $\chi$ Schmidt coefficients. It is easy to prove that for an MPS in canonical form, the reduced density matrix in Fig.(\ref{fig26}.b) is already diagonal. In our laguage, this diagonalization is actually implemented by the matrices $P$ and $P^{-1}$. Introducing these tensors in the TN actually orthonormalizes and truncates the bond indices of the evolved MPS. And this is exactly what the iTEBD algorithm does. Also, it is not difficult to imagine that if we generalize the same procedure to $2d$ systems, then we obtain nothing but a reformulation of the usual iPEPS algortihm but using PEPOs for the time evolution. In such a case, the relevant reduced density matrices can not be computed exactly (since they amount to the contraction of a $2d$ TN) and must be approximated somehow. Simple approximations lead to efficient algorithms, whereas more elaborate approximations are less efficient but also more accurate. 

Let us remark that the reinterpretation that we just found of the iTEBD algorithm in terms of CTMs is in fact similar to the interpretation of DMRG also in terms of CTMs, see Ref.\cite{CTMRG1, CTMRG2}.

\end{document}